\documentclass[twocolumn]{aastex631} 
\usepackage{CLASSY_Preamble}
\newcommand{\mpfit}{MPFIT}
\newcommand{\ciii}{\ion{C}{3}]}
\newcommand{\fciii}{[\ion{C}{3}]}
\newcommand{\civ}{\ion{C}{4}}
\newcommand{\lya}{Ly$\alpha$}
\newcommand{\heii}{\ion{He}{2}}
\newcommand{\oii}{[\ion{O}{2}]}
\newcommand{\oiii}{[\ion{O}{3}]} 
\newcommand{\oiiiuv}{\ion{O}{3}]}
\newcommand{\feiii}{[\ion{Fe}{3}]}
\newcommand{\feii}{\ion{Fe}{2}}
\newcommand{\hii}{\ion{H}{2}}
\newcommand{\hei}{\ion{He}{1}}
\newcommand{\ha}{H$\alpha$}
\newcommand{\hb}{H$\beta$}
\newcommand{\hg}{H$\gamma$}
\newcommand{\hd}{H$\delta$}
\newcommand{\siii}{[\ion{S}{3}]}
\newcommand{\nii}{[\ion{N}{2}]}
\newcommand{\niii}{[\ion{N}{3}]}
\newcommand{\niv}{\ion{N}{4}]}
\newcommand{\sii}{[\ion{S}{2}]}
\newcommand{\fsiiii}{[\ion{Si}{3}]}
\newcommand{\siiii}{\ion{Si}{3}]}
\newcommand{\oi}{[\ion{O}{1}]}
\newcommand{\mgii}{\ion{Mg}{2}}
\newcommand{\cliii}{[\ion{Cl}{3}]}
\newcommand{\ariv}{[\ion{Ar}{4}]}
\newcommand{\ariii}{[\ion{Ar}{3}]}
\newcommand{\neiii}{[\ion{Ne}{3}]}
\newcommand{\cloudy}{Cloudy}
\newcommand{\pyneb}{\texttt{PyNeb}}
\newcommand{\python}{\texttt{PYTHON}}

\begin{document}

 \accepted{09/20/2022 on ApJ}
\shortauthors{Mingozzi et al.}
\shorttitle{CLASSY~IV}

\title{CLASSY IV: Exploring UV diagnostics of the interstellar medium in local high-$z$ analogs at the dawn of the JWST era \footnote{
Based on observations made with the NASA/ESA Hubble Space Telescope,
obtained from the Data Archive at the Space Telescope Science Institute, which
is operated by the Association of Universities for Research in Astronomy, Inc.,
under NASA contract NAS 5-26555.}}

\author[0000-0003-2589-762X]{Matilde Mingozzi}
\affiliation{Space Telescope Science Institute, 3700 San Martin Drive, Baltimore, MD 21218, USA}

\author[0000-0003-4372-2006]{Bethan L. James}
\affiliation{AURA for ESA, Space Telescope Science Institute, 3700 San Martin Drive, Baltimore, MD 21218, USA}

\author[0000-0002-2644-3518]{Karla Z. Arellano-C\'{o}rdova}
\affiliation{Department of Astronomy, The University of Texas at Austin, 2515 Speedway, Stop C1400, Austin, TX 78712, USA}

\author[0000-0002-4153-053X]{Danielle A. Berg}
\affiliation{Department of Astronomy, The University of Texas at Austin, 2515 Speedway, Stop C1400, Austin, TX 78712, USA}

\author[0000-0002-9132-6561]{Peter Senchyna}
\affiliation{Carnegie Observatories, 813 Santa Barbara Street, Pasadena, CA 91101, USA}

\author[0000-0002-0302-2577]{John Chisholm}
\affiliation{Department of Astronomy, The University of Texas at Austin, 2515 Speedway, Stop C1400, Austin, TX 78712, USA}

\author[0000-0003-4359-8797]{Jarle Brinchmann}
\affiliation{Instituto de Astrof\'{i]}sica e Ci\^{e}ncias do Espa\c{c}o, Universidade do Porto, CAUP, Rua das Estrelas, PT4150-762 Porto, Portugal}

\author[0000-0003-4137-882X]{Alessandra Aloisi}
\affiliation{Space Telescope Science Institute, 3700 San Martin Drive, Baltimore, MD 21218, USA}

\author[0000-0001-5758-1000]{Ricardo O. Amor\'{i}n}
\affiliation{Instituto de Investigaci\'{o}n Multidisciplinar en Ciencia y Tecnolog\'{i}a, Universidad de La Serena, Raul Bitr\'{a}n 1305, La Serena 2204000, Chile}
\affiliation{Departamento de Astronom\'{i}a, Universidad de La Serena, Av. Juan Cisternas 1200 Norte, La Serena 1720236, Chile}


\author[0000-0003-3458-2275]{St\'{e}phane Charlot}
\affiliation{Sorbonne Universit\'{e}, CNRS, UMR7095, Institut d'Astrophysique de Paris, F-75014, Paris, France}




\author[0000-0001-6865-2871]{Anna Feltre}
\affiliation{INAF - Osservatorio di Astrofisica e Scienza dello Spazio di Bologna, Via P. Gobetti 93/3, 40129 Bologna, Italy}

\author[0000-0001-8587-218X]{Matthew Hayes}
\affiliation{Stockholm University, Department of Astronomy and Oskar Klein Centre for Cosmoparticle Physics, AlbaNova University Centre, SE-10691, Stockholm, Sweden}

\author[0000-0003-1127-7497]{Timothy Heckman}
\affiliation{Center for Astrophysical Sciences, Department of Physics \& Astronomy, Johns Hopkins University, Baltimore, MD 21218, USA}

\author[0000-0002-6586-4446]{Alaina Henry}
\affiliation{Space Telescope Science Institute, 3700 San Martin Drive, Baltimore, MD 21218, USA}
\affiliation{Center for Astrophysical Sciences, Department of Physics \& Astronomy, Johns Hopkins University, Baltimore, MD 21218, USA}

\author[0000-0003-4857-8699]{Svea Hernandez}
\affiliation{AURA for ESA, Space Telescope Science Institute, 3700 San Martin Drive, Baltimore, MD 21218, USA}



\author[0000-0002-5320-2568]{Nimisha Kumari}
\affiliation{AURA for ESA, Space Telescope Science Institute, 3700 San Martin Drive, Baltimore, MD 21218, USA}

\author[0000-0003-2685-4488]{Claus Leitherer}
\affiliation{Space Telescope Science Institute, 3700 San Martin Drive, Baltimore, MD 21218, USA}

\author[0000-0003-1354-4296]{Mario Llerena}
\affiliation{Instituto de Investigaci\'{o}n Multidisciplinar en Ciencia y Tecnolog\'{i}a, Universidad de La Serena, Raul Bitr\'{a}n 1305, La Serena 2204000, Chile}
\affiliation{Departamento de Astronom\'{i}a, Universidad de La Serena, Av. Juan Cisternas 1200 Norte, La Serena 1720236, Chile}

\author[0000-0001-9189-7818]{Crystal L. Martin}
\affiliation{Department of Physics, University of California, Santa Barbara, Santa Barbara, CA 93106, USA}

\author[0000-0003-2804-0648]{Themiya Nanayakkara}
\affiliation{Swinburne University of Technology, Melbourne, Victoria, AU}



\author[0000-0002-5269-6527]{Swara Ravindranath}
\affiliation{Space Telescope Science Institute, 3700 San Martin Drive, Baltimore, MD 21218, USA}



\author[0000-0003-0605-8732]{Evan D. Skillman}
\affiliation{Minnesota Institute for Astrophysics, University of Minnesota, 116 Church Street SE, Minneapolis, MN 55455, USA}




\author[0000-0001-6958-7856]{Yuma Sugahara}
\affiliation{Institute for Cosmic Ray Research, The University of Tokyo, Kashiwa-no-ha, Kashiwa 277-8582, Japan}
\affiliation{National Astronomical Observatory of Japan, 2-21-1 Osawa, Mitaka, Tokyo 181-8588, Japan}
\affiliation{Waseda Research Institute for Science and Engineering, Faculty of Science and Engineering, Waseda University, 3-4-1, Okubo, Shinjuku, Tokyo 169-8555, Japan}


\author[0000-0001-8289-3428]{Aida Wofford}
\affiliation{Instituto de Astronom\'{i}a, Universidad Nacional Aut\'{o}noma de M\'{e}xico, Unidad Acad\'{e}mica en Ensenada, Km 103 Carr. Tijuana-Ensenada, Ensenada 22860, M\'{e}xico}

\author[0000-0002-9217-7051]{Xinfeng Xu}
\affiliation{Center for Astrophysical Sciences, Department of Physics \& Astronomy, Johns Hopkins University, Baltimore, MD 21218, USA}

\correspondingauthor{Matilde Mingozzi} 
\email{mmingozzi@stsci.edu}


\begin{abstract}
The COS Legacy Archive Spectroscopic SurveY (CLASSY) HST/COS treasury program provides the first high-resolution spectral catalogue of 45 local high-z analogues in the ultra-violet (UV; $1200-2000$~\AA) to investigate their stellar and gas properties.
Here we present a toolkit of UV interstellar medium (ISM) diagnostics, analyzing the main emission lines of CLASSY spectra (\niv~$\lambda\lambda$1483,87, \civ~$\lambda\lambda$1548,51, \heii$\lambda$1640, \oiiiuv$\lambda\lambda$1661,6, \siiii~$\lambda\lambda$1883,92, \ciii~$\lambda$1907,9). Specifically, our aim is to provide accurate diagnostics for reddening $E(B-V)$, electron density $n_e$, electron temperature $T_e$, metallicity 12+log(O/H) and ionization parameter log($U$), taking into account the different ISM ionization zones. We calibrate our UV toolkit using well-known optical diagnostics, analyzing archival optical spectra for all the CLASSY targets.
We find that UV density diagnostics estimate $n_e$ values that are $\sim 1-2$~dex higher (e.g., $n_e$(\ciii$\lambda\lambda$1907,9)~$ \sim 10^4$~cm$^{-3}$) than those inferred from their optical counterparts (e.g., $n_e$(\sii$\lambda\lambda$6717,31)~$ \sim 10^2$~cm$^{-3}$; $n_e$(\ariv$\lambda\lambda$4714,41)~$ \sim 10^3$~cm$^{-3}$).
$T_e$ derived from the hybrid ratio \oiii~$\lambda$1666/$\lambda$5007 proves to be reliable, implying differences in {determining 12+log(O/H) compared to the optical counterpart \oiiiuv~$\lambda$4363/\oiii~$\lambda$5007} within $\sim \pm 0.3$~dex. 
We also investigate the relation between the stellar and gas $E(B-V)$, finding consistent values at high specific star formation rates ($\log(sSFR) \gtrsim -8$~yr$^{-1}$), while at low sSFRs we confirmed an excess of dust attenuation in the gas.
Finally, we investigate UV line ratios and equivalent widths to provide correlations with 12+log(O/H) and log($U$), but note there are degeneracies between the two. With this suite of UV-based diagnostics, we illustrate the pivotal role CLASSY plays in understanding the chemical and physical properties of high-z systems that JWST can observe in the rest-frame UV.

\end{abstract} 




\section{Introduction}\label{sec:intro}
The galaxies that host a substantial fraction of the star formation (SF) in the high-\emph{z} universe ($z\gtrsim6$) and likely play a key role in the reionization era tend to be compact, metal-poor, with a low-mass and large specific star formation rates (e.g., \citealt{wise14,robertson15,madau15,stark16,stanway16}). Deep rest-frame UV spectra of several of these high-redshift galaxies ($z\sim5-7$) already revealed prominent high-ionization nebular emission lines, such as \heii~$\lambda$1640, \oiiiuv~$\lambda\lambda1661,66$, \fciii~$\lambda$1907 and \ciii~$\lambda$1909 (\ciii\ hereafter) and \civ~$\lambda\lambda1548,1551$ (\civ\ hereafter; e.g., \citealt{stark15,mainali17,mainali18}).
In the upcoming era of the James Webb Space Telescope (JWST) and extremely large telescopes (ELTs), the UV spectroscopic frontier will be pushed to higher redshifts than ever before, finally revealing detailed rest-frame UV observations of statistically-significant samples of galaxies in the distant Universe. As such, the time to sharpen our understanding of UV nebular emission and exploit its diagnostic power is upon us.  

Far- and Near-Ultraviolet (FUV, $\sim1200-1700$ \AA; NUV, $\sim1700-2000$ \AA)  spectra can foster our understanding of star-forming galaxies in terms of the stellar populations hosting massive stars and their impact on interstellar medium (ISM) physical conditions, chemical evolution, feedback processes, and reionization. Due to the line production mechanisms alone, nebular UV emission can be use to \textit{directly} calculate the physical and chemical conditions under which they are produced.
For instance, both \ciii\ and \fsiiii~$\lambda$1883, \siiii~$\lambda$1892 (\siiii\ hereafter) doublets are direct tracers of electron density \citep{jaskot16,gutkin16,byler18}, the line intensity ratio \ciii/\oiiiuv$\lambda1666$ can be used to estimate the elemental carbon abundances \citep{garnett95,berg16,perez-montero17,berg18}, and \heii~$\lambda$1640 and the \civ/\ciii\ ratio both have the potential to constrain the level of ionization \citep{feltre16}. Moreover, the combination of all these UV lines can provide information about the nature of the ionizing sources in general \citep{feltre16,gutkin16,jaskot16,nakajima18}. UV emission lines therefore have the capacity to provide the community with a `diagnostic toolkit', with which we can directly diagnose the ISM properties in star-forming galaxies. 

Due to the intrinsic faintness of several UV emission lines, an alternative form of \textit{`in-direct'} diagnostics, which evolve from empirical calibrations between ISM conditions (e.g., metallicity) and properties of the stronger emission line properties (e.g., equivalent widths of \ciii), is also needed. To this end, several past studies have taken big steps forward in the interpretation of UV emission in the local Universe. 
For example, \citet{rigby15} showed that \ciii\ can be used to pick out low-metallicity galaxies with strong bursts of star-formation, whereas \citet{senchyna17} suggested that nebular \heii\ and \civ\ emission has the potential to constrain metallicity. 
Additionally, \citet{senchyna19a} demonstrated that \civ\ emission is ubiquitous in extremely metal poor systems with very high specific star formation rates - albeit with equivalent widths smaller than those measured at high-$z$. 
With regards to the strength of the ionizing radiation, \citet{ravindranath20} found a strong correlation between \ciii\ and \oiiiuv\ emission and the O32 ratio (a proxy for the ionization parameter), confirming that a hard radiation field is required to produce the high-ionization nebular lines. Using two nearby extreme UV emitting galaxies,  \citet{berg19b} showed us that a combination of strong \civ\ and \heii\ emission may identify galaxies that not only produce but also transmit a substantial number of high-energy photons - i.e., potential contributors to cosmic reionization (see also \citealt{schaerer22}). 
While each of these studies provided a significant step-forward in understanding the conditions required for UV emission, these works have been limited to single peculiar objects or small nearby samples and lack the large statistics that we need to interpret the high volume of high-$z$ UV spectroscopy that will arrive in the next decade.

Statistically larger rest-UV spectroscopic studies do exist, typically targeting $2<z<4$ star-forming galaxies.
For instance, using deep VLT/MUSE spectroscopy, \citet{maseda17} collected a sample of 17 unlensed \ciii\ emitters at $1.5\lesssim z \lesssim 4$ which provided an unbiased sample toward the lowest mass, bluest galaxies. 
Stacked spectra of 15 gravitationally lensed galaxies at redshifts $1.68 < z < 3.6$ from project MEGaSaURA by \citet{rigby18a}, produced a new spectral composite of star-forming galaxies at redshift $z\sim2$ which clearly revealed strong \ciii\ and \mgii~$\lambda\lambda$2796,2803 as well as weaker lines, such as \heii, and \siiii.
\citet{llerena21} exploited a broader representative sample of 217 \ciii\ emitters ($\sim 30$\% of the total sample) from the VANDELS survey \citep{mclure18}, collecting main-sequence galaxies at $z\sim 2-4$ to investigate their average properties using the spectral stacking technique. 
Finally, \citet{schmidt21} presented an even larger sample, collecting 2052 spectroscopically confirmed emission line galaxies at $1.5\lesssim z \lesssim 6.4$, providing line properties of the main UV lines, and subsequently confirming the wealth of information and physical properties that rest-frame UV emission features red-wards of \lya\ can probe. 
These works currently represent our most comprehensive rest-FUV spectral datasets at high redshift. However, the majority of them have focused mainly on \ciii\ emitters, since \ciii\ the strongest UV emission line after \lya, and it is extremely challenging to obtain the required high signal-to-noise (S/N) to detect fainter lines even employing the stacking technique. 
Moderate spectral resolution ($R\sim18000)$ and broad wavelength coverage are also necessary to fully investigate the potential of UV diagnostics. Also, the limited wavelength range available for each of these studies has prevented us from carrying out a comparison of multi-wavelength diagnostics for ISM properties within the same targets. 

Indeed, in order for us to derive an accurate and detailed UV toolkit, we not only need to cover the full UV regime, but also optical wavelengths. Historically, ISM tracers have relied heavily on optical diagnostics, and as such they are very well calibrated. A crucial step in understanding the conditions that produce UV emission would therefore be comparing UV line strengths with ISM conditions derived from pre-existing optical diagnostics within the same targets, to effectively calibrate a toolkit that depends solely on UV emission lines.
This aspect is particularly important because the entire optical wavelength range on which our current diagnostic toolkit relies (from \oii$\lambda\lambda$3727,9 to \siii$\lambda$9069]), that is easily accessible in the local Universe, will not be available for sources in the Reionization epoch. Specifically, JWST instruments such as NIRSpec will cover blueward of 7000~\AA\ and 4500~\AA\ only in objects between $z\sim6$ and $z\sim10$, respectively. 
As such, a UV toolkit will be essential for characterizing and interpreting the spectroscopic observations of high-$z$ systems. 

The ideal framework from which a UV toolkit can be built would consist of high $S/N$ spectra with the possibility of extensive wavelength coverage that spans UV to optical wavelengths. Each of these essential elements are offered by local high-$z$ analogs.
In this context, the COS Legacy Spectroscopic SurveY (CLASSY) treasury 
(\citealt{berg22,james22}, \citetalias{berg22} and \citetalias{james22} hereafter) 
represents the first high-quality ($S/N_{1500\mathrm{\AA}}\gtrsim5$ per resolution element, resel), high-resolution ($R\sim15,000$) and broad wavelength range ($\sim1200-2000$ \AA) UV database of 45 nearby ($0.002<z<0.182$) star-forming galaxies. 
These objects were selected to include properties similar to reionization-era systems, in terms of specific star-formation rate, direct gas-phase metallicity, ionization level, reddening, and nebular density (see \citetalias{berg22} for more details). Moreover, optical observations are in-hand for all the galaxies of the sample, allowing us to make detailed comparisons of UV and optical diagnostics.
As such, CLASSY provides the ideal UV atlas with which we can tailor our UV diagnostic toolkit.

This is the first in a series of two CLASSY papers in which we present a FUV-based toolkit and show how this compares to well-known optical diagnostics. Specifically, in this work we provide a detailed calculations of dust attenuation, electron density $n_e$, electron temperature $T_e$, gas-phase metallicity 12+log(O/H) and ionization parameter log($U$), using both UV and optical direct diagnostics, taking into account the different ionization zones of the ISM. Then, from their comparison, we provide a set of diagnostic equations to estimate ISM properties only from UV emission lines.
In Sec.~\ref{sec:sample} we describe the CLASSY sample, covering both the UV and optical data, while in Sec.~\ref{sec:data-analysis} we present the spectroscopic analysis, including stellar continuum and emission line fitting. In Sec.~\ref{sec:methods} we discuss the chemical and physical diagnostics used in our analysis, showing and comparing the derived ISM properties in Sec.~\ref{sec:results}. Then, in Sec.~\ref{sec:discussion}, we introduce and discuss our UV-based toolkit, providing also a comparison with previous works. Finally, in Sec.~\ref{sec:conclusions} we summarize our main findings.

The data presented in this paper were obtained from the Mikulski Archive for Space Telescopes (MAST) at the Space Telescope Science Institute. The specific observations analyzed can be accessed via \dataset[10.17909/m3fq-jj25]{https://doi.org/10.17909/m3fq-jj25}.
All the products of this paper (UV and optical line fluxes; UV and optical $z$; UV-optical flux offsets; ISM properties, i.e., $E(B-V)$, $n_e$, $T_e$, optical and UV 12+log(O/H)) will be provided on the \href{https://archive.stsci.edu/hlsp/classy}{CLASSY MAST webpage} as downloadable tables. In App.~\ref{app:B}, \ref{app:C} and \ref{app:D} we show which information will be provided.
Throughout this paper, we adopt a the solar metallicity scale of \citet{asplund09}, where 12+log(O/H)$_\odot = 8.69$.

\section{Sample presentation} \label{sec:sample}
CLASSY is a sample of 45 star-forming UV-bright ($m_{FUV} < 21 AB$~arcsec$^{−2}$), relatively compact ($GFWHM_{NUV} < 2.5"$) galaxies in the local Universe (0.002~$<$ z $<$~0.182), spanning a wide range of stellar masses ($6.22 < $~log($M_\star$/M$_\odot$)~$< 10.06$), star formation rates ($-2<~$log($SFR$/M$_\odot$yr$^{-1}$)~$<+2$), oxygen abundances ($6.98 < $~12+log(O/H)~$< 8.77$), electron densities ($10 < n_e$/cm$^{-3}$~$< 1120$), degree of ionization ($0.54 < O3O2 < 38.0$, with $O3O2 =$~\oiii~$\lambda$5007/\oii~$\lambda\lambda$3727,9), and reddening values ($0.001 < E(B-V) < 0.673$). This broad sampling of parameter space makes the CLASSY sample representative of star-forming galaxies across all redshifts, with a bias towards more extreme $O3O2$ values, low stellar masses, and high SFRs, typical of high-$z$ systems (see \citetalias{berg22}). 
In \citetalias{berg22}, we presented our sample, explaining in detail the selection criteria and giving an extensive overview of the HST/COS and archival optical spectra. 
To summarize, from the Hubble Spectral Legacy Archive (HSLA), 101 nearby ($z < 0.2$) galaxies were selected on the basis of the high signal-to-noise ($\gtrsim 7$ per 100~km/s resolution element) COS spectroscopy in at least one medium resolution grating (i.e., G130M, G160M, or G185M), applying further selection criteria to assemble a high-quality, comprehensive rest-frame set of FUV spectra for a large and diverse sample of star-forming galaxies. Specifically, any targets with secondary classifications or visually confirmed spectra features of quasi-stellar object (QSO) or Seyfert were removed.
The data reduction has been presented in detail in \citetalias{james22}, including spectra extraction, co-addition, wavelength calibration, and vignetting. 

In this work, we take into account the properties of the CLASSY galaxies in terms of redshift, stellar mass, SFR, 12+log(O/H), and galaxy half-light radius ($r_{50}$), estimated from \citetalias{berg22}, which for completeness we also show in Table~\ref{tab:classyprop}. Specifically, the redshifts have been taken from SDSS where available, $r_{50}$ was estimated from PanSTARRS imaging, while the stellar masses and SFRs have been estimated from the spectra energy distribution (SED) fitting via the BayEsian Analysis of GaLaxy sEds (BEAGLE, \citealt{chevellard16}), as explained in \citetalias{berg22} (see Sec.~4.7). Finally, the calculation of 12+log(O/H) is explained in detail in \citetalias{berg22} Sec.~4.5, and is based on the direct $T_e$ method, using \sii~$\lambda$6717/$\lambda$6731 and \oiii~$\lambda$4363/$\lambda$5007 as electron density and temperature tracers, respectively. The UV redshifts $z_{UV}$ instead are obtained from the analysis of UV emission lines, and are part of the results of this paper.

As described in \citetalias{berg22}, both UV and optical CLASSY spectra have been corrected for the total Galactic foreground reddening along the line of sight of their coordinates using the \python\ \texttt{dustmaps} \citep{green18} interface to query the Bayestar 3D dust maps of \citet{green15}.  The \citet{green15} map was adopted over more recent versions due to its more optimal coverage of the CLASSY sample. The Galactic foreground reddening correction was then applied using the \citet{cardelli89} reddening law.

In the following, we briefly summarize the UV and optical data sample and properties.

\begin{table*}
\begin{center}
\caption{CLASSY sample main properties.}
\label{tab:classyprop}
\begin{tabular}{ccccccccc} 
\hline
\hline  
        &               &               & & Tot. log \Ms\ & log SFR       &             &                 \\
Target & Name  & $z_{\rm lit.}$& $z_{\rm UV}$  & (\Mo)         & (\Mo\ \yr)    & 12+log(O/H) & $r_{50}$   \\
\hline
1. \ \ J0021+0052 &  & 0.09839 &  \nodata                & 9.09$\pm^{0.18}_{0.38}$ & $+1.07\pm^{0.14}_{0.11}$ & 8.17$\pm$0.07 & 0.784 \\
2. \ \ J0036-3333 &  Haro 11 knot  & 0.02060 &  \nodata                & 9.14$\pm^{0.26}_{0.23}$ & $+1.01\pm^{0.19}_{0.21}$ & 8.21$\pm$0.17 & 2.846 \\
3. \ \ J0127-0619 & Mrk~996  & 0.00540 &  $0.00547$    & 8.74$\pm^{0.18}_{0.15}$ & $-0.75\pm^{0.15}_{0.13}$ & 7.68$\pm$0.02 & 2.374 \\
4. \ \ J0144+0453 & UM133   & 0.00520 &  $0.00533$    & 7.65$\pm^{0.24}_{0.29}$ & $-0.81\pm^{0.29}_{0.46}$ & 7.76$\pm$0.02 & 2.851 \\
5. \ \ J0337-0502 & SBS0335-052~E  & 0.01352 &  $0.01346$    & 7.06$\pm^{0.24}_{0.21}$ & $-0.32\pm^{0.07}_{0.11}$ & 7.46$\pm$0.04 & 1.433 \\
6. \ \ J0405-3648 &   & 0.00280 &  \nodata                & 6.61$\pm^{0.28}_{0.28}$ & $-1.81\pm^{0.31}_{0.27}$ & 7.04$\pm$0.05 & 3.557 \\
7. \ \ J0808+3948 &  & 0.09123 &  \nodata                & 9.12$\pm^{0.30}_{0.17}$ & $+1.26\pm^{0.18}_{0.25}$ & 8.77$\pm$0.12 & 1.114 \\
8. \ \ J0823+2806 & LARS9  & 0.04722 &  $0.04741$    & 9.38$\pm^{0.33}_{0.19}$ & $+1.48\pm^{0.15}_{0.32}$ & 8.28$\pm$0.01 & 2.134 \\
9. \ \ J0926+4427 & LARS14  & 0.18067 &  $0.18000$    & 8.76$\pm^{0.30}_{0.26}$ & $+1.03\pm^{0.13}_{0.13}$ & 8.08$\pm$0.02 & 0.889 \\
10. J0934+5514  & I~zw~18~NW    & 0.00250 &  $0.00264$    & 6.27$\pm^{0.15}_{0.20}$ & $-1.52\pm^{0.09}_{0.07}$ & 6.98$\pm$0.01 & 2.606 \\
11. J0938+5428  &    & 0.10210 &  $0.10210$      & 9.15$\pm^{0.18}_{0.29}$ & $+1.05\pm^{0.20}_{0.17}$ & 8.25$\pm$0.02 & 1.095 \\
12. J0940+2935  &    & 0.00168 &  \nodata                & 6.71$\pm^{0.23}_{0.40}$ & $-2.01\pm^{0.42}_{0.37}$ & 7.66$\pm$0.07 & 5.151 \\
13. J0942+3547  &  CG-274, SB 110    & 0.01486 &  $0.01482$    & 7.56$\pm^{0.21}_{0.29}$ & $-0.76\pm^{0.19}_{0.12}$ & 8.13$\pm$0.03 & 1.328 \\
14. J0944-0038 &  CGCG007-025, SB 2     & 0.00478 &  $0.00487$  & 6.83$\pm^{0.44}_{0.25}$ & $-0.78\pm^{0.19}_{0.16}$ & 7.83$\pm$0.01 & 0.984 \\
15. J0944+3442  &    & 0.02005 &  $0.02005$    & 8.19$\pm^{0.40}_{0.23}$ & $-0.01\pm^{0.28}_{0.65}$ & 7.62$\pm$0.11 & 2.458 \\
16. J1016+3754  & 1427-52996-221    & 0.00388 &  $0.00390$  & 6.72$\pm^{0.27}_{0.22}$ & $-1.17\pm^{0.18}_{0.18}$ & 7.56$\pm$0.01 & 1.835 \\
17. J1024+0524  & SB~36    & 0.03319 &  $0.03326$   & 7.89$\pm^{0.37}_{0.24}$ & $+0.21\pm^{0.14}_{0.12}$ & 7.84$\pm$0.03 & 1.325 \\
18. J1025+3622  &    & 0.12650 &  $0.12717$    & 8.87$\pm^{0.25}_{0.27}$ & $+1.04\pm^{0.14}_{0.18}$ & 8.13$\pm$0.01 & 0.843 \\
19. J1044+0353  &    & 0.01287 &  $0.01286$  & 6.80$\pm^{0.41}_{0.26}$ & $-0.59\pm^{0.11}_{0.14}$ & 7.45$\pm$0.03 & 1.204 \\
20. J1105+4444  & 1363-53053-510    & 0.02154 &  $0.02147$    & 8.98$\pm^{0.29}_{0.24}$ & $+0.69\pm^{0.28}_{0.22}$ & 8.23$\pm$0.01 & 2.646 \\
21. J1112+5503  &    & 0.13164 &  \nodata                & 9.59$\pm^{0.33}_{0.19}$ & $+1.60\pm^{0.20}_{0.25}$ & 8.45$\pm$0.06 & 0.920 \\
22. J1119+5130 &     & 0.00446 &  $0.00444$    & 6.77$\pm^{0.15}_{0.28}$ & $-1.58\pm^{0.21}_{0.12}$ & 7.57$\pm$0.04 & 1.870 \\
23. J1129+2034 & SB~179     & 0.00470 &  $0.00467$    & 8.09$\pm^{0.37}_{0.27}$ & $-0.37\pm^{0.38}_{0.56}$ & 8.28$\pm$0.04 & 3.098 \\
24. J1132+5722 &  SBSG1129+576     & 0.00504 &  $0.00504$     & 7.31$\pm^{0.23}_{0.26}$ & $-1.07\pm^{0.27}_{0.35}$ & 7.58$\pm$0.08 & 2.249 \\
25. J1132+1411 & SB~125    & 0.01764 &  $0.01760$     & 8.68$\pm^{0.28}_{0.19}$ & $+0.44\pm^{0.24}_{0.27}$ & 8.25$\pm$0.01 & 7.289 \\
26. J1144+4012 &    & 0.12695 &  $0.12700$      & 9.89$\pm^{0.18}_{0.29}$ & $+1.51\pm^{0.20}_{0.29}$ & 8.43$\pm$0.20 & 1.158 \\
27. J1148+2546 & SB~182     & 0.04512 &  $0.04522$    & 8.14$\pm^{0.34}_{0.24}$ & $+0.53\pm^{0.17}_{0.14}$ & 7.94$\pm$0.01 & 0.874 \\
28. J1150+1501 &  SB~126, Mrk~0750    & 0.00245 &  $0.00246$    & 6.84$\pm^{0.28}_{0.30}$ & $-1.33\pm^{0.29}_{0.23}$ & 8.14$\pm$0.01 & 1.760 \\
29. J1157+3220 &  1991-53446-584     & 0.01097 &  $0.01101$    & 9.04$\pm^{0.32}_{0.18}$ & $+0.97\pm^{0.21}_{0.42}$ & 8.43$\pm$0.02 & 2.894 \\
30. J1200+1343 &     & 0.06675 &  $0.06699$    & 8.12$\pm^{0.47}_{0.42}$ & $+0.75\pm^{0.20}_{0.16}$ & 8.26$\pm$0.02 & 0.908 \\
31. J1225+6109 & 0955-52409-608    & 0.00234 & $ 0.00234$    & 7.12$\pm^{0.34}_{0.24}$ & $-1.08\pm^{0.26}_{0.26}$ & 7.97$\pm$0.01 & 2.596 \\
32. J1253-0312 & SHOC391    & 0.02272 & $0.02267$     & 7.65$\pm^{0.51}_{0.23}$ & $+0.56\pm^{0.15}_{0.15}$ & 8.06$\pm$0.01 & 1.079 \\
33. J1314+3452 & SB~153   	& 0.00288 & $0.00282$     & 7.56$\pm^{0.30}_{0.21}$ & $-0.67\pm^{0.23}_{0.55}$ & 8.26$\pm$0.01 & 1.765 \\
34. J1323-0132  &    & 0.02246 & $0.02246$   & 6.31$\pm^{0.26}_{0.10}$ & $-0.72\pm^{0.08}_{0.09}$ & 7.71$\pm$0.04 & 0.698 \\
35. J1359+5726  &  Ly~52, Mrk~1486   	& 0.03383 & $0.03381$     & 8.41$\pm^{0.31}_{0.26}$ & $+0.42\pm^{0.20}_{0.14}$ & 7.98$\pm$0.01 & 1.395 \\
36. J1416+1223 &     & 0.12316 & \nodata                 & 9.59$\pm^{0.32}_{0.26}$ & $+1.57\pm^{0.21}_{0.25}$ & 8.53$\pm$0.11 & 0.985 \\
37. J1418+2102 &     & 0.00855 & $0.00858$   & 6.22$\pm^{0.49}_{0.35}$ & $-1.13\pm^{0.15}_{0.16}$ & 7.75$\pm$0.02 & 1.130 \\
38. J1428+1653 &     & 0.18167 & \nodata                 & 9.56$\pm^{0.15}_{0.23}$ & $+1.22\pm^{0.26}_{0.19}$ & 8.33$\pm$0.05 & 0.933 \\
39. J1429+0643 &     & 0.17350 & $0.17340$       & 8.80$\pm^{0.35}_{0.21}$ & $+1.42\pm^{0.11}_{0.17}$ & 8.10$\pm$0.03 & 0.859 \\
40. J1444+4237 & HS1442+4250     & 0.00230 & $0.00220$   & 6.48$\pm^{0.17}_{0.17}$ & $-1.94\pm^{0.11}_{0.08}$ & 7.64$\pm$0.02 & 2.760 \\
41. J1448-0110 & SB~61     & 0.02741 & $0.02744$     & 7.61$\pm^{0.41}_{0.24}$ & $+0.39\pm^{0.13}_{0.14}$ & 8.13$\pm$0.01 & 1.070 \\
42. J1521+0759 &    & 0.09426 & \nodata                 & 9.00$\pm^{0.29}_{0.30}$ & $+0.95\pm^{0.16}_{0.17}$ & 8.31$\pm$0.14 & 0.983 \\
43. J1525+0757  &    & 0.07579 & \nodata                 & 10.06$\pm^{0.28}_{0.42}$& $+1.00\pm^{0.69}_{0.24}$ & 8.33$\pm$0.04 & 1.319 \\
44. J1545+0858 &  1725-54266-068     & 0.03772 & $0.03772$   & 7.52$\pm^{0.43}_{0.26}$ & $+0.37\pm^{0.13}_{0.17}$ & 7.75$\pm$0.03 & 1.075 \\
45. J1612+0817  &    & 0.14914 &  \nodata                & 9.78$\pm^{0.28}_{0.26}$ & $+1.58\pm^{0.28}_{0.24}$ & 8.18$\pm$0.19 & 0.878 \\		
\hline
(1)&(2)&(3)&(4)&(5)&(6)&(7)&(8)\\
\hline
\end{tabular}
\end{center}
\tablecomments{
    CLASSY sample properties derived from UV$+$optical photometry and spectra. 
    Columns 1 and 2 indicate the target name used in this work and alternative names, respectively. Columns 3 and 4 give the target redshift from the literature and FUV UV emission-lines, respectively.
    Columns 5 and 6 list the total stellar masses and SFRs derived from 
    \texttt{Beagle} SED fitting in \citetalias{berg22}. Column 7 gives the oxygen abundances derived in \citetalias{berg22}. Column 8 lists the galaxy half-light radius ($r_{50}$), estimated from PanSTARRS imaging, from \citetalias{berg22}.}

\end{table*}

\subsection{UV data}\label{sec:cos}
CLASSY combines 135 orbits of new \HST\ data (PID: 15840, PI: Berg) with 177 orbits of archival \HST\ data, for a total of 312 orbits. 
In order to achieve nearly-panchromatic FUV spectral coverage with the highest spectral resolution possible, CLASSY combines the G130M, G160M, G185M, G225M and G140L gratings, spanning from 1150~\AA\ to 2100--2500~\AA\ to allow synergistic co-spatial studies of stars and gas within the same galaxy.

Each HST/COS grating has a different spectral resolution that must be accounted for when combining data from multiple gratings. This coaddition process is explained in Sec.~2.3 of \citetalias{berg22} and in \citetalias{james22}, which presents all the details of this multi-stage technical process, concerning extracting, reducing, aligning, and coadding the spectra from the different gratings.
This paper focuses on the analysis of all the emission lines (except for \lya) in the range $1150-2000$~\AA. We used the so-called \emph{High Resolution} (HR: G130M$+$G160M; $R\sim10000-24000$) and
\emph{Moderate Resolution} (MR: G130M$+$G160M$+$G185M$+$G225M; $R\sim10000-20000$) co-added spectra with a dispersion of 12.23 m\AA/pixel, and a resolution of 0.073 \AA\ per resolution element (\AA/resel, where 1 resel equates to 6 native COS pixels), and 33 m\AA/pixel and 0.200 \AA/resel, respectively.

For the galaxies J1044+0353 and J1418+2102, instead of the MR co-added spectra, we used the so-called \emph{Low Resolution} (LR: G130M$+$G160M$+$G140L or G130M$+$G160M$+$G185M$+$G225M$+$G140L; $R\sim1500-4000$) coadds, with a nominal point source resolution of 80.3 m\AA/pixel or 0.498 \AA/resel. Additionally, the COS G185M and G225M observations for the J1112+5503 galaxy were impacted by guide star failures and thus we excluded this galaxy from the sample. 

We performed the stellar continuum subtraction with the method described in Sec.~\ref{sec:data-analysis-stellar-cont-uv} from the HR spectra, binned by 15 native COS pixels. 
We also fit the MR coadded spectra, after binning them by 6 native COS pixels, in order to fit all the main emission lines not covered by the HR coadds.
Finally, we doubly rebinned both configurations (i.e., binning the HR and MR coadds by 30 and 12 native COS pixels, respectively), to improve the fit of the faintest emission lines when possible.

\subsection{Optical data}\label{sec:optical}
High-quality optical spectra have been collected for the entire CLASSY sample to ensure uniform determinations of galaxy properties and to allow comparisons between properties derived from optical and UV diagnostics, thus enabling an accurately calibrated suite of UV diagnostics. 

DR7 APO/SDSS spectra with a 3.0" aperture exist for 38 of the CLASSY galaxies \citep{abazajian09}, while for one galaxy, J1444+4237, there are DR13 BOSS spectrograph data with a 2.0" aperture \citep{albareti17,guseva17}. These spectra are in the wavelength range of $3800-9200$\AA\ ($3600-10400$\AA\ for BOSS), with a  spectral resolution of $R \sim1500-2500$ \citep{eisenstein11}.

For the remaining galaxies of the sample (J0036-3333, J0127-0619, J0337-0502, J0405-3648, J0934+5514, and J0144+0453), we used integral field spectroscopy data, when available, or long-slit spectroscopy, instead of SDSS. Specifically, we used
VLT/VIMOS integral field unit (IFU) from \citet{james09} for J0127-0619, MMT Blue Channel Spectrograph spectra from \citet{senchyna19a} for J0144+0453, Keck/KCWI IFU spectra from \citet{rickardsvaught21} for J0934+5514, and VLT/MUSE IFU spectra for the remaining three galaxies.
MUSE spectra are also available for the galaxies J0021+0052 (PI: Göran
Östlin), J1044+0353 and J1418+2102 (PI: Dawn Erb). We used these data to retrieve emission line-ratios involving faint auroral lines, if undetected ($S/N<3$) in SDSS spectra. 
 Finally, for the galaxies J0808+3948, J0944-0038, J1148+2546, J1323-0132 and J1545+0808, Multi-Object Double Spectographs (MODS) data from the LBT telescope, presented in \citealt{arellano-cordova22}, are also available  (\citetalias[Paper~V][]{arellano-cordova22} hereafter).
Information on each of the optical datasets is provided in the following paragraphs.

Concerning the IFU data available for J0021+0052, J0036-3333, J0127-0619, J0337-0502, J0405-3648, J0934+5514, J1044+0353 and J1418+2102, we extracted a spectrum from a 2.5" aperture centred at the same coordinates of COS observations to match the COS aperture (see also \citetalias{berg22} and \citetalias[Paper~V][]{arellano-cordova22}).
Specifically, the integrated VIMOS spectrum of J0127-0619 is obtained combining the high-resolution blue and orange grisms, covering the wavelength range $4150-7400$~\AA\ with a spectral resolution $R\sim1150-2150$ (see \citealt{james09} for more details). The KCWI spectrum of J0934+5514 is in the wavelength range $3500-5600$~\AA\ at a median spectral resolution of $R\sim3600$. Finally, MUSE spectra are in the wavelength range $4300-9300$~\AA\ at a spectral resolution of $R\sim2000-3500$.

Regarding the long-slit data, the MMT spectrum of J0144+0453 was taken with the 300 lines/mm grating with a 10"$\times$1" slit, oriented along the parallactic angle to minimize slit losses (see \citealt{senchyna19a} for more details). The wavelength coverage is $3200-8000$~\AA\ with a resolution $R\sim740$.
For the galaxies J0808+3948, J0944-0038, J1148+2546, J1323-0132 and J1545+0808, instead of SDSS, we took advantage of the MODS data from LBT obtained using the G400L and G670L. MODS long-slit data were taken with a 60"$\times$1" slit, with an extraction aperture of 2.5"$\times$1", and a slit orientation along the parallactic angle (see \citetalias{arellano-cordova22} for more details).  The wavelength coverage extends from 3200 \AA\, to 10000 \AA\, with a moderate spectral resolution of $R \sim 2000$.

\citetalias{arellano-cordova22} compares the SDSS, LBT and MUSE integrated spectra for the galaxies with multiple observations, demonstrating that flux calibration issues or aperture differences do not introduce significant discrepancies in the optical ISM properties in terms of gas attenuation, density, temperature, metallicity and SFRs. This result supports our comparison of the physical properties obtained using these different sets of optical data.
The UV and optical fluxes, highlighting which telescope and instrument was considered for each galaxy, as well as the products of the analysis of this paper will be provided on the \href{https://archive.stsci.edu/hlsp/classy}{CLASSY MAST webpage} as downloadable tables. In App.~\ref{app:B}, \ref{app:C} and \ref{app:D} we show which information will be provided.

\section{Data analysis}\label{sec:data-analysis}
The UV and optical spectra were analyzed making use of a set of customized python scripts in order to first fit and subtract the stellar continuum and then fit the main emission lines with multiple Gaussian components where needed. 
This allowed us to estimate the stellar population properties (i.e., age, metallicity and stellar dust attenuation), emission-line properties (fluxes, velocities, velocity dispersions, and equivalent widths), the UV-optical flux offset (discussed in Appendix~\ref{app:A}), and ISM gas properties. In the following, all the steps are explained in detail.

\subsection{Stellar continuum}\label{sec:data-analysis-stellar-cont}
\subsubsection{UV spectra}\label{sec:data-analysis-stellar-cont-uv}
The analysis of UV spectra relies on a robust stellar continuum fitting procedure both for determining the properties of the stellar population and for accurately measuring nebular UV emission lines, such as \heii~$\lambda$1640 or \civ~$\lambda\lambda$1548,51.
For the purposes of subtracting the UV stellar continuum in the HR spectra\footnote{It is not necessary to fit the stellar continuum in the MR spectra, since the NUV range does not contribute significantly to the stellar population analysis, and in that range there are no significant absorption or resonant emission lines due to stars.} for this paper, we compare the results of two sets of fits which will be described in detail in \citet{senchyna22} (\citetalias{senchyna22} hereafter).
Both fits considered here are based upon a flexible linear combination of spectra of SSPs spanning a wide range of metallicities and ages, as described by \citet{chisholm19}, and assume a \citet{reddy16} attenuation law.
The primary difference between the two sets of results is the stellar population synthesis framework used to generate the basis of SSP spectra.
The first uses the \textsc{Starburst99} theoretical UV predictions described by \citet{leitherer99,leitherer10}, while the second relies on the latest version of the \citet{bruzual03} models \citep[S.\ Charlot \& G.\ Bruzual, in-preparation, hereafter C\&B; see also][]{gutkin16,vidal-garcia17,plat19}.
These population synthesis models adopt different prescriptions for the evolution and atmospheres of massive stars, resulting in particularly significant differences for lines such as \heii~$\lambda 1640$ that can be powered in the dense optically-thick winds of very luminous stars \citep[e.g.][]{senchyna21a}.

In addition to stellar light and dust attenuation, the other crucial constituent of the UV light of star-forming galaxies is the nebular continuum.
Both sets of models include the contribution of the nebular continuum computed in a self-consistent manner and assuming a closed geometry, as described in \citet{leitherer99} and \citet{gutkin16,plat19}, respectively.
The \textsc{Starburst99} predictions do not include variable parameters describing this emission, but the \cloudy-computed nebular continuum for the C\&B models are presented at varying $\log(U)$ which can have an impact in the UV \citep[see e.g.,][]{senchyna22a}.
Our fiducial assumed volume-averaged log($U$)~$=-2.5$ \citep[defined as in][]{gutkin16} represents the median value inferred for the full CLASSY sample from fitting the UV continuum with different log($U$) in the range $[-3;-1]$, and it is also typical of the values inferred from fitting the nebular line emission of similar local star-forming galaxies \citep{plat19,senchyna22a}. 
However, we stress that the choice of fixing this parameter has a minimal impact on the fidelity of the UV continuum fits, with a negligible median difference in the reduced chi-square (i.e., $\lesssim0.01$; \citetalias{senchyna22}). Moreover, \citet{chisholm19} explored variations with the log($U$) and density, finding no changes in the shape of the nebular continuum and in the relative contribution of the nebular/stellar continuum ratio over the expected log($U$) range (see also \citealt{byler17}). 

In most other respects, the fits proceed in a similar manner.
The observed HR spectra are fitted after first rebinning by 15 pixels and after smoothing the models with a Gaussian kernel to best represent the achieved resolution and S/N.
In both cases, we adopt the maximum initial mass function (IMF) upper mass cutoff provided for the models (using \citealt{kroupa01} and \citealt{chabrier03}, for \textsc{Starburst99} and C\&B, respectively); this is 100~$M_\odot$ for \textsc{Starburst99} and 600~$M_\odot$ for C\&B \citep[see e.g.][]{plat19,senchyna22a}.
The uncertainties in the fits were calculated via a Monte Carlo technique, modulating the observed flux with a Gaussian kernel centered on zero with a width equal to the formal estimated error on the flux.
To summarize, the two stellar continuum fits provide independent estimates of the intrinsic stellar $E(B-V)_{UV}$ reddening, and the light-weighted ages and metallicities of the ionizing stellar populations, alongside full fits to the UV continuum. 

For the purpose of this work, the stellar continuum fitting is used to subtract the stellar contribution from the observed UV spectra, thus allowing us to accurately measure the nebular emission lines in the range $1150-2000$~\AA. 
 After carefully checking that the subtraction of either \textsc{Starburst99} or C\&B stellar continuum best-fit gave similar results for our emission line fitting, we ultimately decided to use the C\&B best-fit, since it takes into account the stellar \heii $\lambda1640$ contribution. Moreover, the C\&B models can be extended to wavelengths of $\sim 9000$~\AA\ in the optical, which allowed us to perform accurate flux-scaling between the optical and UV spectra. 
 
The full flux-scaling analysis is described in App.~\ref{app:A}, where we explain our method to properly scale the flux of the optical spectra to the UV. 
In summary, a flux offset between COS and the optical spectra is expected since they have been obtained via different instruments with slightly different apertures and pointing position. 
The median value that we find for this UV-optical flux offset is $0.78 \pm 0.03$, and we report the value obtained for each galaxy on the \href{https://archive.stsci.edu/hlsp/classy}{CLASSY MAST webpage} as shown in App.~\ref{app:C}. 
We multiplied the observed optical spectra of each CLASSY galaxy by its corresponding UV-optical flux offset to correct them.
We highlight that this flux correction has no impact on the properties derived from the flux-ratios within each galaxy, but only when ratios between UV and optical emission lines (e.g., \oiiiuv$\lambda$1666/\oiii$\lambda$5007) are considered.

\subsubsection{Optical spectra}\label{sec:data-analysis-stellar-cont-opt}
Since the UV stellar continuum models were optimized for the young stellar population, it was not feasible to use the UV models to perform a self-consistent fit of the optical wavelength portion of the spectrum, due to the dominant contribution from the older population of stars in this wavelength regime (e.g., \citealt{leitherer99}). 
Thus being, in order to remove any stellar absorption components present in the Balmer emission lines, we model the optical stellar continuum using \texttt{Starlight}\footnote{{\url{www.starlight.ufsc.br}}} spectral synthesis code of  \citet{cidfernandes2005} and the stellar models of \citet{bruzual03} with the IMF of \citet{chabrier03}. The set of the stellar models taken into account comprises 25 ages (1~Myr~--~18~Gyr) and six metallicities ($0.05 <$~Z$\star$/Z$\odot < 2.5$). It should be noted that while the \texttt{Starlight} models do not include a nebular continuum component, the nebular continuum contribution in this wavelength regime is known to be negligible \citep[<10\%;][]{byler17}.

As a preliminary step, we corrected the spectra for the Galactic foreground reddening correction (see Sec.~\ref{sec:cos}), and uniformly sampled the rest-frame wavelength, the flux and the error in steps of $\Delta \lambda =  1$ \AA. For reddening the models, we used the attenuation law of \citet{cardelli89}. The \texttt{Starlight} models are fitted over the wavelength range $3700-9100$~\AA. In Fig.~\ref{fig:optuv_offset}, included in App.~\ref{app:A}, we show our UV and optical stellar continuum best-fit for the galaxy J0021+0052 and J1144+4012 as an example.

\subsection{Emission-lines}\label{sec:emission-lines}
The analysis of the emission lines in the UV and optical spectra (after the subtraction of the best-fit stellar continuum, described in Sec.~\ref{sec:data-analysis-stellar-cont}) was performed separately, but with a similar approach.
We simultaneously fit each spectrum taking into account a set of UV and optical emission lines in the wavelength range $1265-2000$~\AA\ and $3700-9100$~\AA, respectively, with a linear baseline centered on zero and a single Gaussian, making use of the code \mpfit\ \citep{markwardt09}, which performs a robust non-linear least squares curve fitting. 
We list the final fluxes, corrected for dust reddening, of all fitted UV and optical emission lines in the \href{https://archive.stsci.edu/hlsp/classy}{CLASSY MAST webpage}, as shown in  App.~\ref{app:B} and \ref{app:C}.

In our procedure the main Milky Way absorption lines were masked, and the fitting was performed only in windows of 3000~\kms\ centered around each emission line.  We tied together the velocity (i.e., the line center) and in optical spectra also the velocity dispersion (or line width) for all the emission lines, to better constrain weak or blended features, while allowing the line flux to vary freely (in general). An exception was made for the line center of the UV emission-lines \ciii\ and \civ, because they can be significantly shifted in velocity with respect to the others due to their origin, as we will discuss in a forthcoming paper focused on the kinematics and ionization source of the gas in the CLASSY galaxies (\citealt{mingozzi22} in prep.; \citetalias{mingozzi22} hereafter). 

In order to robustly determine uncertainties, we followed the Monte Carlo method where we perturbed $N$ times (with $N=100$) the observed spectra by adding to each spectral element a random value drawn from a Gaussian distribution centred on 0 with a standard deviation equal to the observed spectrum uncertainty. 
We then fitted each configuration with \mpfit\ \citep{markwardt09}, obtaining 100 estimates of the free parameters of the fit, that are flux, velocity and velocity dispersion. 
Finally, we calculated the 50$^{\rm th}$ (i.e., the median) and the (50$^{\rm th}$--16$^{\rm th}$) and (84$^{\rm th}$--50$^{\rm th}$) percentiles of the distributions of the fitted perturbed spectra and of the free parameters of the fit.
The median of each free parameter is considered as best-fit value, with a lower and upper uncertainty given by the sum in quadrature of the (50$^{\rm th}$--16$^{\rm th}$) and (84$^{\rm th}$--50$^{\rm th}$) percentiles, divided by the square root of $N$, and the \mpfit\ error.
The S/N associated to each line is then defined as the ratio between the flux and the flux uncertainty. All emission lines with signal-to-noise higher than 3 are considered to be reliable detections.

\subsubsection{Special constraints}\label{sec:emission-lines-constrains}
In our fitting procedure the line flux of each emission line must be non-negative, but it is left free to vary, apart from the doublets \nii~$\lambda\lambda6548,6584$ and \oi~$\lambda\lambda$6300,64, where we consider the transition probability of the doublets and assumed a fixed line ratio of 0.333 between the fainter and the brighter line \citep{osterbrock89}. 
We did not fix the \oiii~$\lambda\lambda$4959,5007 fluxes because in the KCWI data of J0934+5514 and SDSS data of J1253-0312 the \oiii~$\lambda$5007 line is saturated. For these objects we obtained an estimate of the \oiii~$\lambda5007$ by applying the fixed line ratio of 3 with respect to \oiii~$\lambda$4959 \citep{osterbrock89}. We notice that for J1253-0312 also the \ha\ line is clipped, so we discarded its flux.

Concerning \ciii~$\lambda\lambda$1907,9, we constrained the line ratios \fciii~$\lambda$1907/\ciii~$\lambda$1909 to vary up to 1.6 \citep{osterbrock89}, to avoid non-physical values.
Our procedure allows us to fit this doublet with two well-separated Gaussians, since the distance between the centroids of the two emission lines of the doublet is fixed.
Conversely, the resolution of the COS/G185M ($R\sim20000$@2100\AA) does allow us to resolve the \ciii\ doublet even after the 6-12 native pixel binning (i.e., $\sim40-80$~km/s), as long as the width of the fitted emission lines is smaller than half of their wavelength separation (i.e., $\sim 300$~km/s). Among the galaxies with significant \ciii\ emission ($S/N>3$), the latter condition is not satisfied in J1044+0353 and J1418+2102, because of the lower resolution of COS/G140L.
Concerning \oii~$\lambda\lambda$3727,29 doublet, blended in SDSS, MOD and MMT data, the distribution of flux between these two Gaussians is not reliable enough to derive an accurate line ratio. Therefore, the \oii\ ratio is derived only for the KCWI data of J0934+5514 (the doublet is not covered by the wavelength range of MUSE).

Another aspect we took into account in our fitting procedure is that optical lines such as \oiii~$\lambda$4363 and \ariv~$\lambda$4714 can suffer from contamination due to the \feii~$\lambda$4360 and \hei~$\lambda$4714, respectively (see e.g., \citealt{curti17,arellano-cordova20b}). For instance, \citet{arellano-cordova20b} demonstrated that the use of a contaminated \oiii~$\lambda$4363 could lead differences in metallicity of up to 0.08 dex.
In order to mitigate this problem, we fitted these faint features simultaneously with the other emission lines, tying them to the brighter \hei~$\lambda$4471 and \feii~$\lambda$4288, assuming a ratio of 0.728 and 0.125 (valid at $n_e=100$~cm$^{-3}$ and $T_e=10^4$~K, from \pyneb), respectively\footnote{We fitted the \feii~$\lambda$4288 only for galaxies in which this line is visible, that is at 12+log(O/H)~$\gtrsim7.7$.}. 

\subsubsection{Multi-component fitting}
After careful inspection of the optical spectra, we noticed that the \ha\ profile (in particular) shows a broad component in many CLASSY galaxies.  We therefore performed two-component Gaussian fits to the main optical emission lines (i.e., \hd, \hg, \oiii~$\lambda$4363, \hb, \oiii~$\lambda\lambda$4959,5007, \nii~$\lambda$5755, \oi~$\lambda\lambda$6300,74, \nii~$\lambda\lambda$6548,84, \ha, \sii~$\lambda\lambda$6717,31).
Specifically, we took into account one narrow component, with an observed velocity dispersion $\sigma < 200$~km/s, that is a representative cut-off for the galaxies of our sample, and a broad component ($\sigma <1000$~km/s). Their velocity and velocity dispersion are tied to be the same for all the emission lines.
To understand if the addition of a second component is significant, we calculated the reduced chi-square $\tilde\chi^2$ of the single and double-component fits in the rest-frame wavelength range 6540--6590~\AA\ covering \ha\ (4950--5010~\AA\ for J1253-0312 and J0934+5514, for which \ha\ is unavailable), and chose the model with more components only if the $\tilde\chi^2$ was at least 0.1~dex smaller. This condition is satisfied in 24 out of 44 galaxies of our sample.

In our UV spectra, the S/N is usually not high enough to detect faint broad components in the emission lines of interest here.  However, after a visual inspection we did notice a clear broad profile in the emission lines of J1044+0353 and J1418+2102 (see also \citealt{berg21a}), J1016+3754, J0337-0502, J1323-0132 and J1545+0858. For these objects we fitted the \heii~$\lambda$1640 and \oiii$\lambda\lambda$1661,6 with two components. We tested a two-Gaussian component fitting also on the \ciii\ doublet, without finding a significant improvement in our results. This is due to the very small wavelength separation of the \ciii\ doublet lines, which results in degenerate line centroids that make it difficult to use multiple components.
Interestingly, for J0337-0502, J1044+0353, J1418+2102, J1323-0132 (see Fig.~\ref{fig:uvfitcarbon}) we also observed a doubled-peak profile in the \civ~ doublet. As discussed in \citealt{berg21a}, such profiles are the result of resonant scattering, whereas broadening can be due to radiation transport/scattering. Due to the different line processes responsible for \civ\ emission, it should be noted that the properties of the multi-component fits to this line were not constrained with the same kinematics as the nebular emission lines.

For the purpose of this work, we chose to only consider the narrow (and dominant) component of our emission lines which on average constitutes $>70$\%\ of the total flux. This allows us to maintain the highest accuracy in the emission line diagnostics derived here, since each emission line component originates in gas with different physical conditions \citep[ionization degree, temperature, density, velocity etc, see e.g.][]{james09}. Indeed, broad emission indicates large velocities that can be driven by different mechanisms such as stellar winds, galactic-outflows or turbulence, and possibly linked to different ionization sources, such as photoionization and/or shocks (e.g., \citealt{izotov07,james09,amorin12,bosch19,komarova21,hogarth20}).

It should be noted that we were unable to fit a broad component emission in the UV nebular lines of all the galaxies that displayed broad component emission in the optical due to S/N limitations and the faintness of UV emission lines. For these cases, we are confident that the possible contribution from broad component emission to the single (narrow) component fit is negligible and within the uncertainties of the emission lines. 
We will investigate possible differences of the conditions of the broad component in our next paper focused on the kinematics and ionization mechanisms (\citetalias{mingozzi22}).

\subsubsection{UV emission line detections}\label{sec:emission-lines-detection}
\begin{figure*}
\begin{center}
    \includegraphics[width=0.75\textwidth]{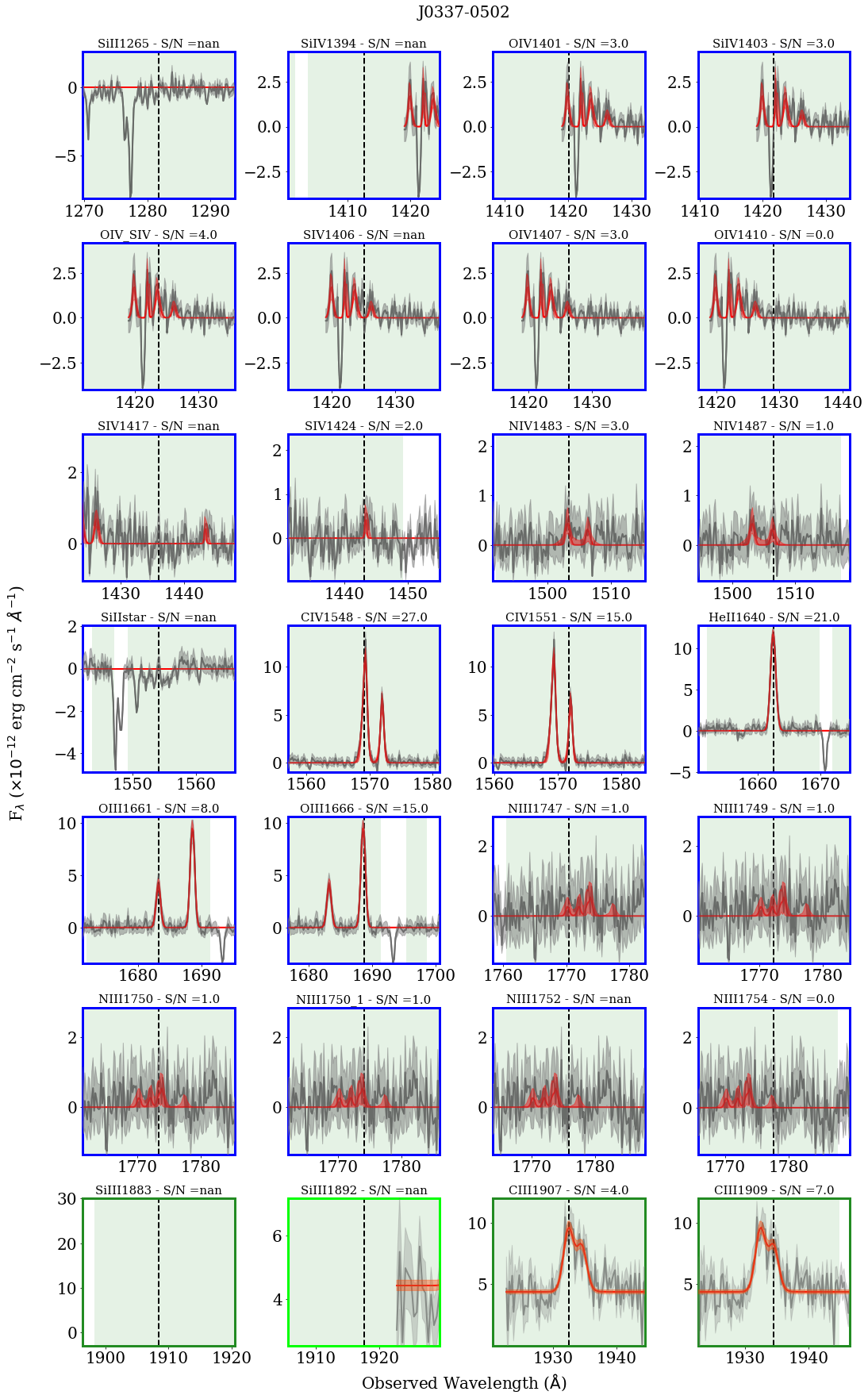}
\end{center}
\caption{UV emission-lines fitted by our fitting routine for the galaxy J0337-0502 (i.e., SBS0335-052~E): the spectrum and the fit are reported in black and red, while the spectrum and fit uncertainties are shown in shaded gray and red, respectively. The id and signal-to-noise of each zoomed line is indicated on top of each panel, whose margins are coloured according to the binning applied to the spectrum before the fitting (HR rebinned of 15 in blue, MR rebinned of 6 in dark green, MR rebinned of 12 in light green). The dashed black vertical lines indicate the observed wavelength of each line according to the redshift of the galaxy $z_{lit}$, reported in Tab.~\ref{tab:classyprop}.}
\label{fig:examplefit}
\end{figure*}
\begin{figure*}
\begin{center}
    \includegraphics[width=1\textwidth]{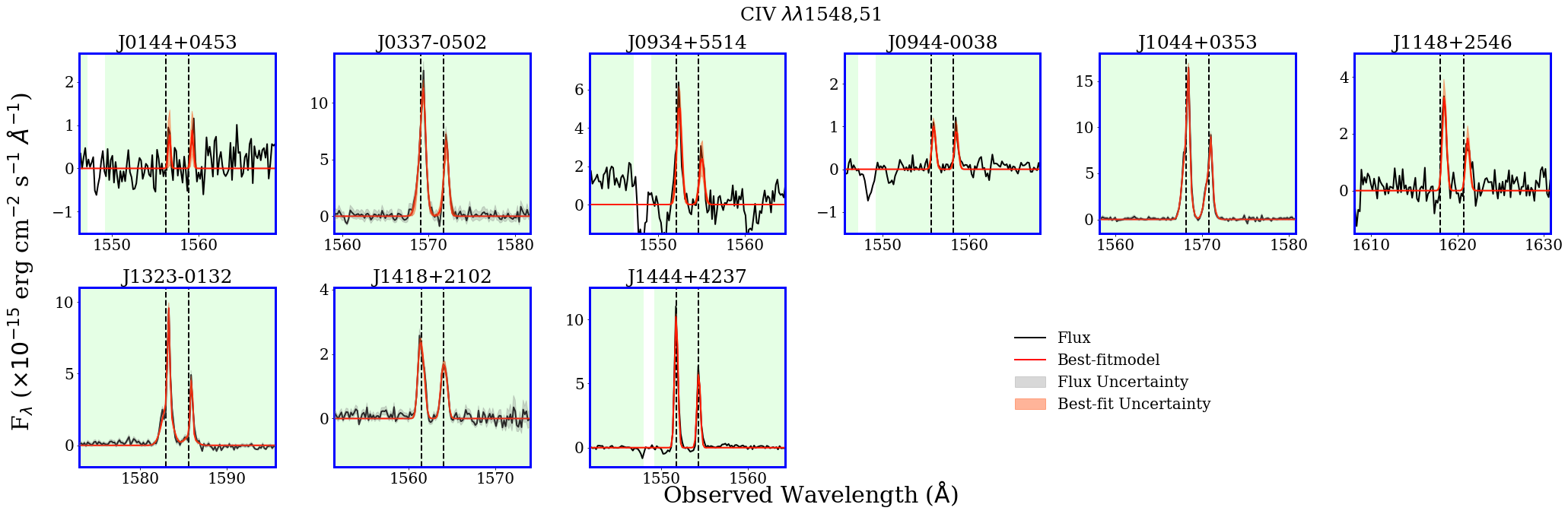}
    \includegraphics[width=1\textwidth]{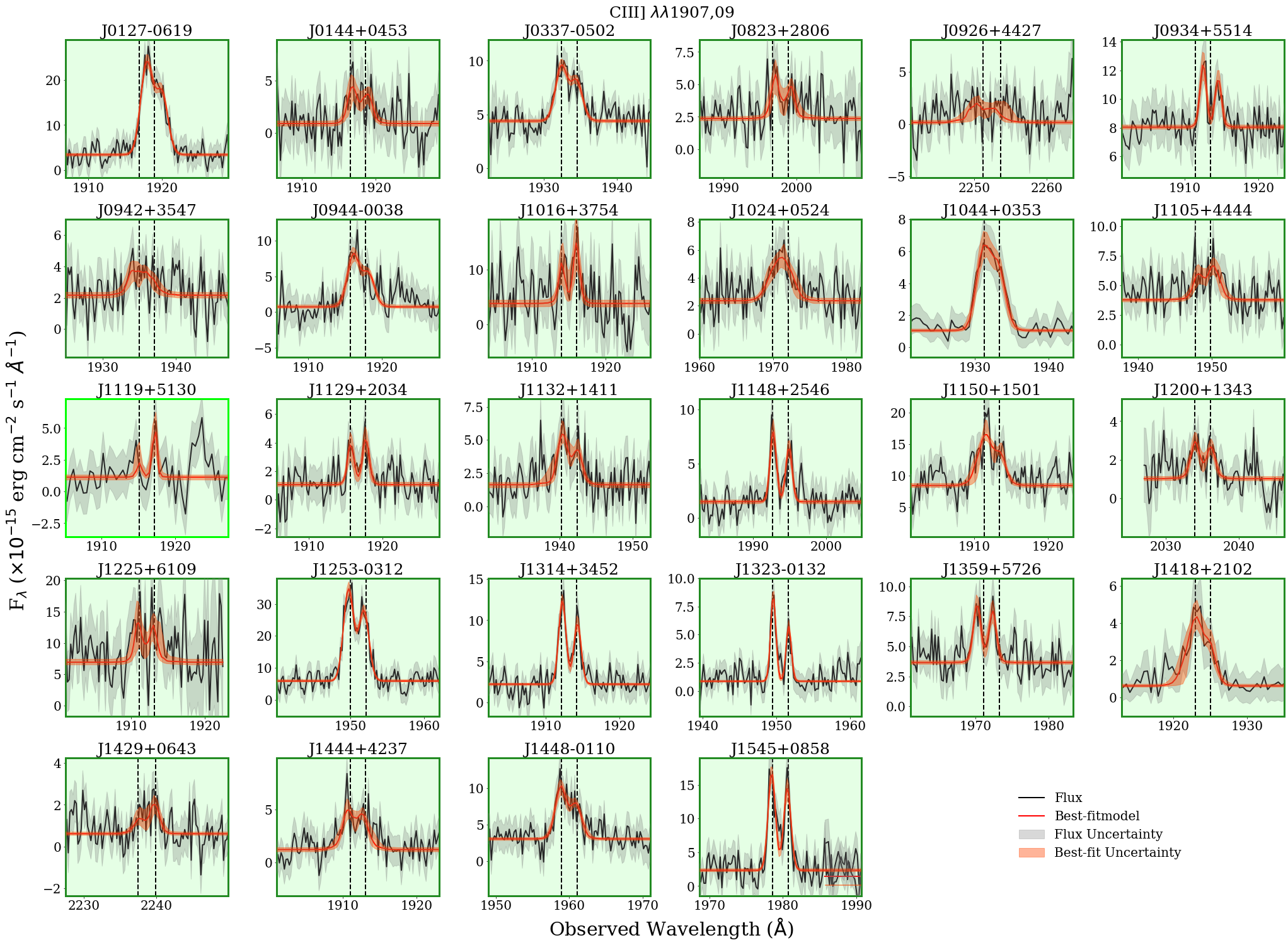}
\end{center}
\caption{Fit of the \civ~$\lambda\lambda$1548,51 and \ciii~$\lambda\lambda$1907,9 emission line doublet with $S/N>3$, visible in 9 and 28 galaxies, respectively. The observed flux and the best-fit model are shown in black and red, respectively, while their uncertainties are given by the gray and red shades. The black dashed vertical lines indicate the line positions, taking into account $z_{\rm lit.}$. The margins are coloured according to the binning applied to the spectrum before the fitting (HR rebinned of 15 in blue, MR rebinned of 6 in dark green, MR rebinned of 12 in light green).}
\label{fig:uvfitcarbon}
\end{figure*}
While there is a plethora of strong optical emission lines that are uniformly detected throughout the sample, UV emission lines can be mostly faint and sometimes not detected at all. It is therefore important for us to highlight in how many galaxies the UV emission lines are clearly detected. 
As explained in Sec.~\ref{sec:cos}, we fitted the HR and MR/LR Coadds spectra, after performing different levels of binning. We consider UV emission lines with $S/N>3$ to be detections.  We took into account the results from the doubly-rebinned spectra (of 30 and 12 pixels for HR and MR/LR Coadds, respectively) only for the emission lines with a $S/N<3$. If the $S/N$ is still lower than the chosen threshold, then we consider the flux as an upper limit, while if the line is not observed at all, as a non-detection. As an example, in Fig.~\ref{fig:examplefit} we show the UV emission-lines fitted by our fitting routine for the galaxy J0337-0502 (i.e., SBS0335-052~E). The spectrum and the fit are reported in black and red, while the spectrum and fit uncertainties are shown in shaded gray and red, respectively. The id and signal-to-noise of each zoomed line is indicated on top of each panel, whose margins are coloured according to the binning applied to the spectrum before the fitting (HR rebinned of 15 in blue, HR rebinned of 30 in cyan, MR rebinned of 6 in dark green, MR rebinned of 12 in light green).
In Fig.~\ref{fig:uvfitcarbon} instead we show the fitted CLASSY COS spectra of the \civ~$\lambda\lambda$1548,51 and \ciii~$\lambda\lambda$1907,9 emission lines for all the galaxies in which the lines are detected with $S/N>3$. In App.~\ref{app:B}, in Fig.~\ref{fig:uvfitniv}--Fig.~\ref{fig:uvfitoiii} we show analogous figures for \niv~$\lambda\lambda$1483,87, \heii~$\lambda$1640, \oiii~$\lambda\lambda$1661,6, \niii~$\lambda\lambda$1747–54 and \siiii~$\lambda\lambda$1893,92 emission lines, respectively, with $S/N>3$. In the following we describe the detection of each of these lines within the CLASSY sample.

{\underline {\civ~$\lambda\lambda$1548,51}} is observed in pure emission with $S/N>3$ in only 9 CLASSY galaxies (see Fig.~\ref{fig:uvfitcarbon}), while in the other galaxies it shows a P-Cygni profile or is only in absorption. Generally, the \civ\ doublet is dominated by a broad P-Cygni profile due to winds of luminous O stars (e.g., \citealt{shapley03,steidel16,rigby18a,llerena21}). Only high resolution spectra such as those of the CLASSY survey can allow to successfully separate the stellar and the nebular components of \civ\ emission (see also \citealt{crowther07,quider09}).
Pure nebular emission in \civ~$\lambda\lambda$1548,51 has been recently detected in $z>6$ targets \citep{stark15,mainali17,schmidt17} and, rarely, in local galaxies \citep{berg16,senchyna17,senchyna19a,berg19a,wofford21,senchyna22a}. 
For this paper, we only take into account only the 9 CLASSY galaxies with \civ\ in pure emission, without considering the galaxies that show a P-Cygni or pure absorption line profile. All these galaxies also show \ciii~$\lambda\lambda$1907,9, \heii~$\lambda$1640 and \oiii~$\lambda$1666, apart from J0934+5514 (i.e., Izw~18), where the \oiii~$\lambda$1666 is undetected because of a MW line contamination.

{\underline {\ciii~$\lambda\lambda$1907,9}}, often the strongest UV nebular emission line, is observed with $S/N > 3$ in 28 CLASSY galaxies (see Fig.~\ref{fig:uvfitcarbon}). Among these, we can see this doublet de-blended in 26 objects (excluding J1044+0353 and J1418+2102; see Sec.~\ref{sec:emission-lines-constrains}). This doublet is a density diagnostic, as we will discuss in Sec.~\ref{sec:results-ne} and Sec.~\ref{sec:discussion-ne}. 

{\underline {\niv~$\lambda\lambda$1483,87}} is observed with $S/N>3$ in only 6 CLASSY galaxies (both doublet lines are observed only in J1044+0353, J1253-0312 and J1545+0858). This doublet has rarely been seen in emission in star-forming galaxies \citep{fosbury03,raiter10,vanzella10,stark14}. 
These lines are probably due to young stellar populations, and, if the source is not hosting an AGN, they could be a signature of massive and hot stars with an associated nebular emission \citep{vanzella10}. 
This doublet is also a density diagnostic \citep{keenan95}, and it traces higher-ionization regions with respect to the \ciii\ and \siiii\ doublets (see Sec.~\ref{sec:results-ne} and Sec.~\ref{sec:discussion-ne}).

{\underline {\heii~$\lambda$1640}} is detected with $S/N>3$ in 19 CLASSY galaxies and generally shows a narrow profile (median velocity dispersion of $\sigma \sim 55$~km/s), indicating its nebular nature. However, in the spectra of J0942+3547, J1129+2034, J1200+1343, J1253-0312 and J1314+3452 the line profile looks broader (with $\sigma$ up to 200~km/s), which suggests the presence of a stellar component residual despite the removal of the C\&B best-fit stellar continuum (see e.g., \citealt{nanayakkara19,senchyna21a}). 

{\underline {\oiiiuv~$\lambda1661,6$}}, one of the strongest UV emission lines, has $S/N>3$ in 22 CLASSY galaxies. In J0127-0619 and J1225+6109, where one of the two lines of the \oiiiuv\ doublet is contaminated by a MW absorption line, we estimated the flux from the other line, using a line ratio measured from \pyneb\ of 0.4 (valid at $n_e=100$~cm$^{-3}$ and $T_e=10^4$~K).
These are auroral lines, similar to the optical \oiii~$\lambda$4363, and thus can be used as a temperature diagnostics in comparison with the optical nebular \oiii~$\lambda\lambda$4959,5007,  as we will discuss in Sec.~\ref{sec:results_temp} and Sec.~\ref{sec:discussion-Te}. 

{\underline {\niii~$\lambda\lambda$1747–54}} is a multiplet (i.e., a blend of emission at 1746.8, 1748.6, 1749.7, 1750.4, and 1752.2~\AA; \citealt{keenan94}). These lines are suggested to have a nebular origin and may be used in the so-called UV-BPT diagrams \citep{feltre16} to discriminate between SF and AGN activity. However, the multiplet is usually revealed in spectra of WN-type stars (e.g., \citealt{crowther97}). Interestingly, only one galaxy of the CLASSY sample, J0127-0619 (i.e., Mrk~996) shows this multiplet in clear emission with S/N~$\sim9$. WR features (mainly late type WN stars) in this galaxy were discovered for the first time by \citet{thuan96}, while their distribution as well as the ISM abundances and kinematics were investigated by \citet{james09}. This could indicate a non-ISM origin of this emission (see also \citetalias{mingozzi22}). A hint of emission with S/N~$\sim3$ is observed also in J1253-0312.

{\underline {\siiii~$\lambda\lambda$1893,92}} is observed with $S/N>3$ in 6 CLASSY galaxies. These lines are generally very faint, but also in many targets one of the two or both fall out of COS observed wavelength range (both doublet lines are observed only in J1044+0353, J1253-0312 and J1448-0110). Similarly to \ciii, this doublet is a density diagnostic. 

Along with these UV emission lines we also fitted the other lines shown in Fig.~\ref{fig:examplefit}, namely: \ion{Si}{2}$\lambda$1265, \ion{Si}{4}$\lambda$1394, \ion{O}{4}$\lambda$1401, \ion{Si}{4}$\lambda$1403, \ion{O}{4}\ion{Si}{4}, \ion{S}{4}$\lambda$1406, \ion{O}{4}$\lambda$1407, \ion{S}{4}$\lambda$1410M \ion{S}{4}$\lambda$1417, \ion{S}{4}$\lambda$1424, \ion{Si}{2}$^*\lambda$1534. Many of these lines can be visible in emission in galaxies that show also \civ\ in pure emission, as shown in Fig.~\ref{fig:examplefit}. Due to their lack of detection throughout the sample with enough $S/N$, we do not consider these lines any further.

\section{Deriving Physical Properties of the ISM}\label{sec:methods}
\begin{table*}
\centering
\caption{Optical ISM diagnostics in the different ionization zones according to the literature and available in this work.}
\label{tab:optdiags}
\begin{tabular}{c|c|c|c|c} 
\hline
\hline
Property            &   \multicolumn{4}{c}{Ionization Zone} \\
    & Low                                   & Intermediate  
            & High                                  & Very-high                     \\
\hline
$E(B-V)$    & \balmer                               & \balmer       
		    & \balmer                               & \balmer                       \\
[2ex]
$n_e$       & [\ion{S}{2}]~\W6717/\W6731              & [\ion{Cl}{3}]~\W5518/\W5538 
		    & [\ion{Ar}{4}]~\W4714/\W4741             & [\ion{Ar}{4}]~\W4714/\W4741     \\ 
		    & [\ion{O}{2}]~\W3729/\W3727              & [\ion{Fe}{3}]~\W4701/\W4659  
		    &                                         &                               \\
[2ex]
$T_e$       & [\ion{N}{2}]~\W5755/\W6584              & [\ion{S}{3}]~\W6312/\W9069
		    & [\ion{O}{3}]~\W4363/\W5007             & [\ion{O}{3}]~\W4363/\W5007     \\ 
            & [\ion{S}{2}]~\W\W4069,72/\W\W6717,31    & 
            &                                       &                               \\
			& [\ion{O}{2}]~\W\W3727,29/\W\W7320,30    & 
			&                                       &                               \\
[2ex]
log$U$      & \multicolumn{2}{c|}{--- --- [\ion{S}{3}]~\W\W9069,9532 / [\ion{S}{2}]~\W\W6717,31 --- ---} & & \\
		    & \multicolumn{3}{c|}{--- --- [\ion{O}{3}]~\W5007 / [\ion{O}{2}]~\W\W3727,9 --- --- } & \\
            & &  \multicolumn{3}{c}{--- --- [\ion{Ar}{4}]~\W\W4714,41 / [\ion{Ar}{3}]~\W7135 --- ---}     \\ 
\hline
\end{tabular}
\tablecomments{ISM diagnostics available in the optical for each ionization zone
in a 4-zone model \citep[see][]{berg21a}.
Specifically, we list diagnostic line ratios for dust attenuation ($E(B-V)$), 
electron density ($n_e$), electron temperature ($T_e$) and ionization parameter (log($U$)).}
\end{table*}


\begin{table*}
\centering
\caption{ISM UV diagnostics in the different ionization zones according to the literature and available in this work.}
\label{tab:uvdiags}
\begin{tabular}{c|c|c|c|c} 
\hline
\hline
Property            &   \multicolumn{4}{c}{Ionization Zone} \\
    & Low                                   & Intermediate  
            & High                                  & Very-high                     \\
\hline
$E(B-V)$    & $\beta$-slope                         & $\beta$-slope
            & $\beta$-slope                         & $\beta$-slope                 \\
[2ex]
$n_e$       & \nodata                               & \ion{C}{3}]~\W1907/\W1909 
            & \multicolumn{2}{c}{\hspace{12ex}--- --- \ion{N}{4}~\W1483/\W1487 --- ---}     \\ 
            &                                       & \ion{Si}{3}]~\W1883/\W1892  
            &                                       &                               \\
[2ex]
$T_e$       & \nodata                               & \nodata
            & \ion{O}{3}]~\W1666/[\ion{O}{3}]~\W5007  & \ion{O}{3}]~\W1666/[\ion{O}{3}]~\W5007 \\ 
[2ex]
log$U$      & & \multicolumn{3}{c}{--- --- { \ion{C}{4}~\W\W1548,51/\ion{C}{3}]~\W\W1907,9} --- ---}   \\ 
		    & & \multicolumn{3}{c}{--- --- \,\,\,\,\,\,\,\,\,\, EW(\ion{C}{4}~\W\W1548,51) \,\,\,\,\,\,\,\,\,\,\,\,\, --- ---}   \\ 
\hline
\end{tabular}
\tablecomments{Same as Table~\ref{tab:optdiags}.}
\end{table*}

\hii\ regions are stratified, with higher ionization species, such as \ariv\ or \oiii, closer to the ionization source and lower ionization species, such as \sii\ or \oii, in the outer parts.
Typically \hii\ regions are modeled by three zones of different ionization: the low-, intermediate- and high-ionization zones. As pointed out in \citet{berg21a,berg22}, high-$z$ systems and their local analogues are characterized by the presence of high-energy UV and optical emission lines due to their low metallicity and thus extreme radiation fields, revealing the presence of an additional `very-high-ionization' zone. It is important to stress that  radiation fields and metallicity are tightly linked such that stellar populations of lower metallicity have harder radiation fields.
This led \citet{berg21a} to extend the classical 3-zone model to a 4-zone model, adding the He$^{+2}$ species necessary to produce the observed \heii\ emission via recombination (ionisation potential $E>54.42$~eV).

Overall, an accurate determination of \hii\ region properties requires reliable tracers for each zone. 
This is because different ions are tracing different conditions of nebulae in terms of density, temperature and ionization, since they are not co-spatial (e.g., \citealt{nicholls20}). 
The COS aperture on CLASSY targets is covering multiple \hii\ regions or even the entire galaxy for the most compact objects.
Hence, we can employ the use of multiple diagnostics both in the optical and in the UV to trace the conditions in the different ionization regions and compare their properties, investigating the ISM structure of our targets with the utmost detail. 
Here we stress that a great advantage of the CLASSY survey is the simultaneous coverage of many optical and UV diagnostic lines. In particular, UV emission lines are coming from higher ionization zones, which gives us access to a wider range of ionization zone tracers than typically available from the optical alone. 

Then, we employed iteratively the \pyneb\ task {\it getCrossTemDen}, that combines a density and a temperature diagnostic, and ultimately converges to a final value of $n_e$ and $T_e$. 
First, we calculated the intrinsic Balmer line ratios using
\pyneb, assuming a Case-B Hydrogen recombination with a starting temperature of $T_e = 1\times 10^4$ K and $n_e = 10^2$ cm$^{-3}$, considered appropriate for typical star-forming regions \citep{osterbrock89,osterbrock06}. 
Then, we iteratively calculated density and temperature, using the reddening value to correct the line ratio used as temperature tracer, and updating at each cycle the \ha\ and \hb\ emissivities (and thus $E(B-V)$), $n_e$ and $T_e$, only if the new value obtained was finite.
Our iterative approach stops once the difference in temperature between two cycles becomes lower than 20~K. To estimate the fiducial values and errors on $n_e$ and $T_e$, we run the {\it getCrossTemDen} task 500 times for each different combination of $n_e$ and $T_e$ diagnostics, taking the median of values and the standard deviation for uncertainties.
Once densities and temperatures are known in each zone of the nebula, it is then possible to calculate the corresponding ionic abundances, with a similar iterative procedure using the \textit{getIonAbundance} \pyneb\ task, and the same method to estimate the uncertainties.

Table~\ref{tab:optdiags} and Table~\ref{tab:uvdiags} show the optical and UV diagnostics investigated for the different ionization zones in this work. 
Unfortunately in this work we lack the \neiii~$\lambda$3342/$\lambda$3868 ratio that \citet{berg21a} used to estimate the temperature of the very-high ionization zone. 
We note, however, that this ratio provided results consistent to the values obtained for the high ionization zone \citep{berg21a}. Our set of UV lines is characteristic of the intermediate and high-ionization zone. 
The comparison between the different properties calculated with optical and UV diagnostics are shown and discussed in Sec.~\ref{sec:results} and Sec.~\ref{sec:discussion}.
In the following sections we provide the details about each calculated quantity.

\subsection{Dust attenuation}\label{sec:method-dust}
Before comparing ratios of emission lines separated in wavelength throughout the UV-optical wavelength regime, the emission lines were corrected for the intrinsic galaxy dust attenuation in terms of $E(B-V)$. $E(B-V)$ was determined comparing the observed relative intensities of the strongest Balmer lines available in our optical spectra (i.e., \ha/\hb, \hb/\hg, \hb/\hd) with their intrinsic values.
These intrinsic values depend on the density and temperature of the gas, which we estimate with the corresponding diagnostics for each ionization zone, as explained in Sec.~\ref{sec:methods-den} and Sec.~\ref{sec:methods-temp}.
The final reddening estimate is an error-weighted average of the
H$\alpha$/H$\beta$, H$\gamma$/H$\beta$, and H$\delta$/H$\beta$ reddening values.

To correct the optical emission lines, we applied the \citet{cardelli89} reddening law with $R_v = 3.1$, which is appropriate for the CLASSY emission line fluxes \citep{berg22}. Indeed, \citet{wild11a} found that the nebular attenuation curve has a slope similar to the MW attenuation curve, rather than that of the SMC \citep{gordon98,gordon03} or the one from \citet{calzetti00}. 
The UV emission lines, instead, were corrected assuming a \citet{reddy16} attenuation curve with $R_v = 2.191$, which represents the first spectroscopic measurement of the shape of the far-UV dust attenuation curve for galaxies at high-redshift ($z\sim3$), i.e., systems that are analogous to our CLASSY sample.

Dust attenuation can be also estimated from comparing the observed slope of the UV spectra in the range $1400-1800$\AA\ (the `$\beta$ slope'; see e.g., \citealt{leitherer99,calzetti94}) with the intrinsic slope of the models used in the best-fit of the stellar populations \citep{calzetti00,reddy16}. This quantity ($E(B-V)_{UV}$ hereafter) is given as an output of the UV stellar continuum fitting described in Sec.~\ref{sec:data-analysis-stellar-cont-uv}. $E(B-V)_{UV}$ represents the stellar attenuation and its relation with the gas $E(B-V)$ derived from the Balmer decrement is not trivial, as it is discussed in Sec.~\ref{sec:discussion-ext}.

\subsection{Density}\label{sec:methods-den}
\begin{figure*}
\begin{center}
    \includegraphics[width=1\textwidth]{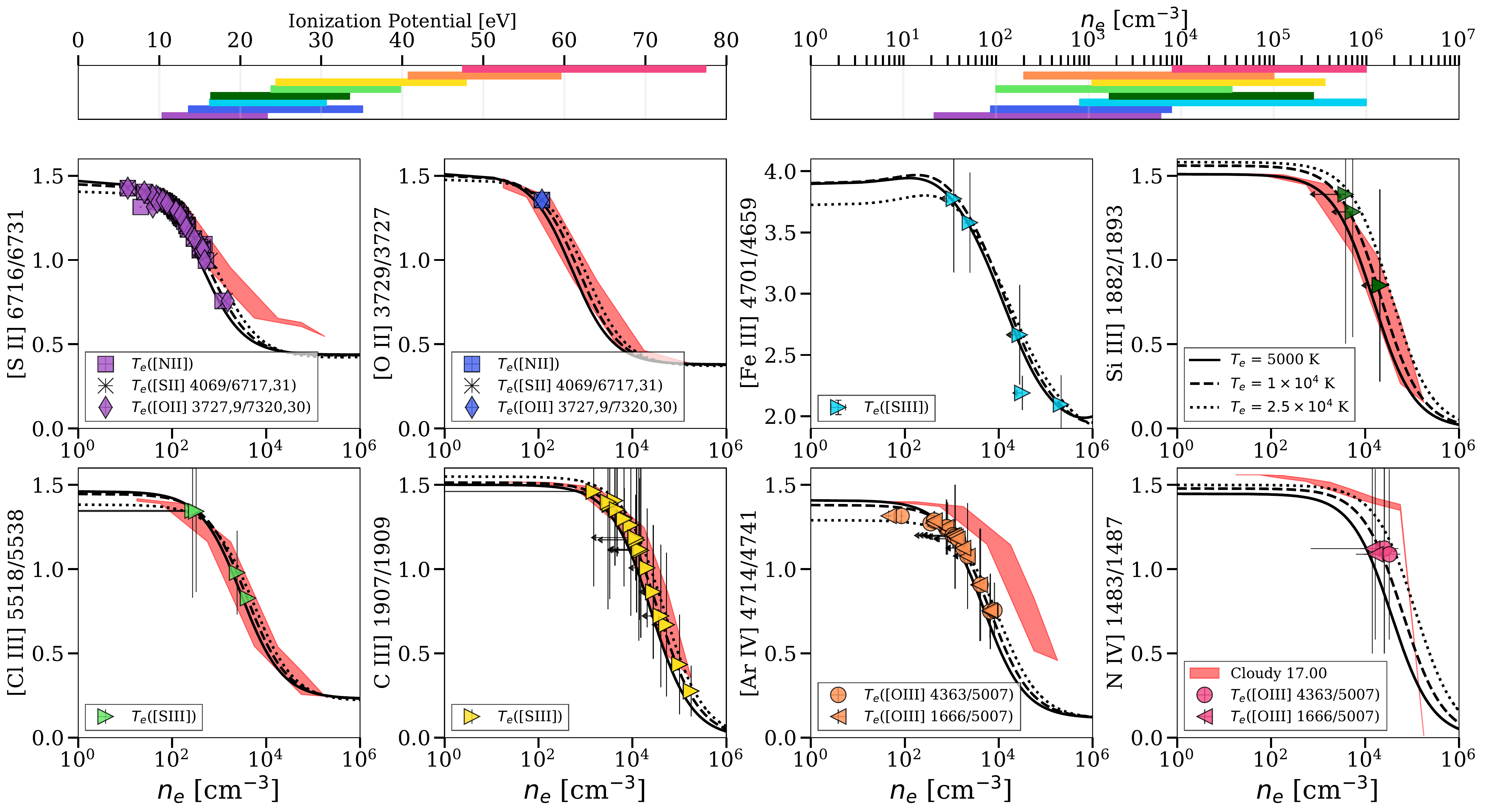}
\end{center}
\caption{The eight square panels show the set of diagnostics to estimate the gas density $n_e$ in the different ionization zones (see also Tab.~\ref{tab:optdiags} and Tab.~\ref{tab:uvdiags}). The minor top left and right panels show the ionization potential and the range of traced densities of the diagnostics taken into account, calculated with \pyneb, using the same color-coding. Specifically, in order of ionization potential, we have: \sii~$\lambda$6717/$\lambda$6731 (purple), \oii~$\lambda$3729/$\lambda$3727 (blue), \feiii~$\lambda$4701/$\lambda$4659 (turquoise), \fsiiii~$\lambda$1883/\siiii~$\lambda$1892 (dark green), \cliii~$\lambda$5518/$\lambda$5538 (green), \fciii~$\lambda$1907/\ciii~$\lambda$1909 (gold), \ariv~$\lambda$4714/$\lambda$4741 (orange) and \niv~$\lambda$1483/$\lambda$1487 (red). The different symbols show the CLASSY calculated values with \pyneb\ using the available temperature diagnostics described in Fig.~\ref{fig:temp_pyneb_cloudy}. The solid, dashed and dotted black curves represent \pyneb\ predictions at $T_e = 5\times10^3$~K,~$1\times10^4$~K and $ 2.5\times10^4$~K, as reported in the legend, while the shaded red regions (very narrow due to the low dependence of density on temperature) show the predictions from our \cloudy\ models of \citet{berg19a,berg21a}.}
\label{fig:den_pyneb_cloudy}
\end{figure*}
The electron density $n_e$ can be derived from intensity ratios of lines emitted by a single ion from two levels with nearly the same energy, but different radiative-transition probabilities or different collisional de-excitation rates  \citep{osterbrock89}.
As a guide to the reader, in Fig.~\ref{fig:den_pyneb_cloudy} we highlight the different combinations of diagnostics considered throughout this study and their characteristics, as introduced in Table~\ref{tab:optdiags} and Table~\ref{tab:uvdiags}. Specifically, in the eight main panels of Fig.~\ref{fig:den_pyneb_cloudy}, we report the measurements of each line ratio used as tracer of $n_e$ for the CLASSY galaxies and its corresponding density calculated with \pyneb. The dots are color-coded according to the ion species and are the same used in the top two panels, where the ionization potential and the traced density range of each ion and line ratio, are reported.
The different symbols indicate which line ratio we assumed to estimate the temperature (described in Fig.~\ref{fig:temp_pyneb_cloudy} and Sec.~\ref{sec:methods-temp}), as reported in the legend. 
The black curves show the variation of each line ratio as a function of the temperature, considering $T_e = 5\times10^3$~K,~$1\times10^4$~K and $ 2.5\times10^4$~K.

Overall, looking at Fig.~\ref{fig:den_pyneb_cloudy}, for the low-ionization zone, the most typical density diagnostics are \sii~$\lambda$6717/$\lambda$6731 and \oii~$\lambda$3729/$\lambda$3727, sensitive in the range $\sim40-5000$~cm$^{-3}$. Moving towards higher ionization potential, other density tracers are \cliii~$\lambda$5518/$\lambda$5538,  sensitive in the range $\sim10^2 - 2\times10^4$~cm$^{-3}$, or \fsiiii~$\lambda$1883/\siiii~$\lambda$1892 and \fciii~$\lambda$1907/\ciii~$\lambda$1909 in the UV, tracing values in the range $\sim10^3 - 2\times10^5$~cm$^{-3}$. 
Moreover, \citet{mendez-delgado21a} proposed the use of \feiii~$\lambda$4701/$\lambda$4659, sensitive in the range $10^3-10^6$~cm$^{-3}$.
At the highest ionization levels ($E>40$~eV), possible diagnostics are \ariv~$\lambda$4714/$\lambda$4741 in the optical and \niv~$\lambda$1483/$\lambda$1487 in the UV, sensitive up to $n_e\sim 1\times10^5$~cm$^{-3}$ and $n_e\sim 1\times10^6$~cm$^{-3}$. respectively. 

In general, the predictions by \pyneb\ accurately represent the line ratios observed within the uncertainties, which can unfortunately be very large for some transitions. In these cases, our measurements can be considered upper limits of the density.
Moreover, from Fig.~\ref{fig:den_pyneb_cloudy}, it is clear that the density diagnostics have generally a very low dependence on $T_e$. 
The highest dependence on temperature is found for the \niv~$\lambda$1483/$\lambda$1487 line ratio, whose derived densities can be different up to $\sim1$~dex, with the highest values derived for the lowest temperatures and vice-versa.

The main aspect highlighted by Fig.~\ref{fig:den_pyneb_cloudy} is that the density range traced by the different diagnostics can vary considerably. This depends on the critical density, that is defined as the density at which collisional transitions are equally probable with radiative transitions \citep{osterbrock89,osterbrock06}.
Hence, transitions with higher critical densities can be used as diagnostics in denser environments. 
Interestingly, we noticed that higher critical densities do not automatically correspond to higher ionization (see upper panels of Fig.~\ref{fig:den_pyneb_cloudy}), which means that the density structure could be not directly related to the ionization structure. 
For instance, \siiii\ and \ciii\ transitions are characterized by a lower ionization potential than \ariv\ or \niv, but overall they can probe higher densities than \ariv\ and similar values to \niv. We will further comment about this in Sec.~\ref{sec:results-ne} and Sec.~\ref{sec:discussion-ne}.
Also, \feiii\ has an ionization potential comparable to \sii\ or \oii, but it is probing electron densities between $10^3$ and $10^6$~cm$^{-3}$. 

Finally, the shaded red regions in Fig.~\ref{fig:den_pyneb_cloudy} show the predictions from the \cloudy~17.00 \citep{ferland13} models from \citet{berg19a,berg21a}, that we used to estimate the ionization parameter, which we discuss in Sec.~\ref{sec:methods-logu}. 
The Cloudy models we took into account are made using Binary
Population and Spectral Synthesis (BPASSv2.14; \citealt{eldridge16, stanway16}) burst models for the input ionizing radiation field. The parameter space covered is appropriate for our sample, including an age range of $1-10$~Myr for young bursts and a range in ionization parameter of
$−4.0 < log(U) < 0$, matching stellar and nebular metallicities ($Z\star = Z_{\rm neb} = 0.001, 0.002, 0.004, 0.008$, corresponding to $0.05, 0.10, 0.20, 0.40$~Z$\odot$).
In particular, \citet{berg19a} used the GASS10 solar abundance ratios (including dust) to initialize the relative gas-phase
abundances, then scaling them to match the observed values for nearby metal-poor dwarf galaxies. 
Specifically, we calculated the median value of Cloudy predictions in the range of densities and temperatures taken into account, and the boundaries of the shaded red regions represent the $\pm3\sigma$ of the distribution. These regions appear narrow because of the very low dependence of density on temperature (see \citealt{berg18,berg19a,berg21a} for all the details).
Overall, from Fig.~\ref{fig:den_pyneb_cloudy} we find good agreement between \cloudy\ and \pyneb, despite minor differences in the default atomic data used by each code.
The main difference that we underline is that \cloudy\ models for \ariv\ are shifted towards higher densities. 
This discrepancy implies that \cloudy\ \ariv\ densities could be $\lesssim 1$~dex higher than those measured with \pyneb. This discrepancy could be due to the different atomic data used by \cloudy\ (see \citealt{dedios17,dedios21} and references within for more details).
Finally, we note that our \cloudy\ models do not include \feiii\ lines.

\subsection{Temperature}\label{sec:methods-temp}
\begin{figure*}
\begin{center}
    \includegraphics[width=1\textwidth]{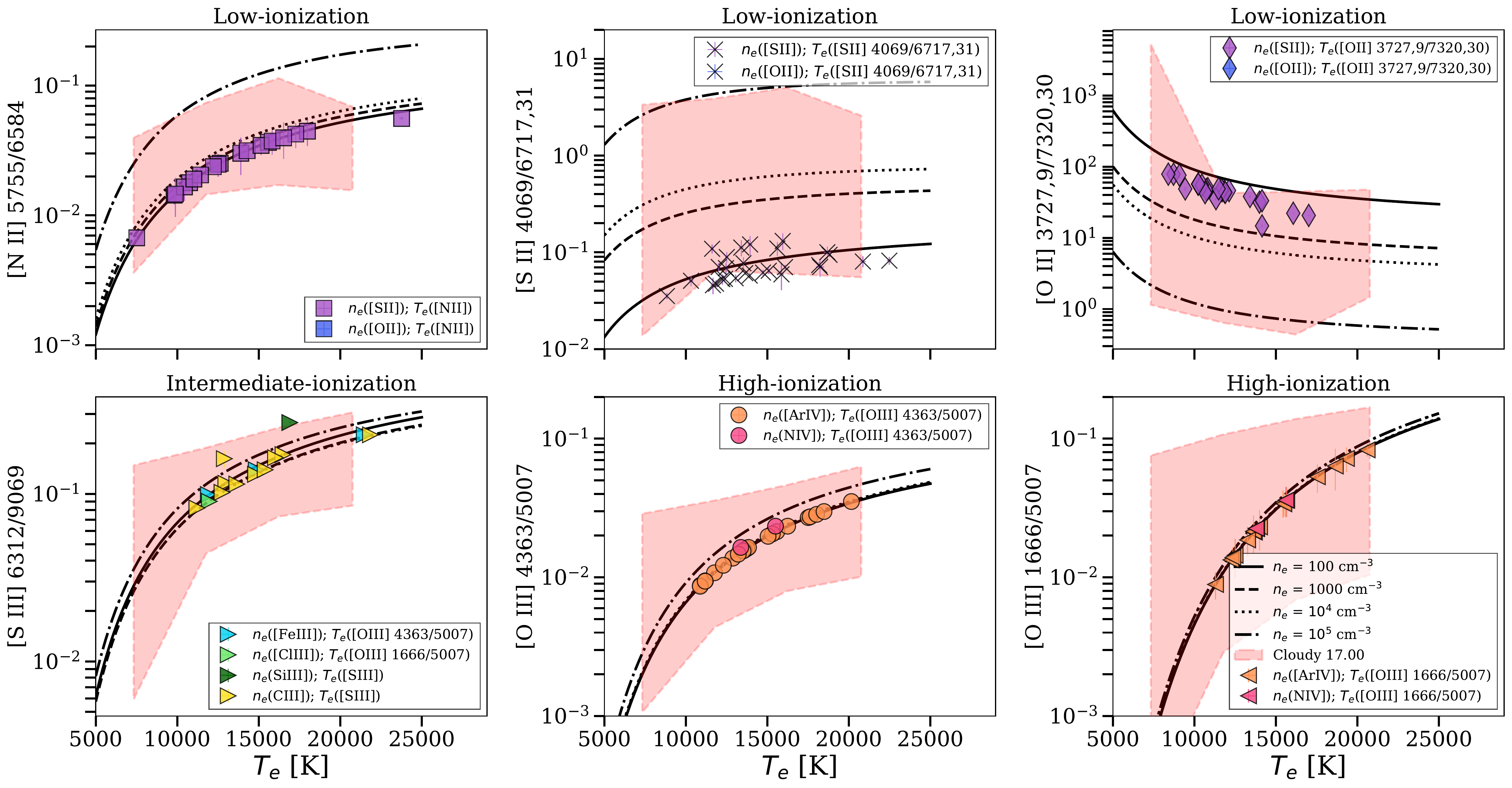}
\end{center}
\caption{Set of diagnostics to estimate the gas temperature in the CLASSY galaxies in the different ionization zones: \nii~$\lambda$5755/\nii~$\lambda$6584, \sii~$\lambda\lambda$4069/\sii~$\lambda\lambda$6717,31 and \oii~$\lambda\lambda$3727,29/\oii~$\lambda\lambda$7320,30 for the low-ionization zone, \siii~$\lambda$6312/\siii~$\lambda$9069, \oiii~$\lambda$4363/\oiii~$\lambda$5007 and \oiiiuv~$\lambda$1666/\oiii~$\lambda$5007. The different filled symbols show the CLASSY calculated values with \pyneb, as reported in the legend. 
The black curves represent \pyneb\ predictions at densities $n_e = 100$~cm$^{-3}$,~$5000$~cm$^{-3}$,~$10^4$~cm$^{-3}$ and ~$10^5$~cm$^{-3}$, while the shaded red regions show the predictions from out \cloudy\ models.}
\label{fig:temp_pyneb_cloudy}
\end{figure*}
The temperature $T_e$ can be determined via the intensity ratios of particular emission line doublets, emitted by a single ion from two levels with considerably different excitation energies \citep{osterbrock89}.
To guide the reader, we show the available temperature diagnostics used within this study in the six panels of Fig.~\ref{fig:temp_pyneb_cloudy}, as introduced in Table~\ref{tab:optdiags} and Table~\ref{tab:uvdiags}. 
In Fig.~\ref{fig:temp_pyneb_cloudy}, we also report the measurements of each line ratio for the CLASSY galaxies and its corresponding temperature, using symbols and colors consistent with Fig.~\ref{fig:den_pyneb_cloudy}, to indicate which density diagnostic we used while calculating the temperature using the {\it getCrossTemDen} task, as reported in the legend. 

Overall, the most used optical $T_e$ tracers are given by \nii~$\lambda$5755/$\lambda$6584, \siii~$\lambda$6312/$\lambda$9069 and \oiii~$\lambda$4363/$\lambda$5007 for the low-, intermediate- and high-ionization zones, respectively.
Indeed, \nii\ emission is stronger in the outer parts of \hii\ regions, where the ionization is lower and O mostly exists as O$^+$ \citep{osterbrock89}.
With respect to the low ionization zone in particular, other available temperature indicators are the \sii~$\lambda\lambda$4069/$\lambda\lambda$6717,31 and \oii~$\lambda\lambda$3727,29/$\lambda\lambda$7320,30 line ratios. 
Their main drawback is the wide separation in wavelength, that introduces larger relative uncertainties via the dust attenuation correction. Also, \sii~$\lambda$4069 is usually very faint, while \oii~$\lambda\lambda$7320,30 lines can be affected by telluric absorption depending on the redshift of the galaxy.

Unfortunately, \nii~$\lambda$5755/$\lambda$6584 and \siii~$\lambda$6312/$\lambda$9069 lines are not available for all the targets, given the very faint nature of the \nii~$\lambda$5755 and \siii~$\lambda$6312 auroral lines, and the fact that \siii~$\lambda$9069 can fall out of the observed wavelength range, depending on the redshift of the source. 
In these cases, to estimate the temperature of the low- and intermediate-ionization regions, in our iterative procedure we used the \citet{garnett92} relations that link $T_e$(\nii) and $T_e$(\siii) to $T_e$(\oiii), :
\begin{equation}\label{eq:g92nii}
    T_e({\rm [N~II])/[K]} = 0.70 \times T_e({\rm [O~III]}) +3000
\end{equation}
\begin{equation}\label{eq:g92siii}
    T_e({\rm [S~III])/[K]} = 0.83 \times T_e({\rm [O~III]}) +1700
\end{equation}
These derived values are not reported in Fig.~\ref{fig:temp_pyneb_cloudy} for the sake of clarity, but do show good agreement with the \pyneb\ curves.
For high-redshift targets the \oiiiuv~$\lambda$1666/$\lambda$5007 has been explored as temperature diagnostics, since this ratio can be helpful for those cases where the optical auroral line is not available (weak, undetected lines, or outside of the observed wavelength range; \citealt{villar-martin04,james14,steidel14,berg16,vanzella16,kojima17,perez-montero17,patricio18,sanders20}). 

Comparing the fifth and sixth panels of Fig.~\ref{fig:temp_pyneb_cloudy}, we notice that in general the \oiiiuv~$\lambda$1666/$\lambda$5007 dependence on temperature is steeper than \oiii~$\lambda$4363/$\lambda$5007, suggesting that in principle \oiiiuv~$\lambda$1666/$\lambda$5007 could represent a better $T_e$ diagnostic (see also \citealt{kojima17,nicholls20}). 
However, in practice it is worth noting that several issues arise in deriving ratios from optical and UV emission lines. Firstly, there can be flux matching issues and mismatched aperture effects, if observations are taken with different instruments, as we discuss in detail in App.~\ref{app:A}.
A second drawback to take into account is the large uncertainties resulting from reddening estimates derived over such a large wavelength window. 
Finally, there can be an intrinsic effects due to the density, temperature and ionization structure of star-forming regions. Indeed, if the ISM is patchy, the UV light is visible only through the less dense and/or less reddened regions along the line of sight, while the optical may be arising also from denser and/or more reddened regions.
To further discuss this, in Sec.~\ref{sec:results_temp} and Sec.~\ref{sec:discussion-Te}, we show the comparison of temperatures derived with \oiiiuv~$\lambda$1666/$\lambda$5007 and \oiii~$\lambda$4363/$\lambda$5007, and the resulting difference in deriving 12+log(O/H).

Finally, Fig.~\ref{fig:temp_pyneb_cloudy} highlights a very low dependence of the intermediate- and high-ionization temperature diagnostics on $n_e$. Concerning the low-ionization temperature diagnostics, the dependence on density is higher, but the comparison with the observed line ratios used as diagnostics indicate that only $n_e <10^4$~cm$^{-3}$ are feasible. This is in-line with the fact that these line ratios are tracing the external and more diffuse regions of nebulae.

\subsection{Ionization level}\label{sec:methods-logu}
Another important parameter of the ISM is the ionization parameter, log($U$), defined as the ratio of the number of ionizing photons to the density of hydrogen atoms. Empirically, this property is best determined by ratios of emission lines of the same element with a different ionization stage, such as O3O2~$=$ ~\oiii~$\lambda$5007/\oii~$\lambda\lambda$3727,29 and S3S2~$=$ ~\siii~$\lambda\lambda$9069,9532/\sii~$\lambda\lambda$6717,31 line ratios (e.g., \citealt{kewley19}). 
Usually, O3O2 is the most widely used proxy in the optical range because these oxygen lines lie in a wavelength range accessible to many different instruments and are among the strongest line in the optical range. 
Also, they nicely span the entire energy range of an \hii\ region \citep{berg21a}. 
However, O3O2 is strongly dependent on metallicity \citep{kewley02,kewley19}.
S3S2 is less commonly used because the near-infrared (NIR) \siii~$\lambda\lambda$9069,9532 lines are weaker than their oxygen counterparts and lie at wavelengths that are less frequently covered together with \sii\ lines.
Moreover, the NIR wavelength range suffers more from telluric absorption and sky line contamination. Nevertheless, given the redder wavelengths of the sulphur emission lines, and consequently their lower excitation energies with respect to oxygen, S3S2 is less affected by metallicity and also insensitive to ISM pressure \citep{dopita86,kewley02,kewley19,mingozzi20}. 
The lower excitation energies of \sii\ and \siii\ also imply that this ratio is tracing the ionization parameter of the low-ionization regions of nebulae (see e.g., Fig.~4 in \citealt{berg21a}).
Finally, \citet{berg21a} introduced the Ar4Ar3~$=$~\ariv~$\lambda\lambda$4714,71/\ariii~$\lambda$7138 ratio as a ionization parameter tracer of the very high-ionization region.

In our analysis, in order to calculate log($U$), we used the calibration of \citet{berg18} for O3O2 (see their table~3) and the one of \citet{berg21a} for S3S2 and Ar4Ar3 (see their table~4). 
These calibrations relate these line ratios and log($U$) as a function of the gas metallicity, and are obtained using the set of \cloudy\ models described in Sec.~\ref{sec:methods-den} and Sec.~\ref{sec:methods-temp}, where we showed their agreement both with our observed line ratios and \pyneb\ predictions. 
To calculate the corresponding log($U$) of each CLASSY galaxy, we assumed the gas-phase metallicity reported in Table~\ref{tab:classyprop}.

Given that \siii$\lambda\lambda9069,9532$ is outside the observed wavelength range for many CLASSY galaxies, and \ariv$\lambda\lambda9069,9532$ and \ariii$\lambda$7138 lines can be faint and thus below the required signal-to-noise threshold of 3, we can measure the S3S2 and Ar4Ar3 line ratios only in 20 and 28 CLASSY galaxies, respectively. On the other hand, we can calculate the O3O2 line ratios for all the galaxies of our sample, apart from J1444+4237. For the 18 CLASSY galaxies with \oii$\lambda\lambda$3727,9 outside the observed wavelength range, we estimated O3O2 using the emissivities of \oii$\lambda\lambda$3727,9 and \oii$\lambda\lambda$7320,30 obtained with \pyneb, where calculations were performed using $n_e$ and $T_e$ associated to the low-ionization emitting zone. It is thus possible to recover \oii$\lambda\lambda$3727,9 by multiplying the calculated empirical ratio with the observed \oii$\lambda\lambda$7320,30 line fluxes (see also \citetalias{arellano-cordova22}).
In Sec.~\ref{sec:results_logu} we describe our results, while in Sec.~\ref{sec:discussion-metlogu} we discuss how the log($U$) optical tracers relate to potential UV analogs.

\section{Results: Comparing UV and optical Physical Properties of the ISM}\label{sec:results}
\subsection{Dust attenuation diagnostics}\label{sec:results_ext}
\begin{figure*}
\begin{center}
    \includegraphics[width=0.45\textwidth]{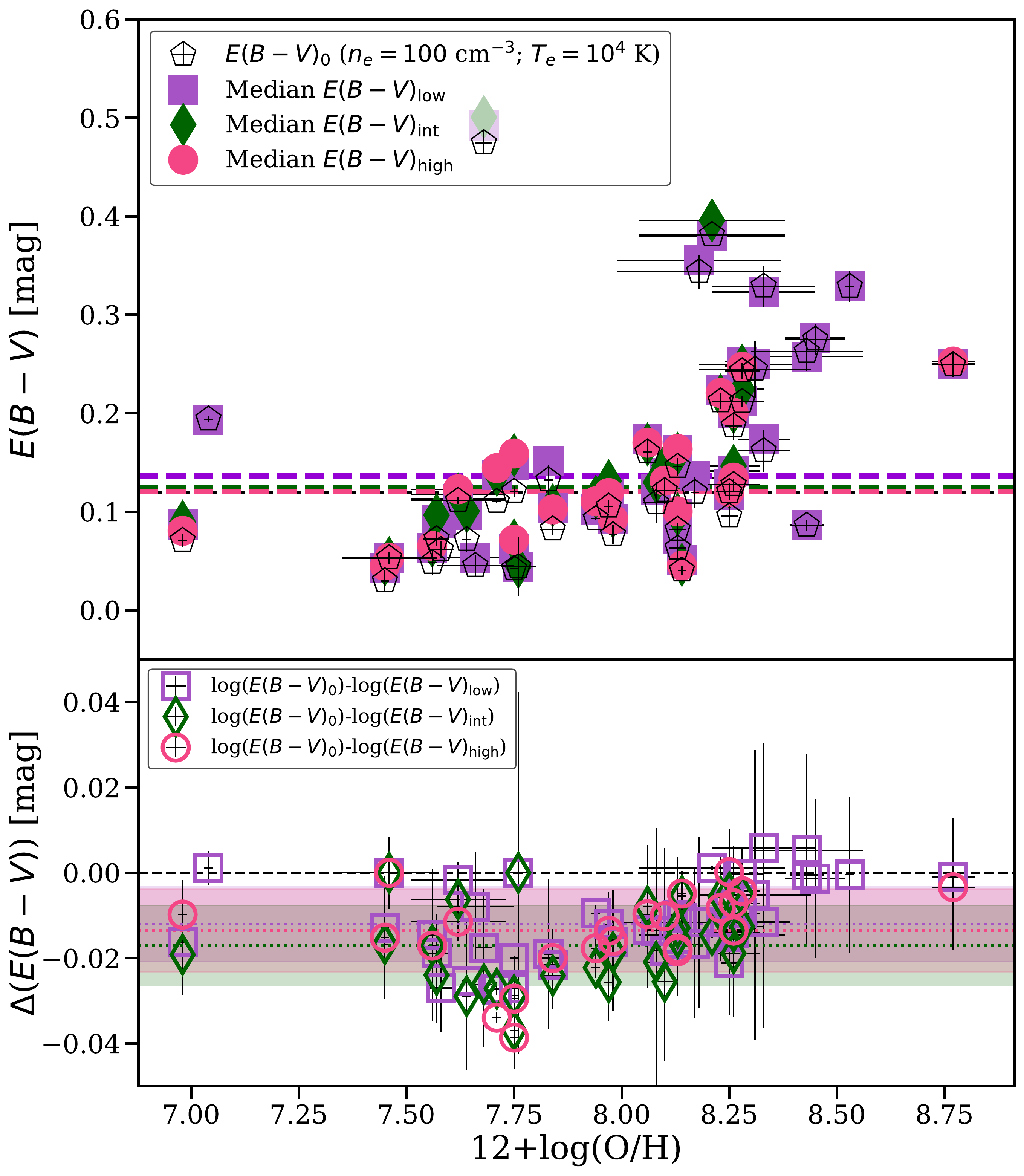}
    \includegraphics[width=0.45\textwidth]{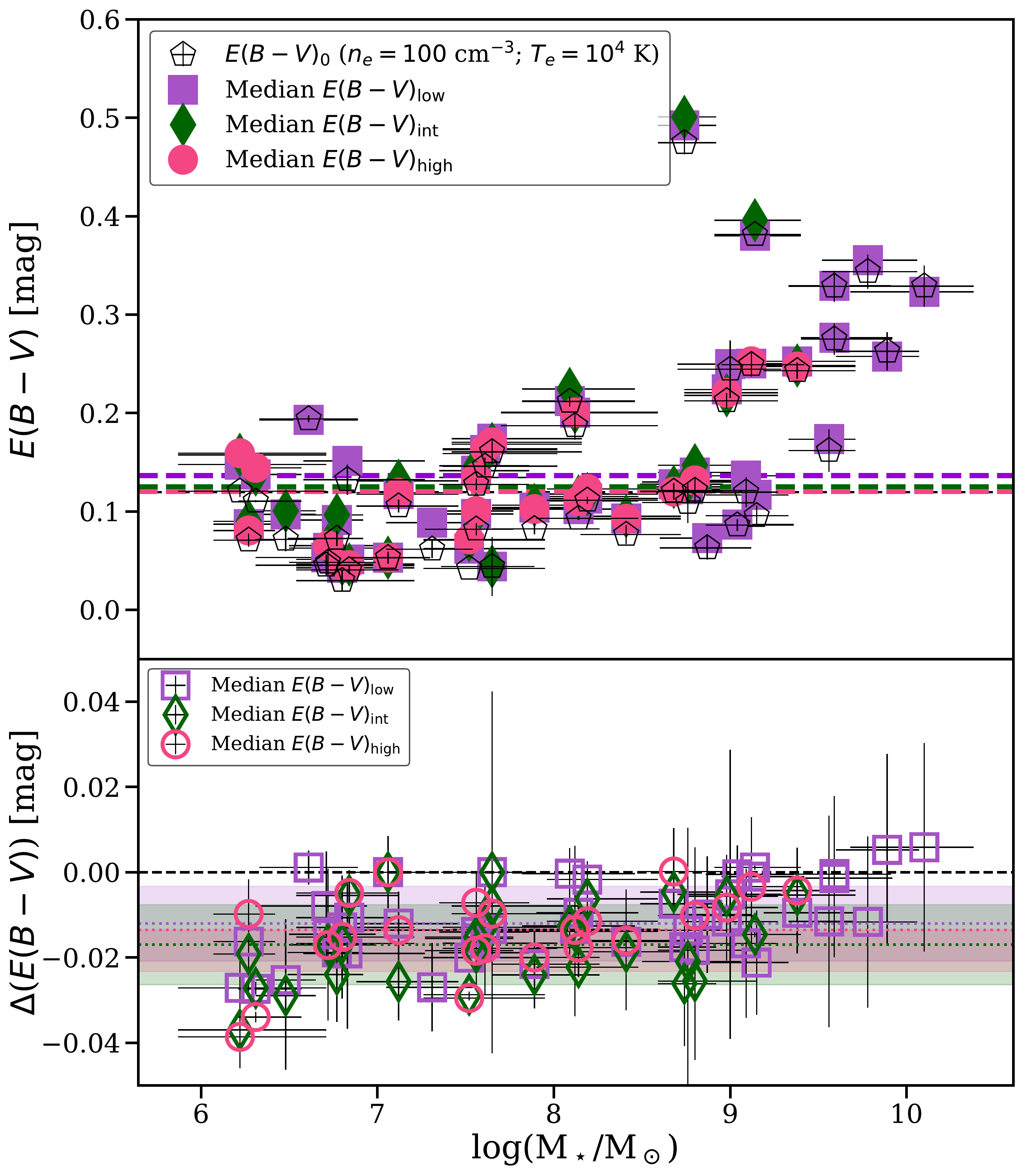}
\end{center}
\caption{Top panels: comparison of the $E(B-V)$ estimated assuming $T_e = 1\times 10^4$ K and $n_e = 10^2$ cm$^{-3}$ ($E(B-V)_0$; empty black pentagon) with the value obtained from low- (purple squares), intermediate- (green diamonds) and high-ionization regions (red circles), as a function of 12+log(O/H) and M$\star$. The uncertainties of the displayed quantities are shown in both axis. Bottom panels: difference between $E(B-V)_0$ and the $E(B-V)$ measurements of each ionization zone, keeping the same symbols and colors as in the top panel. The colored dashed and dotted lines indicate the median values of the shown distributions. The shaded regions (partially overlapped) in the bottom panel represent the 68\% intrinsic scatter of each distribution, color-coded accordingly. At 12+log(O/H)~$\lesssim 7.75 $ or stellar mass log(M$\star$/M$\odot$)~$\lesssim6.5$, $E(B-V)$ is down to $\sim-0.04$~mag lower than $E(B-V)_0$.}
\label{fig:dustext}
\end{figure*}
We calculated $E(B-V)$ for the CLASSY galaxies as described in Sec.~\ref{sec:method-dust}, with the iterative method explained in Sec.~\ref{sec:methods}, finding values in the range $E(B-V)\sim0-0.5$~mag. To explore the systematic uncertainties on $E(B-V)$ due to the effect of density and temperature variations on the Balmer decrement, we obtained a value for each combination of $T_e$ and $n_e$ estimates for each ionization zone (described in Fig.~\ref{fig:den_pyneb_cloudy} and Fig.~\ref{fig:temp_pyneb_cloudy}).
The top panels of Fig.~\ref{fig:dustext} display the $E(B-V)$ values obtained for the low- (pink squares), intermediate- (red diamonds), high-ionization region (blue circles) densities and temperatures as a function of 12+log(O/H) and stellar mass, on the left and right, respectively. 
We also show $E(B-V)_0$, which we define assuming $n_e = 10^2$~cm$^{-3}$ and $T_e = 1\times 10^4$~K (empty black pentagons), usually considered appropriate conditions for star-forming regions \citep{osterbrock89,osterbrock06}. 
The dashed lines colored accordingly indicate the median values. 
The Pearson correlation factors are $R\sim0.50$ for 12+log(O/H) and $R\sim0.63$ for stellar mass, with pvalue of $p\sim1\times10^{-3}$ and $p\sim5\times10^{-6}$, respectively.

The lower panels show the difference between $E(B-V)_0$ and the values calculated assuming the coherent density and temperature of the different ionization zones, expressed in [mag]: $$\rm \Delta(E(B-V))= E(B-V)_0-E(B-V) $$ 
where $E(B-V)_0$ is the value derived with $n_e = 10^2$~cm$^{-3}$ and $T_e = 1\times 10^4$~K). The median values and the 68\% intrinsic scatter of the distributions are shown by the dotted lines and shaded regions (mostly overlapped), color-coded correspondingly. 

Even though the difference of the colored dots with respect to $E(B-V)_0$ looks small, with an overall median value around $\sim-0.01$~mag and similar intrinsic scatter, there are a few objects with discrepancies down to $\sim - 0.04$~mag, at 12+log(O/H)~$\lesssim 7.75 $ or stellar masses log(M$\star$/M$\odot$)~$\lesssim6.5$.
Specifically, $E(B-V)$ is generally larger than $E(B-V)_0$ (of $\sim-0.01$~mag on average, or $\sim-0.05$~dex), while at higher 12+log(O/H) their difference tends to 0.
Therefore, assuming $T_e = 1\times 10^4$~K and $n_e = 10^2$~cm$^{-3}$ can lead to underestimates in the dust attenuation, especially at low 12+log(O/H), where $T_e$ is higher (see Sec.~\ref{sec:results_temp}). 
This is due to the slight dependence of recombination lines on temperature and density, that still causes the Balmer decrement to drop from $\sim2.86$ to $\sim2.70$ \citep{osterbrock89,osterbrock06}.
The fact that the low, intermediate and high-ionization $E(B-V)$ values are consistent within their uncertainties indicates that a simpler approach, considering a single value of density and temperature that represents all zones, can be used to correct the emission lines without introducing a significant bias in the results. 
Hence, for our analysis we used the weighted average of these values (defined generically as $E(B-V)$ hereafter), to correct both UV and optical emission lines as explained in Sec.~\ref{sec:method-dust}. 

\subsection{Density diagnostics}\label{sec:results-ne}
\begin{figure*}
\begin{center}
    \includegraphics[width=.9\textwidth]{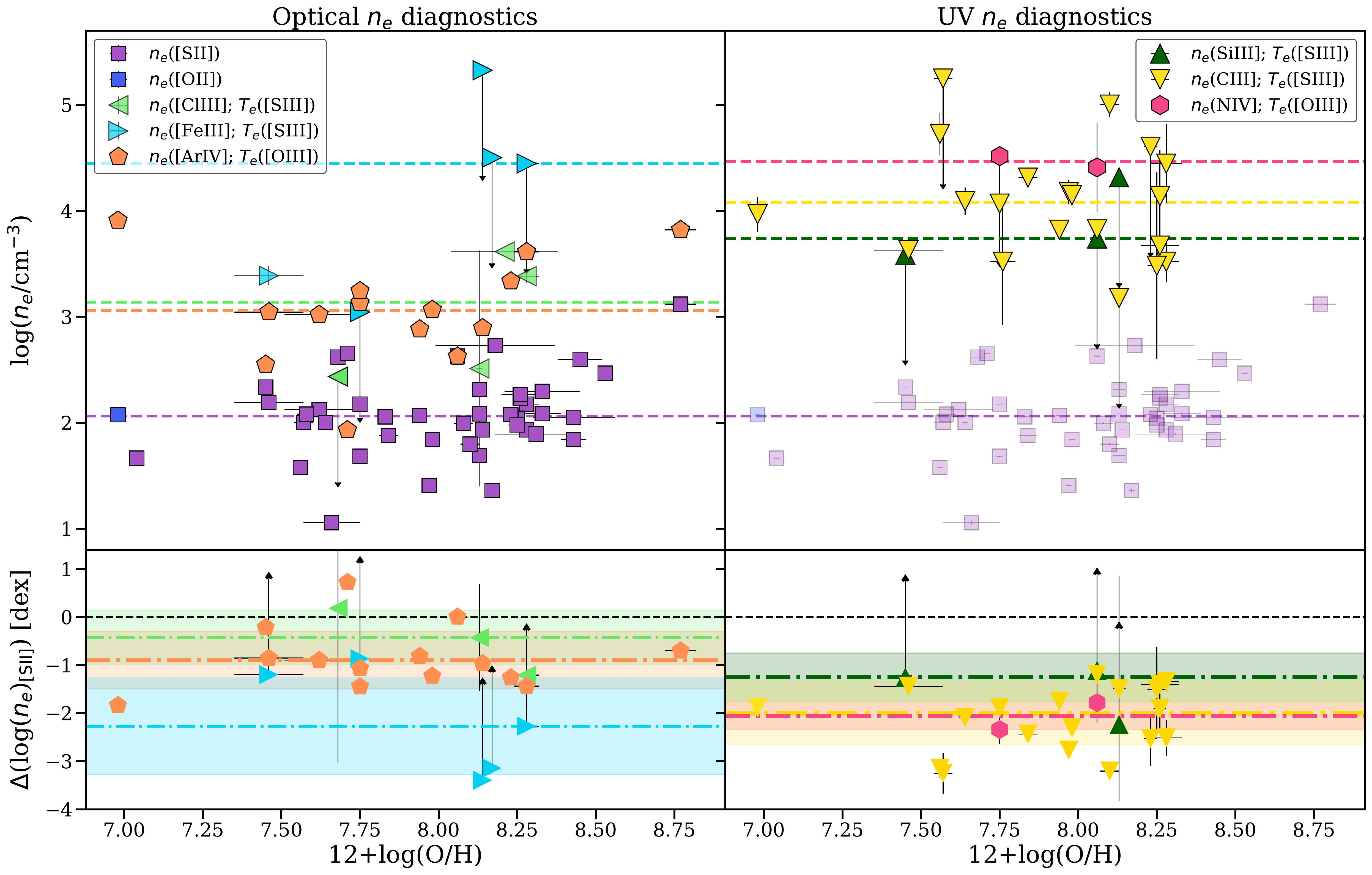}
\end{center}
\caption{Top panels: comparison of $n_e$ estimated from low- ($n_e$(\sii) or $n_e$(\oii) for J0934+5514; squares), intermediate- ($n_e$(\cliii), $n_e$(\feiii), $n_e$(\siiii), $n_e$(\ciii); triangles) and high-ionization regions ($n_e$(\ariv, $n_e$(\niv); pentagons, hexagons), as a function of 12+log(O/H). For clarity reasons the optical and UV diagnostics are shown on the left and right, respectively, while the optical $n_e$(\sii) or $n_e$(\oii) are shown in both panels. The dots are color-coded consistently with Fig.~\ref{fig:den_pyneb_cloudy}, as indicated in the legend. The uncertainties of the displayed quantities are shown in both axis. Bottom panels: difference in dex between the low-ionization density (purple squares) and the other $n_e$ measurements, keeping the same symbols as in the top panel. The colored dashed and dotted lines indicate the median values. The shaded regions (partially overlapped) in the bottom panel represent the 68\% intrinsic scatter of each distribution, color-coded accordingly.  Excluding $n_e$(\feiii) for which we have mainly upper limits, clearly the highest discrepancies with respect to $n_e$(\sii) are found for UV tracers, that predict on average densities $\sim 1-2$~dex higher.}
\label{fig:den_diag}
\end{figure*}
In the left and right upper panels of Fig.~\ref{fig:den_diag} we show the optical and UV density diagnostics as a function of 12+log(O/H). In both panels we also display the low-ionization density obtained through \sii~$\lambda$6717/$\lambda$6731 (purple squares) as a reference. For the galaxy J0934+5514, we estimate the low-ionization density from \oii~$\lambda$3729/$\lambda$3729 (blue square), since KCWI data do not cover \sii~$\lambda\lambda$6717,31 lines.
For the optical, we show the intermediate-ionization density derived from \cliii~$\lambda$5518/$\lambda$5538 (green left-pointing triangles) and \feiii~$\lambda$4701/$\lambda$4659 (turquoise left-pointing and blue right-pointing triangles), and the high-ionization density from \ariv~$\lambda$4714/$\lambda$4741 (orange pentagons). For the UV, instead, we show the intermediate-ionization density derived from \fsiiii~$\lambda$1883/\siiii~$\lambda$1892 (darkgreen up-pointing triangles) and \fciii~$\lambda$1907/\ciii~$\lambda$1909 (gold down-pointing triangles), and the high-ionization density from \niv~$\lambda$1483/$\lambda$1487 (red hexagons) values.
Overall, the values obtained span in $n_e \sim 30$~cm$^{-3}$ to $n_e \sim 10^5$~cm$^{-3}$. The dashed lines represent the median value of density given by each diagnostic, and are color-coded accordingly. 

In the bottom panels of Fig.~\ref{fig:den_diag}, we show the difference in dex between the low-ionization zone density and the other values ($\Delta$(log($n_e$)$_{\rm [SII]}$)), keeping the same color-coding of the main panel.  
The dotted lines show the median values of the offsets in dex, that are of $\sim-0.8$, $\sim-2.3$ and $\sim-0.9$~dex, for \cliii, \feiii\ and \ariv, and $\sim-1.3$, $\sim-2.0$, and $\sim -2.1$~dex for \siiii, \ciii\ and \niv, respectively, sorting the optical and UV diagnostics as a function of the increasing ionization potential. 
The 68\% intrinsic scatter of each distribution is shown by the shaded regions, color-coded correspondingly. The partial overlaps indicate a similar behavior of the distributions.

We note that \ariv\ densities are slightly larger than \sii\ densities with values on average around $\sim 1000$~cm$^{-3}$, while \sii\ densities are always lower than $\sim 1000$~cm$^{-3}$. This is expected because \ariv\ densities have a higher critical density than \sii, which instead traces the low-ionization and diffuse gas within nebulae (see Fig.~\ref{fig:den_pyneb_cloudy}).
Interestingly, we find that in the optical, the \ariv\ densities that trace the high-ionization regions are somewhat lower than expected when compared to their UV counterparts \niv, and are instead consistent with \cliii\ values, which trace the intermediate-ionization regions. 
Also, we note that when comparing the UV and optical diagnostics, both the \ariv\ and \cliii\ densities are lower than those derived from \ciii\ and \siiii\ in the UV, which are both tracers of intermediate-ionization regions.
On the other hand, \feiii, which is characterized by an ionization potential lower than \cliii, traces similar densities to the UV diagnostics, but the values can only be considered upper limits due to the large error bars.
Excluding \feiii, clearly the highest discrepancies with respect to $n_e$(\sii) are found for UV tracers, that predict on average densities around $\sim 10^4$~cm$^{-3}$.

We will further comment about this in Sec.~\ref{sec:discussion-ne}.
As expected, we find no correlation between log($n_e$) (and also $\Delta$(log($n_e$)$_{\rm [SII]}$)) and 12+log(O/H), as well as the other galaxy properties, such as stellar mass, stellar metallicity, stellar age, or SFR. 

\subsection{Temperature diagnostics}\label{sec:results_temp} 
\begin{figure*}
\begin{center}
    \includegraphics[width=0.45\textwidth]{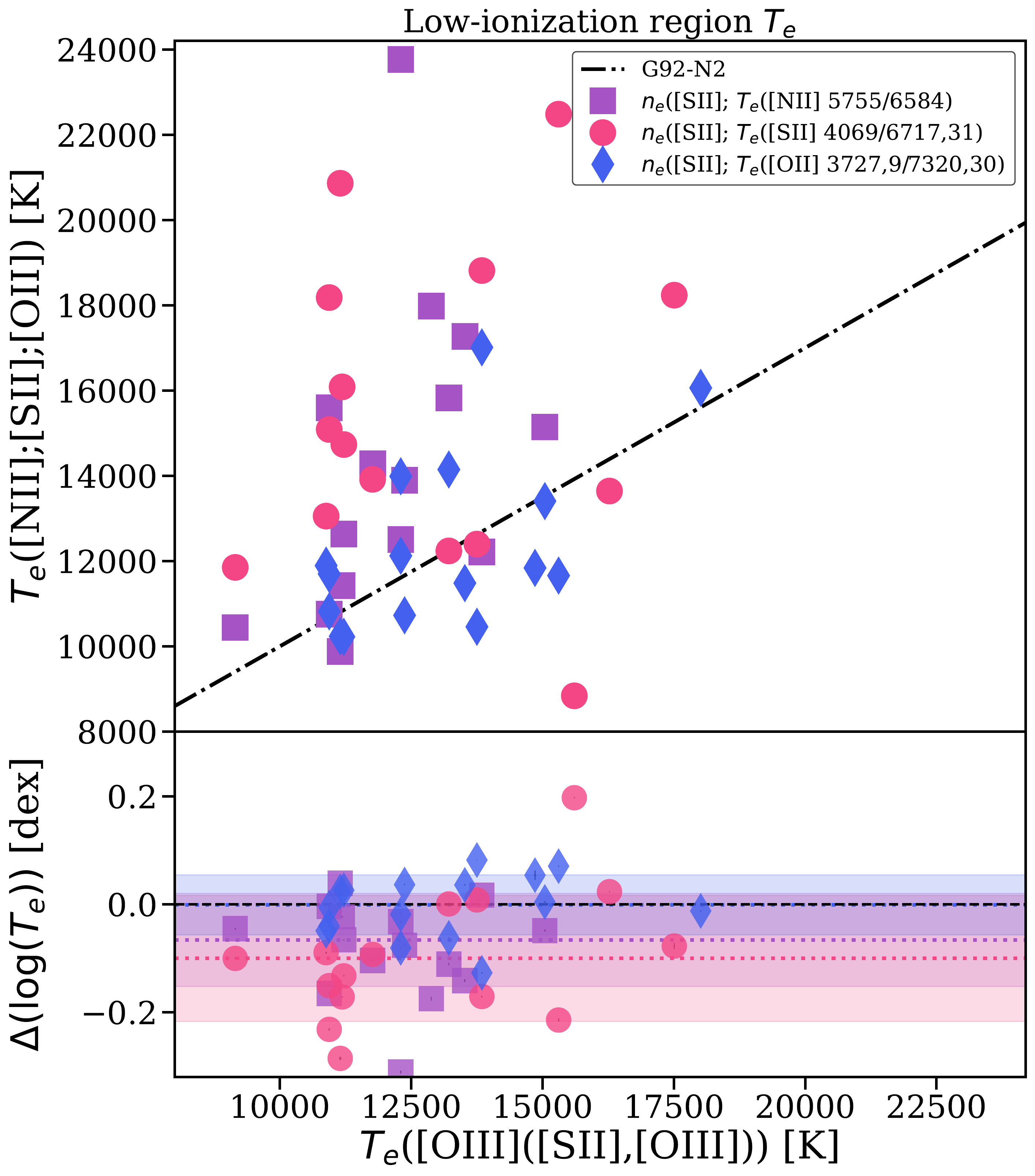}
    \includegraphics[width=0.45\textwidth]{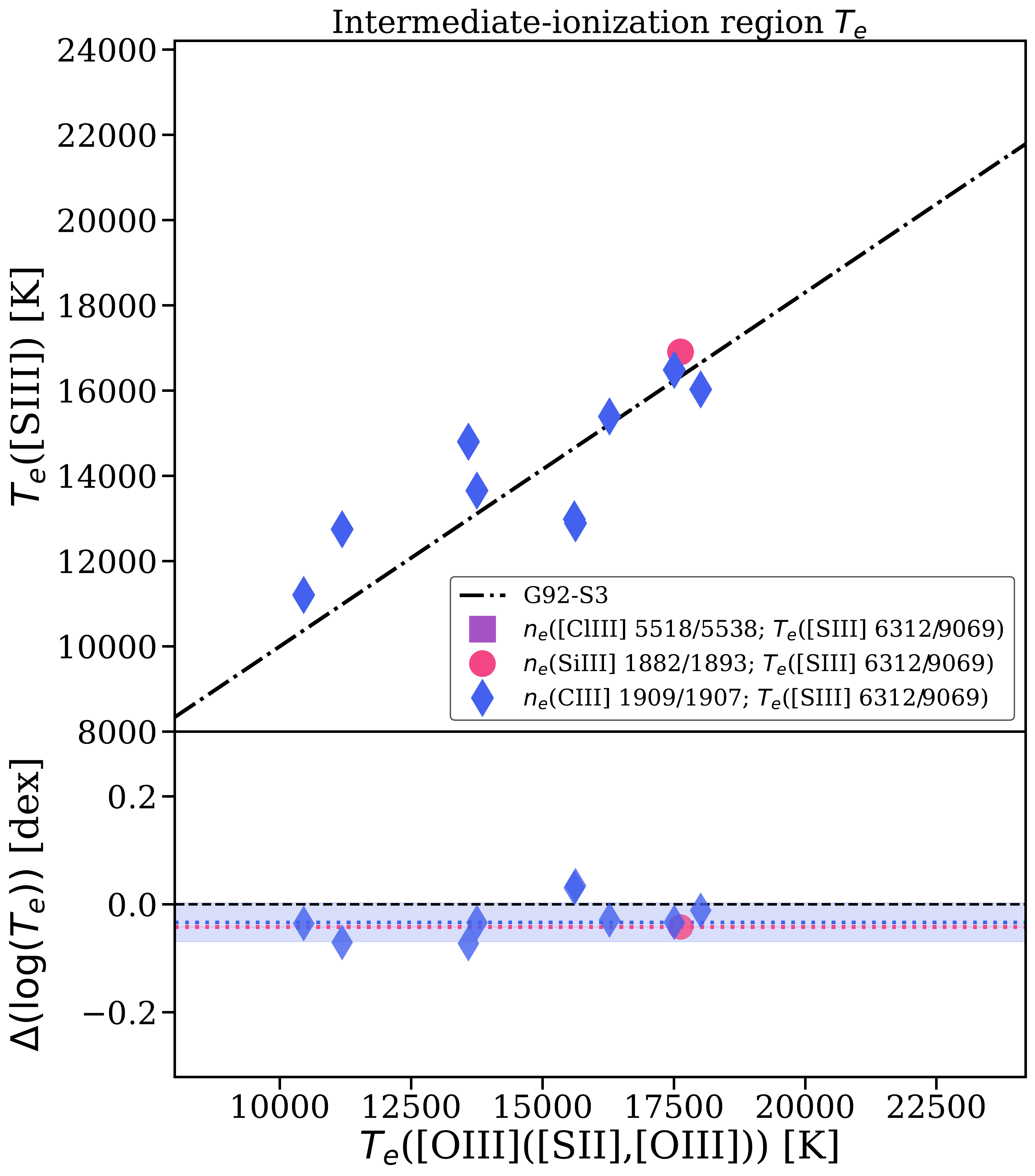}
\end{center}
\caption{Upper panels: Comparison of the three $T_e$ estimates of the low-ionization (on the left) and intermediate-ionization (on the right) with the temperature diagnostics \nii~$\lambda$5755/$\lambda$6584, \sii~$\lambda\lambda$4069/$\lambda\lambda$6717,31, \oii~$\lambda\lambda$3727,29/$\lambda\lambda$7320,30 and \siii~$\lambda$6312/$\lambda$9071, as a function of the high-ionization temperature T(\oiii(\sii,\oiii)), taking into account the available density tracers. The uncertainties of the displayed quantities are shown in both axis. The dot-dashed lines in the left and right panels indicate Eq.~\ref{eq:g92nii} and Eq.~\ref{eq:g92siii}, respectively. On the right panels there are no estimates of T(\siii(\cliii,\siii)), because we could not find finite values with \pyneb, and only one measurement for T(\siii(\siiii,\siii)).
Lower panels: difference in dex between the prediction of Eq.~\ref{eq:g92nii} and Eq.~\ref{eq:g92siii}, and the calculated values.
The shaded regions (overlapped) in the bottom panel represent the 68\% intrinsic scatter of each distribution, color-coded accordingly. Overall, the median values of the computed differences are $\sim -0.1$~dex and $\sim -0.05$~dex (dotted lines in the minor panels) for the low-ionization and intermediate-ionization $T_e$, respectively.}
\label{fig:den_temp_diag_lowint}
\end{figure*}
\begin{figure}
\begin{center}
    \includegraphics[width=0.5\textwidth]{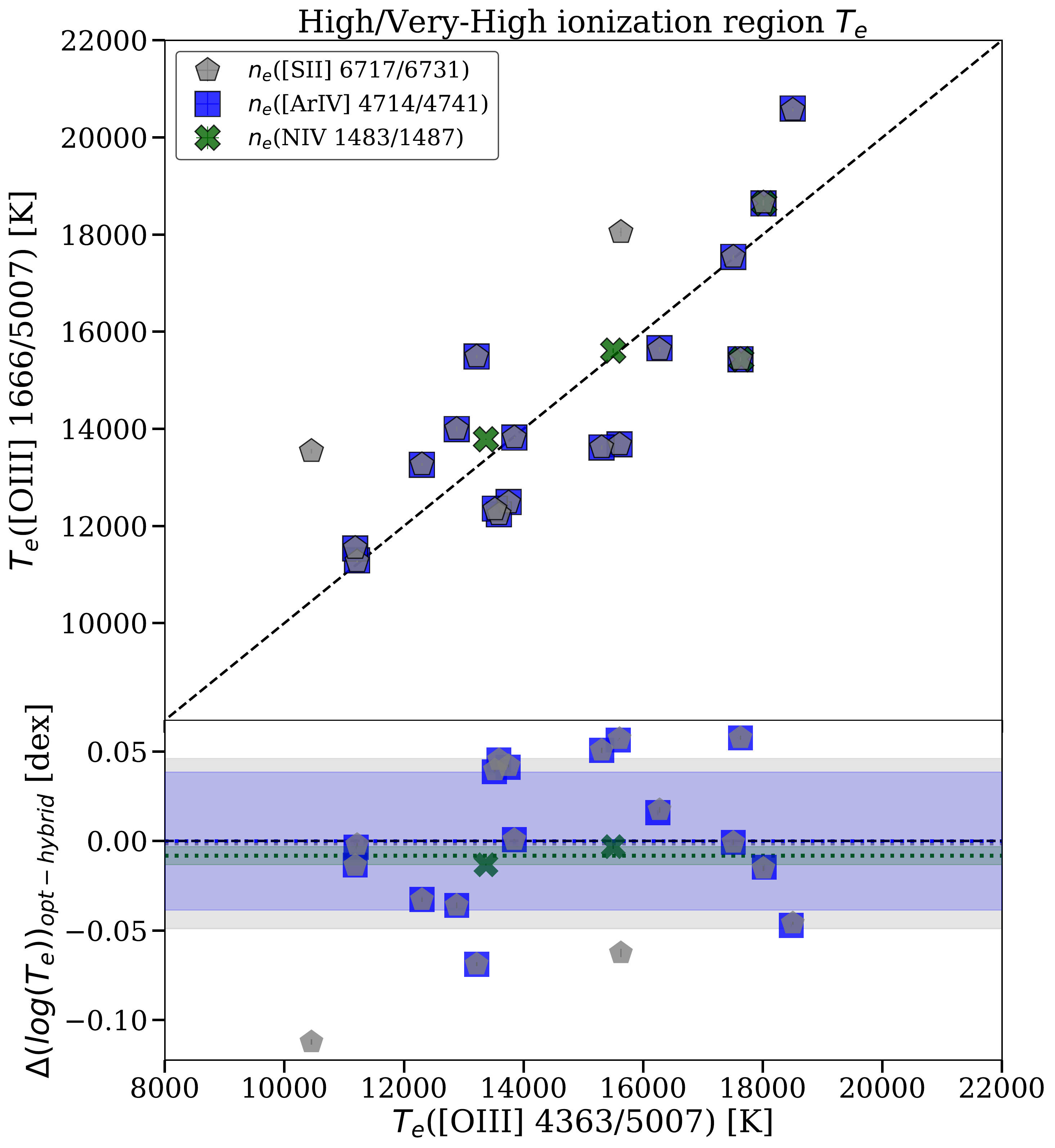}
\end{center}
\caption{Top panels: comparison of the high-ionization regions $T_e$(\oiiiuv~$\lambda$1666/\oiii~$\lambda$5007) estimated from the the hybrid UV-optical ratio, using $n_e$(\ariv) (blue squares on the left) or $n_e$(\niv) (green crosses on the right), as a function of $T_e$(\oiii~$\lambda$4363/\oiii~$\lambda$5007). The uncertainties of the displayed quantities are shown in both axis. In general, the values are roughly in agreement with the dashed black line, which represents the 1:1 relation. Bottom panels: difference in dex between $T_e$(\oiii~$\lambda$4363/\oiii~$\lambda$5007) and $T_e$(\oiiiuv~$\lambda$1666/\oiii~$\lambda$5007), keeping 
same symbols and colors as in the top panel. 
The colored dashed and dotted horizontal lines in the main and minor panels respectively show the median values of the displayed quantities.}
\label{fig:temp_diag}
\end{figure}
The left and right upper panels of Fig.~\ref{fig:den_temp_diag_lowint} show the comparison between the low- and intermediate-ionization region temperatures inferred through the \citet{garnett92} relations (Eq.~\ref{eq:g92nii} and Eq.~\ref{eq:g92siii}, respectively) as a function of the high-ionization zone temperature T$_e$(\oiii~4363/5007), to see to what extent they give consistent results.
Specifically, in the left panel of Fig.~\ref{fig:den_temp_diag_lowint} we report the three different low-ionization $T_e$ estimates made with \nii~$\lambda$5755/$\lambda$6584 (purple squares), \sii~$\lambda$4069/$\lambda\lambda$6717,31 (pink dots) and \oii~$\lambda\lambda$3727,29/$\lambda\lambda$7320,30 (blue diamonds), using the \sii~$\lambda\lambda$6717,31 doublet as density tracer (and the \oii~$\lambda\lambda$3727,29 doublet for J0934+5514). 
On the right panels there are no estimates of T(\siii(\cliii,\siii)), because we could not find finite values with \pyneb, and only one galaxy for which we measured T(\siii(\siiii,\siii)).
In general, the low-ionization temperatures obtained are in the range $T_e\sim8000~-24000$~K. 
The bottom panels of Fig.~\ref{fig:den_temp_diag_lowint} report the differences in dex between the values obtained with the relations from \citet{garnett92} and the $T_e$ inferred through the different temperature diagnostics.  We note that the median values (dotted lines) in these subpanels are close to zero (the median values are $\sim - 0.1$~dex), but systematically below, with a better agreement with Eq.~\ref{eq:g92nii} for $T_e$(\sii~$\lambda$4069/$\lambda\lambda$6717,31) and $T_e$(\oii~$\lambda\lambda$3727,29/$\lambda\lambda$7320,30) with respect to $T_e$(\nii~$\lambda$5755/$\lambda$6584). Also the 68\% intrinsic scatter of the distributions, shown by the shaded regions color-coded correspondingly, show a similar behaviour (i.e., mostly overlapped).
However, we note that overall, Eq.~\ref{eq:g92nii} tends to underestimate the temperature up to $\sim 0.2$~dex.

Fig.~\ref{fig:den_temp_diag_lowint} shows that the intermediate-ionization $T_e$ derived with \siii~$\lambda$6312/$\lambda$9069 are better in agreement with \citet{garnett92}'s relation with respect to the low-ionization $T_e$ discussed in the previous paragraph. 
To summarise, several authors found significant differences with \citet{garnett92}'s relations using large samples of star-forming galaxies (e.g. \citealt{kennicutt03a,binette12,berg15}). 
A possible interpretation of the larger discrepancy that we find for Eq.~\ref{eq:g92nii} than for Eq.~\ref{eq:g92siii} could be explained by the large absorption cross section of low energy ionizing photons \citep{osterbrock89}, which are thus preferentially absorbed in the \hii\ regions with respect to higher energy ones, leading to a hardened spectrum (e.g., \citealt{hoopes03}). 
An implication would be that the low-ionized regions of the \hii\ regions could have higher temperatures than the high-ionized regions, as observed. 
However, galaxies characterised by a very high excitation such as those covered by the CLASSY survey are expected to have minimal contributions from the low-ionization lines, and thus little dependence on the low-ionization zone temperatures for the oxygen abundance \citep{berg21a}. 
In this respect, we feel confident that for those galaxies in our sample for which we used the \citet{garnett92} relation (i.e., 23 using Eq.~\ref{eq:g92nii} and 13 using Eq.~\ref{eq:g92siii}), the derived $T_e$ values are reliable estimates.

Concerning the high-ionization temperature, the main panel of Fig.~\ref{fig:temp_diag} compares the values obtained iteratively with either \oiii~$\lambda$4363/$\lambda$5007 or the hybrid UV-optical ratio \oiiiuv~$\lambda$1666/$\lambda$5007, and the \sii, \ariv\ and \niv\ density diagnostics, as reported in the legend. We note differences lower than $\lesssim 750$~K using either \sii, \ariv\ or \niv, confirming again the low dependence of temperature on density diagnostics, even when the difference in log($n_e$) can be as large as $\sim2$~dex. 
In general, the values are roughly in agreement with the dashed black line that indicates the 1:1 relation. 
To better evaluate this, the bottom panel shows the difference between $T_e$(\oiii~$\lambda$4363/$\lambda$5007) and  $T_e$(\oiiiuv~$\lambda$1666/$\lambda$5007) in dex ($\Delta$log($T_e$)$_{opt - hybrid}$), keeping the same symbols and colors of the main panel. 
$\Delta$log($T_e$)$_{opt - hybrid}$ is in median $\sim -0.025$~dex ($\sim 1000$~K), with the highest temperatures measured with \oiii$\lambda$1666/$\lambda$5007. 
This trend is consistent with other works in the literature who made the same comparison in smaller samples (e.g., \citealt{berg16}). 
We comment more on the offset $\Delta$log($T_e$)$_{opt - hybrid}$ and  its impact on the estimate of 12+log(O/H) in Sec.~\ref{sec:discussion-Te}.

\subsection{Ionization parameter diagnostics}\label{sec:results_logu}
\begin{figure}
\begin{center}
    \includegraphics[width=0.48\textwidth]{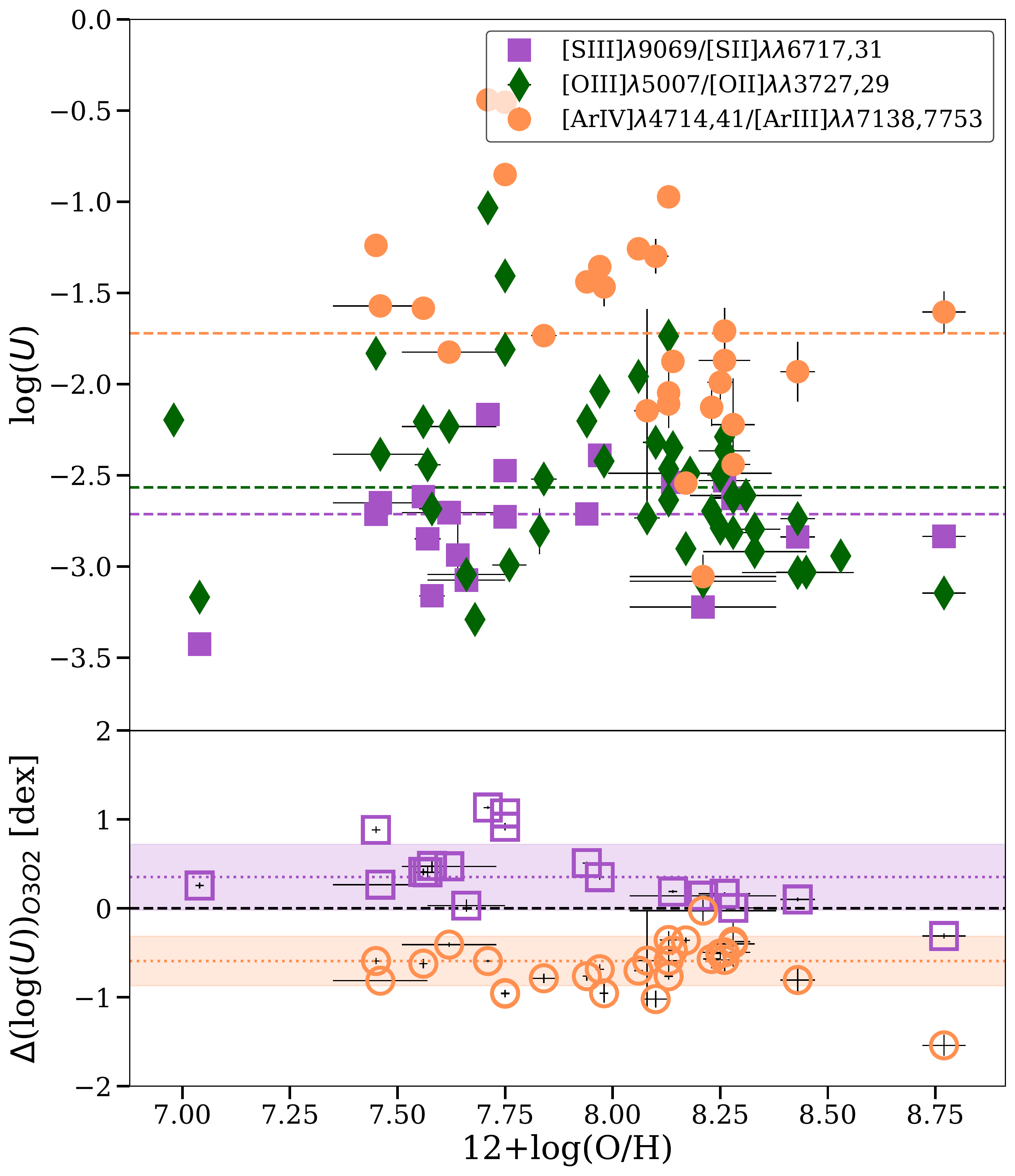}
\end{center}
\caption{Top panel: comparison of the $\log(U)$ estimated from low- ($S2S3$; purple squares), intermediate- ($O3O2$; green diamonds) and high-ionization regions ($Ar4Ar3$; orange circles), as a function of 12+log(O/H). The uncertainties of the displayed quantities are shown in both axis. Bottom panel: differences $\Delta$(log$U$)$_{O3O2-S3S2}$, in purple, and $\Delta$(log$U$)$_{O3O2-Ar4Ar3}$, in orange (we refer to this quantity as $\Delta$(log$U$)$_{O3O2}$), revealing significant discrepancies in the range $\pm 1$~dex. The colored dashed (in the main panel) and dotted (in the bottom panel) lines indicate the median values. The shaded regions in the bottom panel represent the 68\% intrinsic scatter of each distribution, color-coded accordingly. This scatter of log($U$) values is in line with the presence of a clear ionization structure in the nebular environments of our targets.}
\label{fig:ionu_diag}
\end{figure}

The upper panel of Fig.~\ref{fig:ionu_diag} shows the comparison of  log($U$) estimated from low- (purple squares), intermediate- (green diamonds) and high-ionization regions (orange circles) as a function of 12+log(O/H), using the S3S2, O3O2 and Ar4Ar3 line ratios, respectively, as explained in Sec.~\ref{sec:methods-logu}. 
Fig.~\ref{fig:ionu_diag} demonstrates that in our sample there is not an evident anti-correlation between log($U$)(O3O2) and metallicity ($r\sim-0.3$, $p\sim0.05$), which is instead predicted by the theoretical relation presented in \citet{dopita86} and \citet{dopita06}. This would be expected as a consequence of the enhanced opacity of stellar winds and higher level of scattering at increasing metallicity. The former factor would lead to a decrease of the ionizing photons in the surrounding \hii\ region, while the latter to a greater conversion efficiency from luminous energy flux to mechanical energy flux, hence decreasing log($U$). 
On the other hand, we notice a slight anti-correlation with log(M$\star$) ($r\sim-0.5$, $p\sim1\times10^{-4}$), as suggested in some previous works on local star-forming galaxies (e.g., \citealt{dopita06,brinchmann08}). We also find no correlation with SFR, which instead has been revealed from some works based on spatially resolved optical spectra (e.g., \citealt{dopita14,kaplan16}). 
Finally, in agreement with \citet{kewley15,bian16,kaasinen18,mingozzi20}, we find a strong correlation between log($U$)(O3O2) and log(sSFR) ($r\sim0.7$, $p\sim1\times10^{-8}$). We comment further about this in Sec.~\ref{sec:discussion-metlogu}.
We stress that similar results hold for log($U$)(S3S2) and log($U$)(Ar4Ar3).

Interestingly, from Fig.~\ref{fig:ionu_diag} we note that using these three different diagnostics we obtain a scatter of values of log($U$), in the range $-3.5-0.$, with the lowest values derived from S3S2 and the highest ones with Ar4Ar3. 
The lower panel of Fig.~\ref{fig:ionu_diag} illustrates better this scatter, showing the differences $\Delta$(log$U$)$_{O3O2-S3S2}$ and $\Delta$(log$U$)$_{O3O2-Ar4Ar3}$ in purple and orange, respectively (we refer to this quantity as $\Delta$(log$U$)$_{O3O2}$ hereafter), revealing significant discrepancies in the range $\pm 1$~dex. 
The median values and the 68\% intrinsic scatter of the distributions are shown by the dotted lines and shaded regions, color-coded correspondingly. 
Specifically, in our sample log($U$(O3O2)) is in median $\sim0.4$~dex higher than log($U$(S3S2)) and $\sim-0.6$~dex lower than log($U$(Ar4Ar3)).
This is in line with the presence of a clear ionization structure in the nebular environments of our targets, as described in Sec.~\ref{sec:methods} and in \citet{berg21a} for the CLASSY galaxies J1044+0353 and J1418+2102.

Among the available UV emission lines, only the \civ/\ciii\ line ratio involves emission originating from different ionization states of the same element, and thus could constrain the ionization parameter. We will further discuss this in Sec.~\ref{sec:discussion-metlogu}.

\section{The UV Toolkit} \label{sec:discussion}
In this section we compare the optical and UV diagnostic results detailed in Section~\ref{sec:results} and discuss correlations between gas-phase properties and UV emission lines. Using these relationships, we provide a set of diagnostic equations that can be used to estimate gas-phase E(B-V), electron density, electron temperature, metallicity and ionization parameter. Each of these equations relies only on emission lines observed at rest-frame UV wavelengths and thus represents a \emph{UV Toolkit} for deriving the chemical and physical conditions of star-forming galaxies.

\subsection{UV Diagnostics for E(B-V)}\label{sec:discussion-ext}
\begin{figure*}
\begin{center}
    \includegraphics[width=0.45\textwidth]{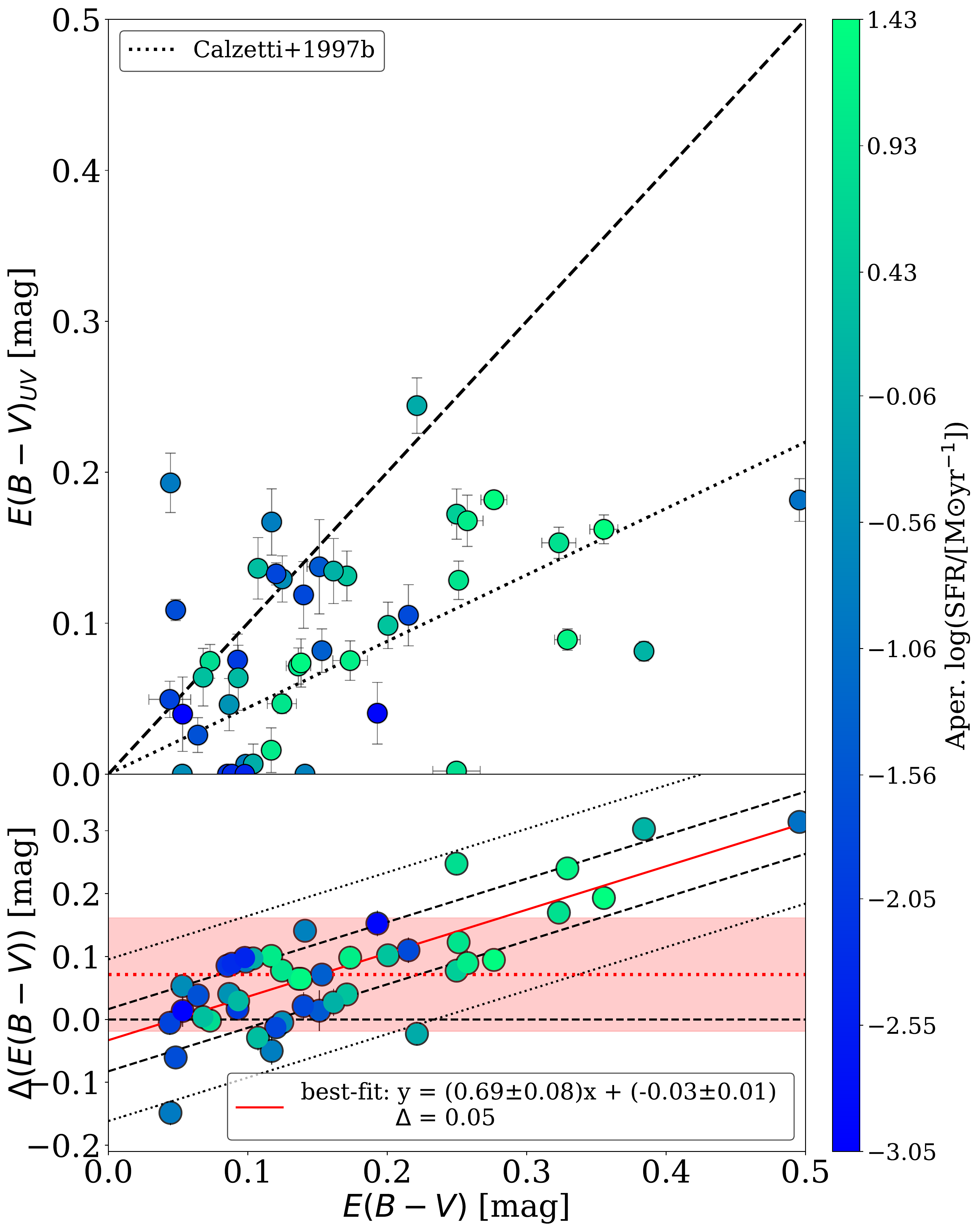}
    \includegraphics[width=0.45\textwidth]{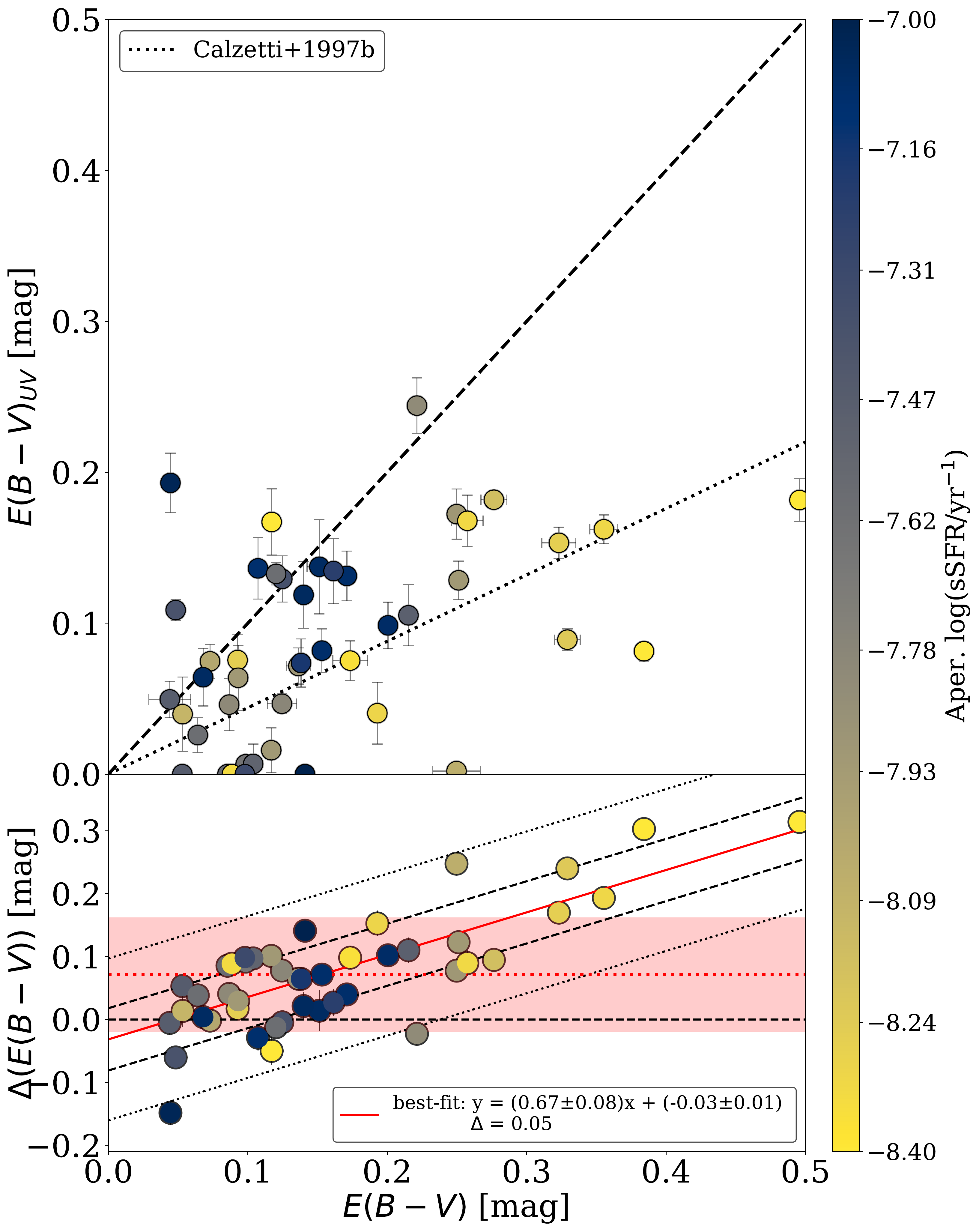}
\end{center}
\caption{Upper panels: comparison between the stellar $E(B-V)_{UV}$ with respect to the gas $E(B-V)$, color-coded as a function of $\log(SFR)$ (on the left) and $\log(sSFR)$ (on the right) within the COS aperture. The dashed black line represents the 1:1 relation, while the dotted line is the \citetalias{calzetti97b} empirical relation between stellar and gas $E(B-V)$ (i.e., $E(B-V)_{UV} = (0.44\pm 0.03)\times E(B-V)$). The uncertainties of the displayed quantities are shown in both axis. Lower panels: $\Delta(E(B-V))$ defined as $E(B-V) - E(B-V)_{UV}$. $\Delta(E(B-V))$ increases linearly with $E(B-V)$, and the linear best-fit is indicated by the red solid lines, reported on the upper right, while the dashed and dotted lines represent the $1\sigma$ and $2.6\sigma$ around the best-fit. The red horizontal dotted line represents the median $\Delta(E(B-V))$ while the shaded red region shows the 68\% intrinsic scatter of the distribution. $\Delta(E(B-V))$ increases at increasing $\log(SFR)$ and decreasing $\log(sSFR)$}. Overall, $E(B-V)\sim E(B-V)_{UV}$ at $\log(SFR/[$M$_\odot$yr$^{-1}$])~$ \lesssim -1.5$ and $\log(sSFR/yr^{-1}) \gtrsim -8$, whereas at higher SFRs and lower sSFRs the stellar vs. gas E(B-V) relation follows the \citetalias{calzetti97b} relation.
\label{fig:ebv-gas-stars}
\end{figure*}

Here, we aim to explore the behaviour of the gas $E(B-V)$ (obtained with the Balmer decrement and the \citealt{cardelli89} attenuation law, as described in Sec.~\ref{sec:method-dust}), and the stellar $E(B-V)_{UV}$ (obtained from the $\beta$ slope of stellar continuum fitting and the \citealt{reddy16} attenuation law, as described in Sec.~\ref{sec:data-analysis-stellar-cont-uv}).
The upper panels of Fig.~\ref{fig:ebv-gas-stars} show the comparison between the stellar $E(B-V)_{UV}$ and the gas $E(B-V)$, color-coded as a function of the SFR and specific SFR (sSFR), calculated considering the stellar mass and SFR within the COS aperture (see Table~6 in \citetalias{berg22}). The black dashed line shows the 1:1 relation, while the black dotted line shows the empirical relation between 
stellar and gas $E(B-V)$ according to \citet{calzetti97b} (\citetalias{calzetti97b} hereafter): $E(B-V)_{UV} = (0.44\pm 0.03)\times E(B-V)$ (or $E(B-V) = (2.27\pm0.15)\times E(B-V)_{UV}$).

From Fig.~\ref{fig:ebv-gas-stars}, we note that there is just a mild correlation between $E(B-V)_{UV}$ and $E(B-V)$ (Pearson factor of $\sim 0.43$). We notice that majority of the CLASSY galaxies lies along the 1:1 relation or are within that and the \citetalias{calzetti97b} line. 
The most significant correlation that we observe is between the gas $E(B-V)$ and the difference (in mag) $\Delta(E(B-V)) = E(B-V)-E(B-V)_{UV}$, characterized by a Pearson factor of $\sim+0.78$, as shown in the bottom panel of Fig.~\ref{fig:ebv-gas-stars}. 
The linear best-fit of the correlation between $\Delta(E(B-V))$ and $E(B-V)$ is indicated by the solid red lines and reported on the upper right of the bottom panels, and translates into:
\begin{equation}\label{eq:ebv}
  \begin{aligned}
    E(B-V)_{UV} = (0.33 \pm 0.08)\times E(B-V) + (0.03 \pm 0.01)\\
    E(B-V) = (3.00 \pm 0.70)\times E(B-V)_{UV} - (0.09 \pm 0.04)
    \end{aligned}
\end{equation}
with scatter of 0.05, while the dashed and dotted lines represent the $1\sigma$ and $2.6\sigma$ around the best-fit. To infer the best-fit (here and in the following sections) we used the {\it LtsFit} package developed by \citet{cappellari2013a} which allows us to perform a robust linear fit taking into account the uncertainties on both axis and the intrinsic scatter. This method also clips outliers, using the robust Least Trimmed Squares (LTS) technique by \citet{rousseeuw2006}. As explained in Section 3.2.2 of \citet{cappellari2013a}, their algorithm adopts an initial guess and performs a first least-squares fit. Then it computes the standard deviation of the residuals, selecting all data point deviating no more than $2.6\sigma$ from the fitted relation, performing a final fit for the selected points. The errors on the coefficients are computed from the covariance matrix.
 
Interestingly, Fig.~\ref{fig:ebv-gas-stars} shows that $E(B-V)$ and $E(B-V)_{UV}$ have similar values at low $\log(SFR)$ and high low $\log(sSFR)$ (and low gas-phase metallicity, since 12+log(O/H) correlates with $E(B-V)$; see Fig.~\ref{fig:dustext}), and are in fact close to the 1:1 relation (dashed black line) until $\sim 0.2$~mag. At increasing $\log(SFR)$ and decreasing $\log(sSFR)$, the stellar $E(B-V)_{UV}$ remains slightly constant with values always below $\sim0.2$~mag, while the gas $E(B-V)$ increases, reaching $\sim 0.5$~mag. This confirms the presence of an excess of dust attenuation in the gas with respect to the stars at high $\log(SFR)$ and low $\log(sSFR)$, in line with the empirical relation found by \citetalias{calzetti97b}. This relation implies that stars are on average a factor two less reddened than the ionized gas, which is related to the fact that the covering factor of the dust is larger for the gas than for the stars (\citetalias{calzetti97b}).
This topic is quite debated, with works suggesting a similar nebular and stellar reddening (e.g., \citealt{erb06,reddy10,pannella15,shivaei15b,puglisi16}) and others finding a higher nebular reddening (e.g., \citealt{calzetti97b,calzetti00,wild11a,kashino13,price14,reddy15,qin19,theios19,shivaei20}), as nicely summarised in Table~1 in \citet{shivaei20}.

As explained in \citetalias{calzetti97b}, the nebular emission requires the presence of the ionizing stars, that remain relatively close to their (dusty) place of birth during their short lifetime, while the UV stellar continuum is contributed also by non-ionizing stars, that have the time to move to regions of lower dust density. Hence, ionizing and non-ionizing stars are not expected to be co-spatial, implying that a correlation is not expected between stellar continuum and nebular emission, and thus between their reddening, in contrast with the empirical result found by \citetalias{calzetti97b}. A possible interpretation to explain this differential attenuation is a two-component dust model, with diffuse dust attenuating light from all stars, and the birth cloud dust component only attenuating light originating from the star-forming regions \citep{wild11b}. In line with this interpretation, \citet{wild11b} found a dependence of this extra-attenuation and sSFR. As nicely illustrated and explained in Fig.~5 of \citet{price14}, at high sSFR the continuum light is dominated by young, massive stars located in the birth clouds, so both the continuum and emission lines are attenuated by both dust components, and stellar and nebular $E(B-V)$ are similar; at decreasing sSFR, the discrepancy between stellar and nebular $E(B-V)$ increases, since less massive stars, generally residing outside the birth clouds, have a higher contribution to the continuum emission, while the emission lines are still attenuated by both dust components. This is in line with what we find in Fig.~\ref{fig:ebv-gas-stars}. An important thing to underline is that both \citet{wild11b} and \citet{price14} take into account galaxies with $\log(sSFR/yr^{-1}$) in the range $\sim$~[-10,-8.5], while CLASSY galaxies are systematically shifted at higher $\log(sSFR/yr^{-1}$) in the range $\sim$~[-8.4,-7] (see \citetalias{berg22}). Indeed, CLASSY galaxies were selected to be compact and UV bright, and thus have significantly enhanced sSFRs with respect to their star-forming main-sequence counterparts at $z\sim0$, and are more comparable to the $z\sim2$ galaxy population \citepalias{berg22}.
This could explain why we find no correlation of the sSFR with $E(B-V)_{UV}$, and just a mild anti-correlation with the gas $E(B-V)$ (Pearson factor of $\sim -0.36$), and with their difference (Pearson factor of $\sim -0.40$). 
Overall, we conclude that $E(B-V) \sim E(B-V)_{UV}$ at $\log(SFR/$[M$\odot$yr$^{-1}$])~$ \lesssim -1.5$ and $\log(sSFR/yr^{-1}) \gtrsim -8$, whereas at higher SFRs and lower sSFRs the stellar vs. gas E(B-V) relation follows that of \citetalias{calzetti97b}. 
This relationship has the potential for us to derive the gas $E(B-V)$ in galaxies for which rest-frame UV spectra are available but the optical wavelengths are not accessible, as it will happen for the reionization galaxies that JWST will reveal. 

\subsection{UV Diagnostics for Electron Density}\label{sec:discussion-ne}
As shown in Sec.~\ref{sec:results-ne}, the electron density estimated from \fciii~$\lambda$1907/\ciii~$\lambda$1909 is overall $\sim2$~dex higher than the one obtained from \sii~$\lambda$6717/\sii~$\lambda$6731, despite the large errors on \ciii\ densities. 
Unfortunately, many previous works have been limited in determining \ciii\ densities by the low resolution of the spectra (e.g., \citealt{ravindranath20}). 
However, this discrepancy has also been found in a handful of cases at $z\sim1-3$ where both the rest-frame optical and UV density diagnostics were available \citep{hainline09,quider09,christensen12,bayliss14,james14,maseda17,james18,berg18,acharyya19,schmidt21}, and in few local galaxies as well (e.g., \citealt{berg16}). We also find a consistent offset between $n_e$(\sii) and the density values from \fsiiii~$\lambda$1883/\siiii~$\lambda$1892 for the three galaxies for which it is available (i.e., J1044+0353, J1253-0312 and J1448-0110) and an even larger offset for the two galaxies with \niv\ densities (i.e., J1253-0312 and J1545+0858).

A possible reason for this UV-optical $n_e$ discrepancy, as explained in Sec.~\ref{sec:methods}, is that the optical \sii\ is tracing the {\it low-ionization} regions while the UV doublets originate in {\it intermediate-ionization} and {\it high-ionization} regions, closer to the ionizing source \citep{maseda17,berg18,james18}. 
Indeed, the ionization potential of these UV diagnostics ($25-75$~eV) and the range of densities for which they are $n_e$-sensitive (up to $10^5$~cm$^{-3}$) reach higher values than \sii\ doublet ($E\sim 10-20$~eV and $n_e<10^4$~cm$^{-3}$), as shown in the upper panels of Fig.~\ref{fig:den_pyneb_cloudy}.
In contrast to our findings, it should be noted that \citet{patricio16} measured consistent densities using \ciii\ and \oii\ density diagnostics for a $z=3.5$ star-forming galaxy, finding $n_e \sim 100$~cm$^{-3}$, typical of local \hii\ regions \citep{osterbrock89,osterbrock06}. 
However, they did find higher values of $n_e$ using the UV \niv\ and \siiii\ diagnostics, suggesting that there must be other ISM properties that are playing a role.

As introduced in Sec.~\ref{sec:methods}, a more self-consistent comparison between UV and optical $n_e$ diagnostics would be achieved by comparing UV $n_e$ diagnostics with the \cliii\ and \ariv\ optical line ratios, which similarly originate in the {\it intermediate-} and {\it high-ionization} regions, respectively. 
As shown in Fig.~\ref{fig:den_pyneb_cloudy}, \cliii\ traces a slightly lower range of densities (with $n_e$ up to $\sim5\times10^4$), while \ariv\ a similar range to \ciii. 
If the density structure were following the ionization structure, going from the diffuse to the densest gas, looking at the ionization potential of each density diagnostic tracer, the order would be: $n_e$(\sii) < $n_e$(\oii) < $n_e$(\feiii) < $n_e$(\siiii) < $n_e$(\cliii) < $n_e$(\ciii) < $n_e$(\ariv) < $n_e$(\niv). 
This would be true assuming that nebulae are formed of uniform shells or rings, with density decreasing inside-out according to $1/r^2$ \citep{stanghellini89}. 
However, this density structure is not clearly confirmed by observations, with many works showing contradicting results.
For instance, \citet{maseda17} found a consistent median density of log($n_e$)~$\sim 3.2$ derived from \ciii\ for a sample of 17 $z>1.5$ \ciii-emitting galaxies and densities obtained from \ariv\ in a control-sample of SDSS $z\sim0$ objects.
On the contrary, as detailed in Sec.~\ref{sec:results-ne}, we found that in the optical, the \ariv\ densities are consistent with \cliii\ values, with a median value $n_e\sim 10^3$~cm$^{-3}$, whereas both are lower than the UV diagnostics by $\sim 1$~dex, despite the fact that \ariv\ should trace a higher ionization zone. UV densities are found to be consistent with \feiii\ densities (and \feiii\ has a lower ionization potential than \cliii), but these are just upper limits given the large error bars of the measured \feiii\ line ratios.
Indeed, the relation between the ionization and density structure is not trivial, since the internal structure of \hii\ regions is shaped both by winds and radiation pressure even when they are not dynamically important (see e.g., Fig~2 \citealt{geen20}).
Also, the apertures of our UV and optical observations are not allowing us to observe singular \hii\ regions, and because of the differential dust attenuation with wavelength, UV and optical tracers may not be tracing the same regions along the line of sight, with optical tracers coming also from regions completely extincted in the UV.

A possible scenario that explains why $n_e$(\ariv) densities are not higher than $n_e$(\ciii) would be an ionization front due to the central source that produces a rarefication at the inner boundary of the nebula, and thus high-ionization zone density tracers such as \ariv\ would be produced in a central hole \citep{giuliani81,stanghellini89}. 
However, if this were the case, 
we should see a significant difference in the kinematics of \ariv\ lines with respect to lower-ionization region density tracers (which we do not, \citetalias{mingozzi22} in prep.). 
Hence, we cannot find evidence of an hydrodynamic effect at the basis of the density structure from our data. 
In this context, very high spectral resolution optical data, especially integral-field spectroscopy data that allows to map the density structure within a galaxy, could help understand this offset.
An alternative explanation for the $n_e$(\ariv)-$n_e$(\ciii) discrepancy could be related to the atomic parameters taken into account \citep{morisset20,dedios21}. 
Indeed, \pyneb\ and \cloudy\ adopt different atomic data for the ions considered in this work, and in Fig.~\ref{fig:den_pyneb_cloudy}, we showed that in general \cloudy\ predictions for densities were in agreement with \pyneb, apart from \ariv, for which \cloudy\ would predict densities up to $\sim 1$~dex higher. This corresponds to the discrepancy with the UV tracers that we find in this work. 

\begin{figure}
\begin{center}
    \includegraphics[width=0.5\textwidth]{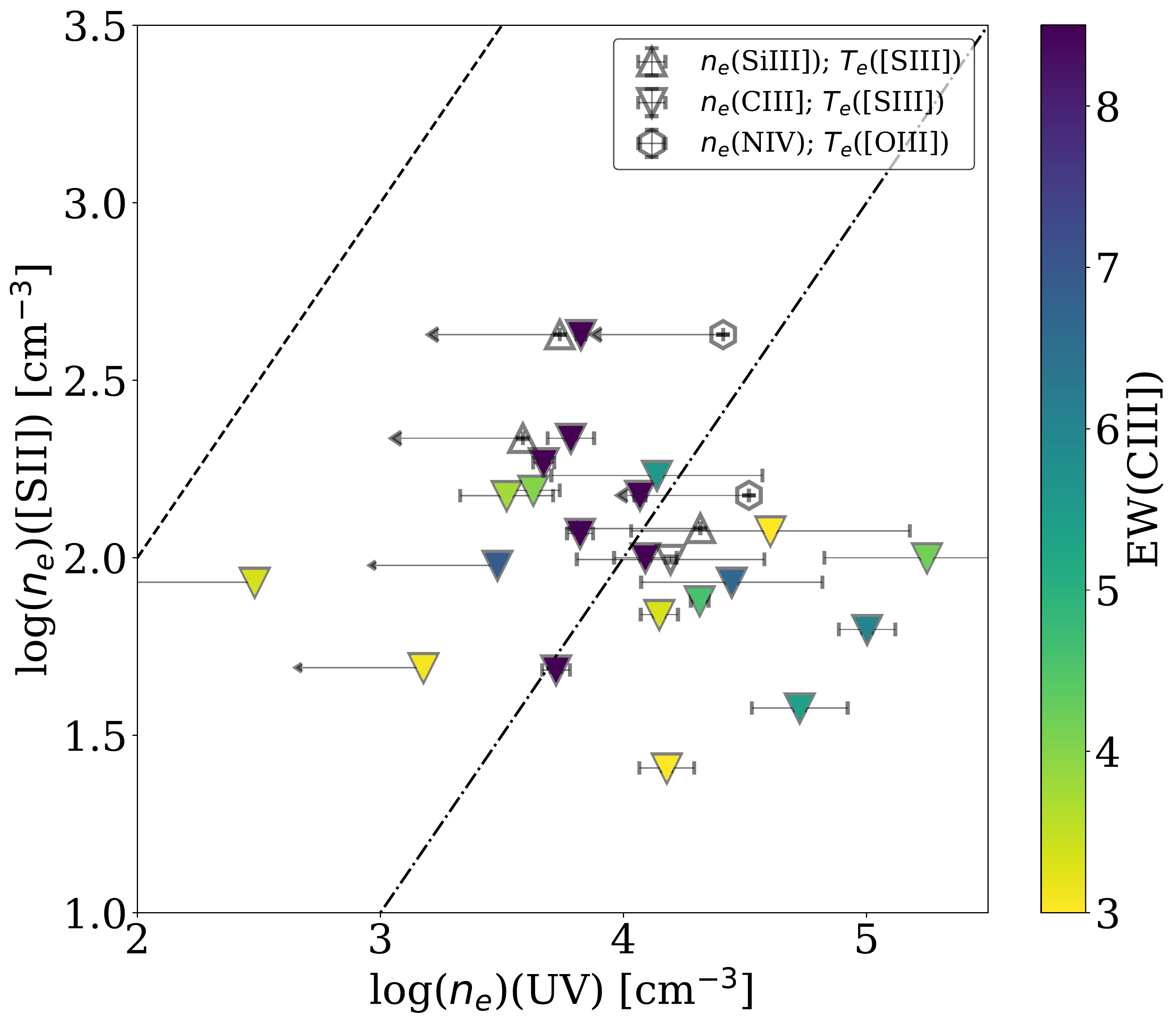}
\end{center}
\caption{The $n_e$(\sii) in logarithm as a function of $n_e$(\siiii) (up-pointing triangles) and $n_e$(\niv) (hexagons), as well as $n_e$(\ciii) (down-pointing triangles), color-coded as function of EW(\ciii). The uncertainties of the displayed quantities are shown in both axis. The black dot-dashed line shows the 1:1 relation, while the dot-dashed black line represents the same relation shifted of 2~dex, that we added to highlight the level of the offset.} 
\label{fig:den_siiciii_comp}
\end{figure}
To further investigate the relation between optical and UV densities, Fig.~\ref{fig:den_siiciii_comp} shows $n_e$(\sii) as a function of $n_e$(\siiii) (up-pointing triangles), $n_e$(\ciii) (down-pointing triangles) and $n_e$(\niv) (hexagons), colour-coded as a function of the EW(\ciii). The dashed black lines shows the 1:1 relation, while the dot-dashed black line represents the 1:1 relation shifted of 2~dex, that we added to highlight the level of the offset.
We tested several quantities, but a weak correlation was only observed between log($n_e$(\sii$\lambda\lambda$6717,31)) and EW(\ciii) (Pearson: $r \sim0.40$, $p \sim0.02$), which can be appreciated by the color-coding of Fig.~\ref{fig:den_siiciii_comp}.
A slight increase of EW(\ciii) with density would be expected according to \citet{jaskot16} models. Indeed, the higher density nebula are characterized by a higher incident ionizing flux, because of the definition of the ionization parameter (see Eq.~1 in \citealt{jaskot16}). Therefore, a higher density implies higher temperatures, as well as higher collision rates, which, in combination with the abundance of C$^{2+}$ ions, ultimately determine the \ciii\ emission strength \citep{jaskot16}.
However, the strength of \ciii\ emission is also affected by the gas-phase metallicity, the ionization parameter or the stellar age, with increasing equivalent widths predicted to reach values $\gtrsim 10-15 $~\AA\ at ages $<3$~Myr and log($U$)~$>-2$ \citep{jaskot16,nakajima18,ravindranath20}.
In CLASSY galaxies, we find an anti-correlation between log(EW(\ciii)) and 12+log(O/H) (Pearson: $r = -0.43$, $p=0.03$), a slight correlation with log($U$) (Pearson: $r = 0.3$, $p=0.4$), and an anti-correlation with stellar age (Pearson: $r = -0.4$, $p=0.04$), confirming that the \ciii\ strength of the emission is determined by many factors that are not yet fully understood \citep{ravindranath20}. 
We will comment more on the complex relation between \ciii, metallicity and ionization parameter in Sec.~\ref{sec:discussion-metlogu}. 

Overall, the CLASSY galaxies show that $n_e$(\ciii) is on average $\sim2$~dex higher than $n_e$(\sii), and this offset can be used to determine the low-ionization zone electron density in similar galaxies if only the intermediate-ionization zone electron density is available from rest-frame UV spectra covering \ciii\ lines.

\subsection{UV Diagnostics for Electron Temperature}\label{sec:discussion-Te}
\begin{figure*}
\begin{center}
    \includegraphics[width=1\textwidth]{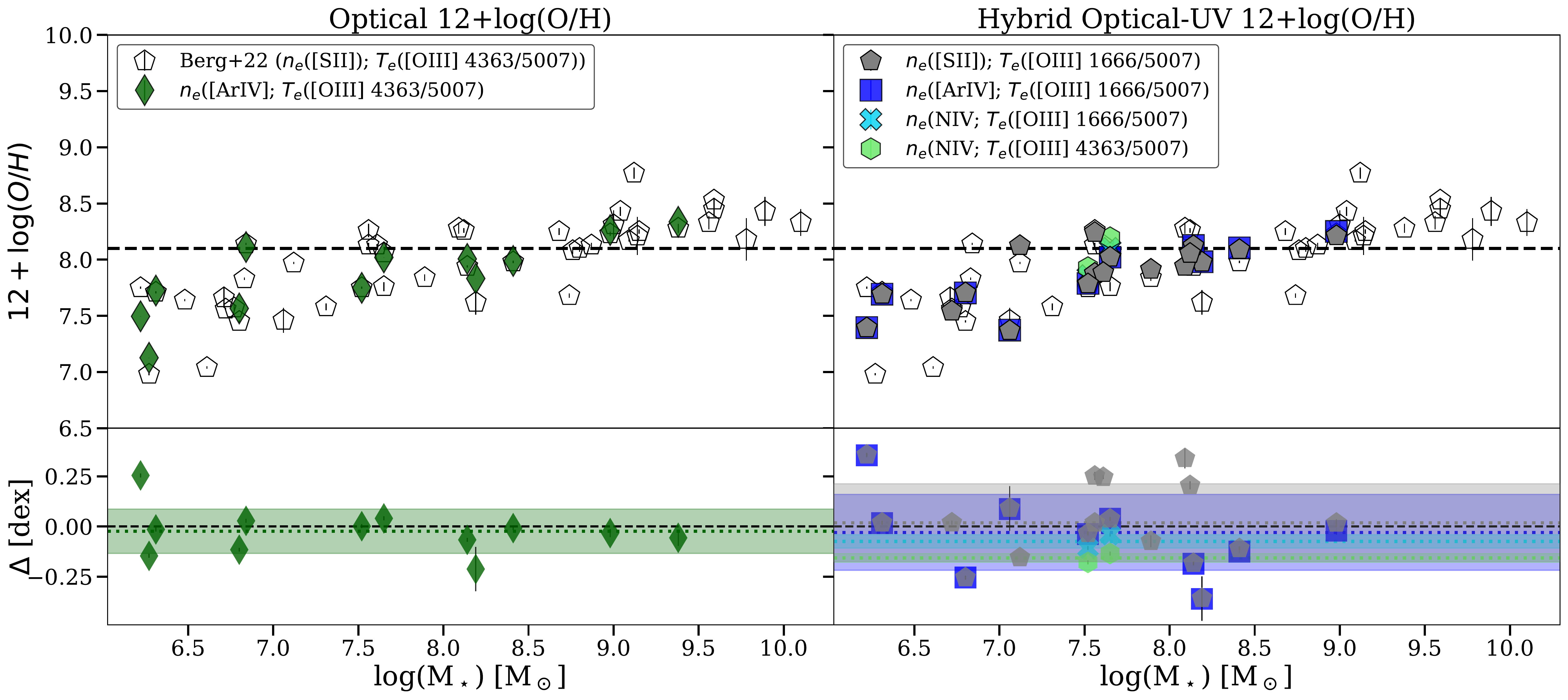}
\end{center}
\caption{Top panels: comparison of 12+log(O/H) from \citetalias{berg22}, with the estimates obtained using the combination of with either $n_e$(\ariv) or $n_e$(\niv), and either $T_e$(\oiii~$\lambda$4363/$\lambda$5007) or $T_e$(\oiii~$\lambda$1666/$\lambda$5007), as indicated in the legend, as a function of the stellar mass. For clarity reasons, the optical and hybrid optical-UV measurements of 12+log(O/H) are shown on the left and right, respectively, while 12+log(O/H) from \citetalias{berg22} is shown in both panels. The uncertainties of the displayed quantities are shown in both axis.
Bottom panels: difference in dex between \citetalias{berg22} 12+log(O/H) and the other estimates, keeping the same symbols and colors as in the top panel. }
\label{fig:tempmet_diag}
\end{figure*}

Here we have the important (and rare) opportunity to investigate if the hybrid UV-optical ratio \oiiiuv~$\lambda$1666/$\lambda$5007 ($T_{e,hybrid}$) is a reliable tracer of $T_e$ with respect to the most generally used optical ratio, \oiii~$\lambda$4363/$\lambda$5007 ($T_{e, opt}$). 
In Sec.~\ref{sec:results_temp} we commented that the median difference in temperature between $T_{e,opt}$ and $T_{e,hybrid}$ is on average $\Delta$log($T_e$)$_{opt - hybrid} \sim -1000$~K, with higher values found for $T_{e,hybrid}$.
This difference could be due to several factors. Indeed, it is not trivial to compare \oiiiuv~$\lambda$1666 and \oiii~$\lambda$5007, given that different instruments are used, with different apertures and calibration accuracies (see also App.~\ref{app:A}), as well as the large difference in wavelength, that can lead to higher uncertainties in the dust attenuation correction. Moreover, as commented in Sec.~\ref{sec:methods-temp}, in a patchy ISM, the UV light could be visible only through the less dense and/or less reddened regions along the line of sight, while the optical may trace also from denser and/or more reddened regions, leading to an intrisic difference between the line ratios, and thus in the derived temperatures. 

Another reason to explain the discrepancy could be related to an effect of the gas pressure, as proposed by \citet{nicholls20}. In their work, \citet{nicholls20} list this effect under the  intrinsic caveats and limitations of the direct method to derive temperatures. Indeed, they commented that \oiii~$\lambda$5007 is quenched more rapidly than  \oiii~$\lambda$4363 as the gas pressure increases, meaning that \oiii~$\lambda$4363/$\lambda$5007 increases at increasing pressure, emulating a higher temperature as it happens in Active Galactic Nuclei (AGN; e.g., \citealt{nagao01}). This occurs because of  \oiii~$\lambda$4363 has a significantly higher critical density than \oiii~$\lambda$5007 ($n_{\rm crit}\sim 3.3 \times 10^7$~cm$^{−3}$ versus $n_{\rm crit}\sim 7.0 \times 10^5$~cm$^{−3}$; \citealt{osterbrock89,osterbrock06}).
They specify that this effect can especially be a problem in unresolved high luminosity \hii\ regions. \oiiiuv~$\lambda$1666 has an even higher critical density ($n_{\rm crit}\sim 4 \times 10^{10}$~cm$^{−3}$; \citealt{zheng88}) than \oiii~$\lambda$4363.
Therefore, \citet{nicholls20} conclude that higher values of $T_e$(\oiiiuv~$\lambda$1666/\oiii~$\lambda$5007) compared to $T_e$(\oiii~$\lambda$4363/$\lambda$5007) may be due to pressure bias, under suitable conditions of pressure and temperature. 

All the factors discussed in the previous paragraphs could contribute to the median offset of $\sim -1000$~K between $T_{e,opt}$ and $T_{e,hybrid}$ that we observe. However, the fundamental aspect to understand is to what extent this offset impacts the estimate of 12+log(O/H).
We explore this in the top panels of Fig.~\ref{fig:tempmet_diag}, where we compare the 12+log(O/H) values that \citetalias{berg22} calculated using $n_e$(\sii) and \oiii~$\lambda$4363/$\lambda$5007 (black pentagons), with the values we calculate using either $n_e$(\ariv) and \oiii~$\lambda$4363/$\lambda$5007 ($T_{e,opt}$; darkgreen diamonds) or $n_e$(\sii), $n_e$(\ariv) and \oiiiuv~$\lambda$1666/$\lambda$5007 ($T_{e,hybrid}$; gray pentagons and blue squares), as a function of the stellar mass. 
We also tested the combinations $n_e$(\niv) and \oiii~$\lambda$4363/$\lambda$5007 (green hexagons), or $n_e$(\niv) and \oiiiuv~$\lambda$1666/$\lambda$5007 (turquoise crosses). 
For clarity reasons, we display the optical 12+log(O/H) on the left panels and the Hybrid/UV 12+log(O/H) on the right panels, showing as a reference the \citetalias{berg22} measurements.
The lower panels of Fig.~\ref{fig:tempmet_diag} show the difference in dex between the various estimates and the values from \citetalias{berg22}.
The dashed and dotted horizontal lines in the main and minor panels show the median values of the plotted quantities, respectively, while the shaded regions in the bottom panels represent the 68\% intrinsic scatter of each distribution, color-coded accordingly. 

While there is an offset between 12+log(O/H) from \citetalias{berg22} and the different estimates of 12+log(O/H), it is mainly within $\pm0.3$~dex.
In general, we find that this offset (the median value and its distribution, shown in the bottom panels of Fig.~\ref{fig:tempmet_diag}) is consistent between the value derived using $T_{e,opt}$ and the one using $T_{e,hybrid}$, when measured with different density diagnostics. 
As commented in Sec.~\ref{sec:methods}, if possible it is important to take into account diagnostics of the same ionization zone because the lack of co-spatiality in estimating ISM properties could lead to incorrect estimates of physical and chemical properties. 
Despite this fact, the method based on the \sii\ and the optical \oiii\ ratios used in \citetalias{berg22} is the most common, even though \sii\ emission is not co-spatial with \oiii\ emission (e.g., see Fig.~4 from \citealt{nicholls20}). 
Using for instance \ariv\ or \niv\ diagnostics to measure the density would be more ideal, but in practice these alternative optical density diagnostic lines are faint, and thus not available for all the CLASSY targets.

\begin{figure}
\begin{center}
    \includegraphics[width=0.48\textwidth]{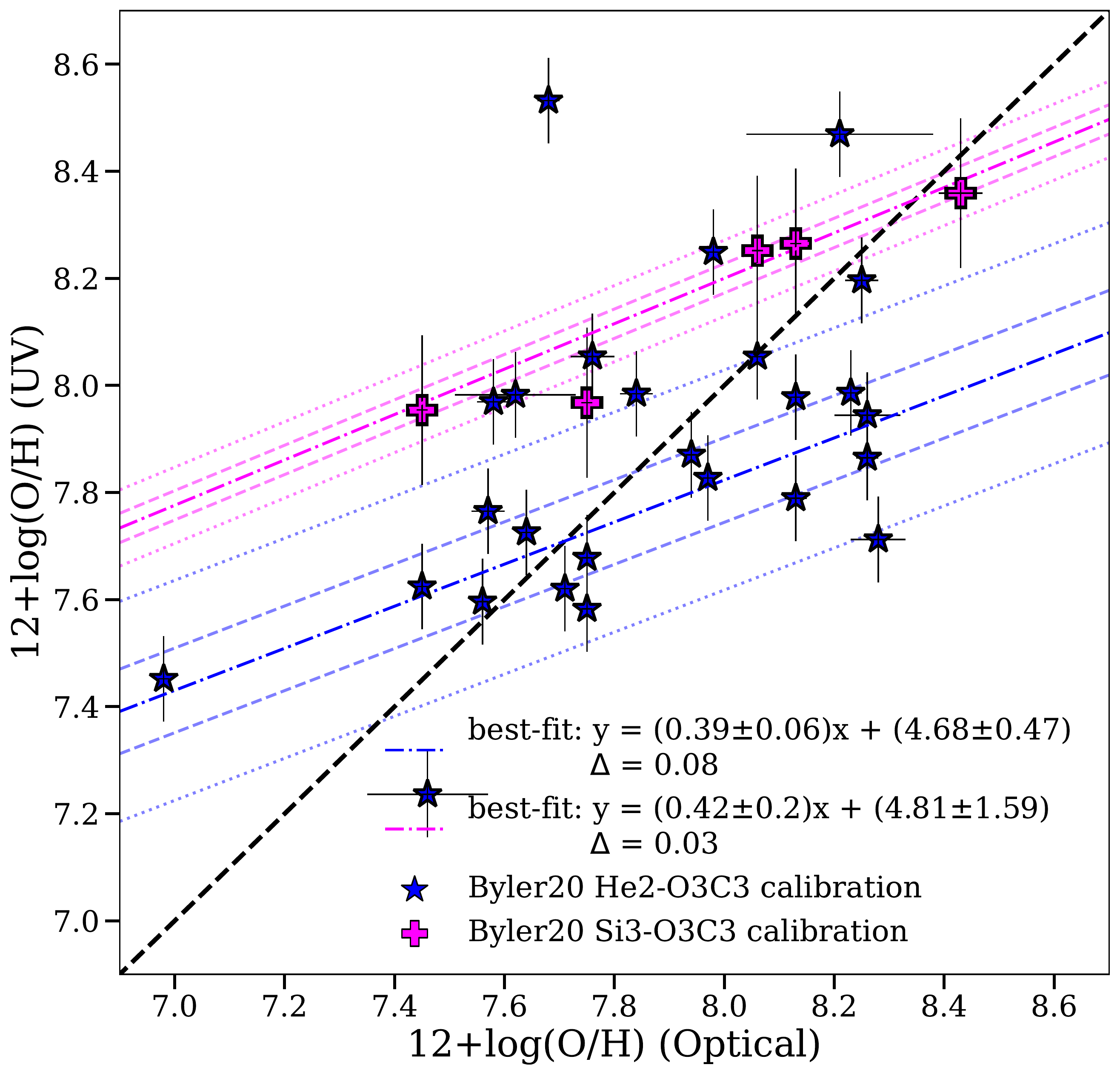}
\end{center}
\caption{Comparison of optical and UV metallicities, comparing \citetalias{berg22} 12+log(O/H) with the exclusively UV He2-O3C3 and Si3-O3C3 calibrations proposed by \citetalias{byler20}, shown as blue stars and magenta pluses. The uncertainties of the displayed quantities are shown in both axis.
The dashed black line indicates the 1:1 relation. The dot-dashed blue and magenta lines show the best-fits of the 12+log(O/H)$_{UV}$ as a function of the optical 12+log(O/H), as reported in the legend. The blue and magenta dashed and dotted lines represent the 1$\sigma$ and 2.6$\sigma$ around each best-fit.}
\label{fig:bylermet}
\end{figure}

Overall, we conclude that the hybrid UV/optical temperature diagnostic, $T_{e,hybrid}$, is not dominated by potential systematic uncertainties, and can be used as a reliable tracer of the electron temperature in the systems covered here. 
We also conclude that the hybrid UV/optical $T_e$ allows to estimate the gas-phase metallicity with a scatter of $\pm0.3$~dex with respect to the optical 12+log(O/H), again without showing systematic discrepancies. 
This is particularly important as in our sample we find that the \oiiiuv~$\lambda$1666 flux is on average $\sim$0.3~dex brighter than the \oiii~$\lambda$4363, and thus it could be more easily observed in high-$z$ galaxies than the weaker and less detectable \oiii~$\lambda$4363 emission.

While we have shown that \oiiiuv~$\lambda$1666 can allow us to reliably and directly estimate the metallicity of the ionized gas, the $T_{e,hybrid}$ method still depends on us having access to the optical \oiii$\lambda\lambda$4959,5007 emission.
To the best of our knowledge the only work in the literature that proposed promising 12+log(O/H) diagnostics using exclusively UV emission lines is \citet{byler20} (\citetalias{byler20} hereafter). 
Using \cloudy\ modeling, these authors derived calibrations from \heii$\lambda$1640/\ciii$\lambda\lambda$1907,9 versus \oiiiuv$\lambda$1666/\ciii$\lambda\lambda$1907,9 (He2-O3C3 hereafter) and \siiii$\lambda\lambda$1883,92/\ciii$\lambda\lambda$1907,9 versus \oiiiuv$\lambda$1666/\ciii$\lambda\lambda$1907,9 (Si3-O3C3 hereafter). The typical statistical errors for these relations are $\pm0.08$~dex and $\pm0.14$~dex, respectively \citep{byler20}. 
In Fig.~\ref{fig:bylermet}, we compare the optical-based 12+log(O/H) values derived in \citetalias{berg22} with the values we derive here from the He2-O3C3 and Si3-O3C3 calibrations, shown as blue stars and magenta pluses, respectively. 
A clear disagreement can be seen between the 12+log(O/H) values derived from the UV diagnostics and optical direct method. 
The comparison with the 1:1 relation shows the difference between optical and UV 12+log(O/H).
The Si3-O3C3 values are systematically higher with an offset up to $\sim 0.6$~dex with respect to the optical 12+log(O/H) from \citetalias{berg22}, which decreases at increasing 12+log(O/H). However, we do not have enough \siiii$\lambda\lambda$1883,92 measurements to robustly evaluate the level of this trend.
The He2-O3C3 estimates instead show a larger scatter of $\sim \pm 0.5$~dex, over-predicting the metallicity at 12+log(O/H)~$<8$, and under-predicting it at 12+log(O/H)~$>8$.
This is in agreement with what found by \citet{rigby21}, who compared UV and optical 12+log(O/H) diagnostics for a single gravitationally lensed source, finding that the UV estimates from \citetalias{byler20} are $\sim0.5−0.8$~dex lower than the optical estimates, found in the range 12+log(O/H)~$\sim8.20-8.60$.

In order to investigate this offset further, we tested whether the optical and UV-diagnostic 12+log(O/H) offset correlates with any optical or UV quantities that we measured here, or with CLASSY galaxy properties in general. We only found correlations with the optical 12+log(O/H), as it can be seen in Fig.~\ref{fig:bylermet}. 
The dot-dashed blue and magenta best-fit relations in Fig.~13 allow us to essentially re-calibrate the \citetalias{byler20} equations and to obtain the optical 12+log(O/H) from UV-only quantities, which we show in the following expressions:
\begin{equation}\label{eq:metcorr1}
\begin{aligned}
    12+\log(O/H) & = (0.39 \pm 0.06)\times (12+\log(O/H))_{\rm He2} & \\ - (4.68 \pm 0.47)
\end{aligned}
\end{equation}
\begin{equation}\label{eq:metcorr2}
\begin{aligned}
    12+\log(O/H) & = (0.42 \pm 0.20)\times (12+\log(O/H))_{\rm Si3} & \\ - (4.81 \pm 1.59)
\end{aligned}
\end{equation}
with a scatter of 0.08 and 0.03~dex, respectively, while the dashed and dotted lines represent the 1$\sigma$ and 2.6$\sigma$ around the best-fits.

\subsection{UV diagnostics for metallicity and ionization parameter}\label{sec:discussion-metlogu} 
There have been many attempts to infer the gas-phase metallicity using a combination of photoionization models and UV emission lines (e.g., \citealt{fosbury03,jaskot16,gutkin16,feltre16,chevellard16,perez-montero17,byler18,nakajima18}) and/or measuring metallicity directly with UV \oiiiuv\ auroral line (e.g., \citealt{erb10,berg18, james18}).
However, few 12+log(O/H) calibrations have been proposed using exclusively UV emission lines (e.g., \civ$\lambda\lambda$1548,51, \heii~$\lambda$1640, \oiiiuv~$\lambda$1666 and \ciii$\lambda\lambda$1907,9; see \citealt{byler20,perez-montero17}), because UV lines are also strongly sensitive to other properties, such as the ionization parameter and shape of the ionizing continuum \citep{gutkin16,feltre16,nakajima18,maiolino19}.
The parameter space coverage of CLASSY and the high frequency of UV line detections throughout the sample allow us to explore how UV lines vary with the gas-phase metallicity 12+log(O/H) and ionization parameter, log($U$), which we infer from reliable optical diagnostics.

In order to fully explore the utility of UV diagnostic lines as possible tracers for these ISM properties, in this work we investigated a suite of derived quantities including: \oiiiuv~$\lambda$1666/\heii~$\lambda$1640; \ciii~$\lambda\lambda$1907,9/\heii~$\lambda$1640; \ciii~$\lambda\lambda$1907,9/\oiiiuv~$\lambda$1666; \civ~$\lambda\lambda$1548,51/\ciii~$\lambda\lambda$1907,9; EW(\heii~$\lambda$1640); EW(\oiiiuv~$\lambda$1666); EW(\ciii~$\lambda\lambda$1907,9); EW(\civ~$\lambda\lambda$1548,51).
Since after \lya, \ciii~$\lambda\lambda$1907,9 and \civ~$\lambda\lambda$1548,51 are among the brighest UV emission lines in star-forming galaxies (e.g., \citealt{ravindranath20}), it is particularly useful, if possible, to exploit them to find reliable calibrations to infer metallicity and ionization parameter for galaxies with properties comparable to those of the CLASSY sample.

\begin{figure*}
\begin{center}
    \includegraphics[width=1\textwidth]{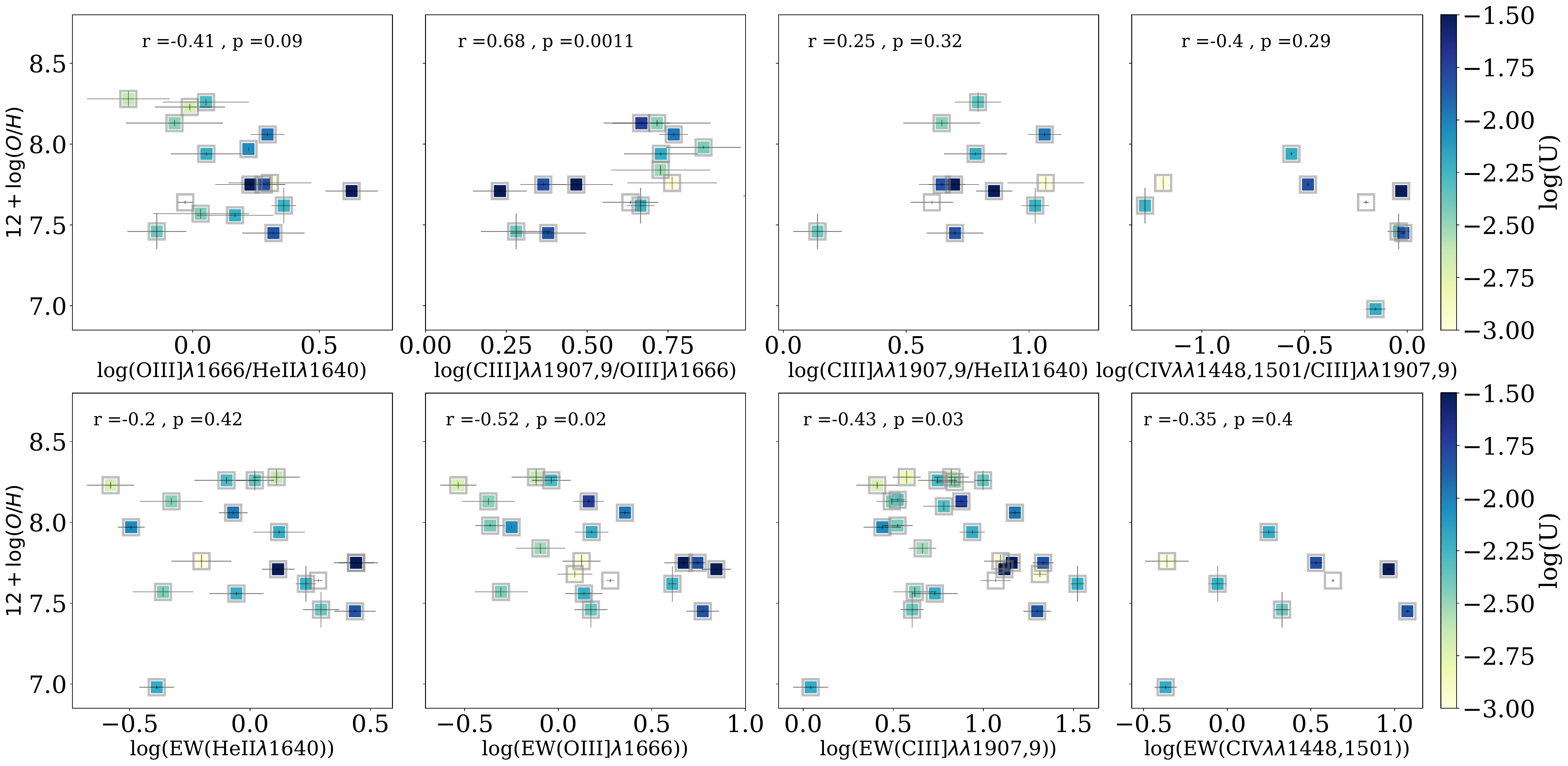}
    \includegraphics[width=1\textwidth]{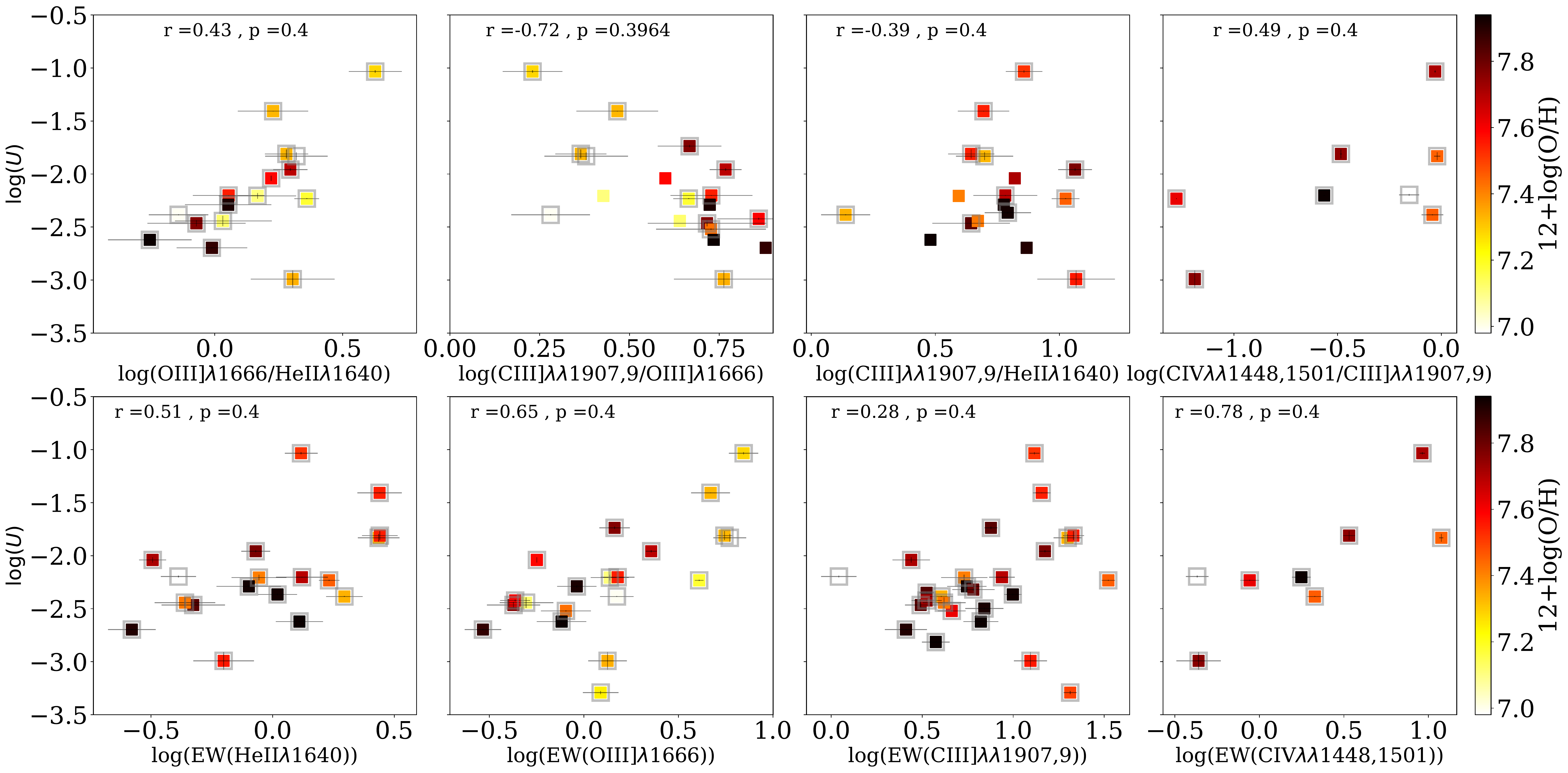}
\end{center}
\caption{Top panel: 12+log(O/H) as a function of \oiiiuv~$\lambda$1666/\heii~$\lambda$1640, \ciii/\oiiiuv~$\lambda$1666, \ciii/\heii~$\lambda$1640 and \civ/\ciii\ line ratios (upper panels), and EW(\heii~$\lambda$1640), EW(\oiiiuv~$\lambda$1666), EW(\ciii) and EW(\civ) (lower panels), color-coded as a function of the ionization parameter log(U) derived from O3O2. The uncertainties of the displayed quantities are shown in both axis. All the quantities are in logarithm, and the Pearson factor and pvalue are shown in each panel. Overall the most promising relations are between \ciii/\oiii~$\lambda$1666,  EW(\oiiiuv~$\lambda$1666)~$<4$~\AA\ and EW(\ciii) with 12+log(O/H), but there is scatter in the plots, partly due to a degeneracy with the ionization parameter, as it can be appreciated from the color-coding. Bottom panel: Same for log($U$), color-coding as a function of 12+log(O/H) from \citetalias{berg22}.}
\label{fig:temp-sn-all}
\end{figure*}

In Fig.~\ref{fig:temp-sn-all} we show how these quantities vary as a function of  12+log(O/H) (derived with the direct method, \citetalias{berg22}), color-coded as a function of log($U$) derived from O3O2 (top panel), and vice-versa (bottom panel). In particular, we show only the measurements for which the S/N of the lines used to derive the quantities is higher than the chosen threshold ($S/N>3$).
In each panel we also provide the Pearson correlation factor and relative pvalue. 
Among these quantities, we find that log(\ciii/\oiiiuv), log(EW(\oiiiuv)) and log(EW(\ciii)) show the most promising relation with 12+log(O/H), with the highest Pearson factor found for \ciii/\oiiiuv\ (Pearson correlation factor $\sim0.68$ and p-value~$\sim0.001$). 
\oiiiuv~$\lambda$1666/\heii~$\lambda$1640 is clearly increasing as metallicity decreases but with a turn-over at 12+log(O/H)~$\sim7.7$. 
We also notice that the presence of nebular \civ\ in pure emission can be used as well as a metallicity indicator, since it comes uniquely from 12+log(O/H)$\lesssim8$ (see also \citealt{senchyna19a,berg21a,senchyna22a,schaerer22}).
A slight anti-correlation between log(\civ/\ciii), log(EW(\civ)) and log(EW(\heii)) with metallicity (Pearson: $r\sim-0.40$, $r\sim-0.35$ and $r\sim-0.20$, respectively) could be due instead to residuals in the subtraction of stellar features, which would be most prominent at increasing metallicities. 

\begin{figure*}
\begin{center}
    \includegraphics[width=0.9\textwidth]{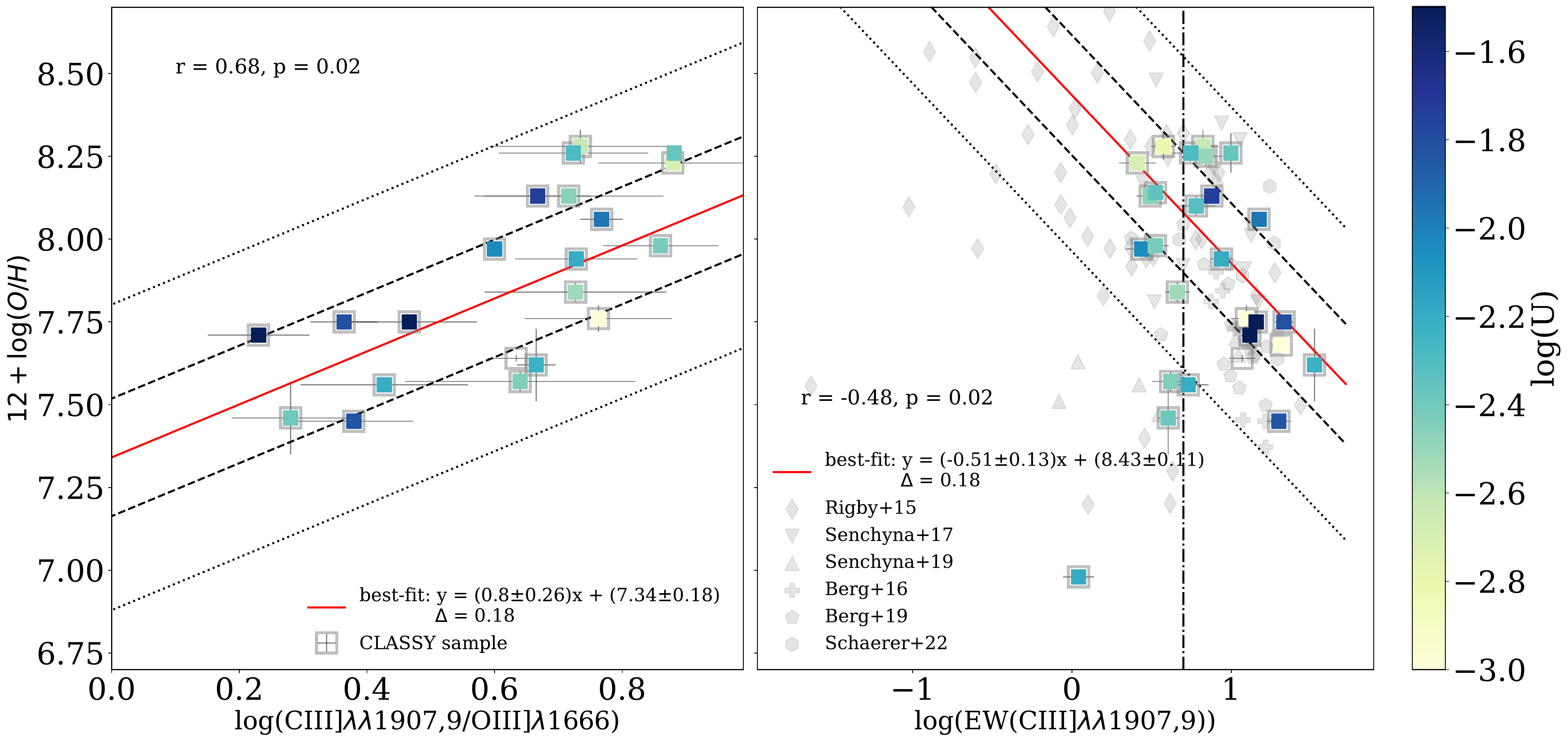}
\end{center}
\caption{Gas phase metallicity, 12+log(O/H), as a function of \ciii/\oiiiuv~$\lambda$1666 (left-panel) and  EW(\ciii) (right-panel), color-coded as a function of log($U$). The uncertainties of the displayed quantities are shown in both axis.
Our best-fit linear relations are shown by the solid red lines, while the dashed and dotted black lines represent the 1$\sigma$ and 2.6$\sigma$ around the best-fit. The scatter in both panels can be partly explained by a secondary dependence on the ionization parameter. In the right panel we also include values found for star-forming galaxies studied in recent works \citep{rigby15,berg16,senchyna17,senchyna19a,berg19b,schaerer22}, in which 12+log(O/H) was computed with the direct method, comparably to this work. This figure shows the most promising UV tracers of metallicity.}
\label{fig:temp-sn}
\end{figure*}

In the left panel of Fig.~\ref{fig:temp-sn}, we highlight the \ciii/\oiiiuv\ as a function of the 12+log(O/H), color-coded with respect to log($U$). 
\ciii/\oiiiuv\ clearly increases at increasing metallicities, following the C/O abundance ratio (e.g., \citealt{garnett90,izotov99,garnett04,erb10,guseva11}). 
Indeed, \ciii/\oiiiuv\ is considered a good tracer of C/O, because of the similar excitation and ionization potentials of C$^{+2}$ and O$^{+2}$ (i.e., $6-8$~eV and $24.8-35.1$~eV, respectively), and its minimal uncertainty due to reddening, given that the interstellar attenuation curve is nearly flat over the wavelength $1600-2000$~\AA\ \citep{garnett95,berg16}.

The metallicity dependence of C/O was also shown in \citet{jaskot16}, who attribute it either to the weaker winds of low-metallicity massive stars or to the longer timescales to enrich the ISM for lower mass stars (see also, \citealt{maeder92,henry00,chiappini03}). 
However, this relation is also characterized by a scatter that largely increases at low metallicities (12+log(O/H)~$<8$; see Fig.~6a of \citealt{berg16}, Fig.~3 of \citealt{amorin17} or Fig.~11 of \citealt{berg19a}) because at low metallicity, C/O can be sensitive to different physical conditions, including star formation histories, star formation efficiency, inflow rates, and variations in the IMF \citep{skillman98,berg16,amorin17,berg19a}. 
In agreement with \citet{stark14}, and the predictions of \citet{jaskot16},
\ciii/\oiiiuv~$>1$ for all the CLASSY galaxies (the lowest value is \ciii/\oiiiuv~$\sim 1.70$ for J1323-0132). 
\citet{jaskot16} predict that \ciii\ should only become comparable to \oiiiuv\ in photoionization models with metallicity of $Z<0.004$, only if both C/O ratios are low (C/O~$\leq 0.12$) and the ionization parameter is high (log($U$)~$\geq-2$), which is consistent with the values of 12+log(O/H) and log($U$) we measure in this work. 

The best-fit relation between \ciii/\oiiiuv\ and the direct-method metallicity 12+log(O/H), shown as the  red solid line in the left-panel of Fig.~\ref{fig:temp-sn}, is represented by:
\begin{equation}\label{eq:ciiioiii-met}
\begin{aligned}
    12+\log(O/H) = (0.80\pm0.26) & \times \log(CIII]1907,9/OIII]1666) \\
    & + (7.34\pm 0.18)
    \end{aligned}
\end{equation}
with an intrinsic scatter of 0.18~dex. The dashed and dotted black lines indicate the 1$\sigma$ and $2.6\sigma$ scatter from the relation, respectively.
The scatter in this relation, which is expected as we commented in the previous paragraph, can also be partly explained by a secondary dependence to log($U$), given the difference of the ionization potentials of the C and O lines \citep{jaskot16}. We will investigate in detail the behavior of \ciii/\oiiiuv\ with the C/O ratio and its scatter with respect to O/H in CLASSY in a forthcoming paper, focused on detailed photoionization modeling of the CLASSY galaxies.

The right panel of Fig.~\ref{fig:temp-sn} highlights the anti-correlation between EW(\ciii) and 12+log(O/H). Here we also include values found for local star-forming galaxies studied in recent works \citep{rigby15,berg16,senchyna17,senchyna19a,berg19b,schaerer22}, in which 12+log(O/H) was computed with the direct method, as in this work.
As found by \citet{rigby15,nakajima18,ravindranath20} low-metallicity galaxies tend to have a stronger EW(\ciii), showing a linear anti-correlation down to EW(\ciii)~$>5$~\AA\ for galaxies with 12+log(O/H)$~\lesssim 8.25$. A similar transition at EW(\ciii)~$>5$~\AA\ at 12+log(O/H)$~\lesssim 8.4$ is found and discussed in \citet{senchyna17,senchyna19a}. However, as shown in Fig.~\ref{fig:temp-sn}, for galaxies with EW(\ciii)~$<5$~\AA\ there is clearly a large scatter towards lower metallicities, due to the fact that galaxies with 12+log(O/H)~$\lesssim 7.5$ can show EW(\ciii)~$<5$~\AA\ as well. As such, the EW(\ciii)-12+log(O/H) relation is double-branched with a turn-over point at 12+log(O/H)~$\sim 7.5$. 

Indeed, as discussed in \citet{jaskot16}, metallicity sets the shape of the ionizing spectral energy distribution (SED) and defines the nebular gas temperature, which enhances \ciii\ emission at low metallicity due to the larger amount of ionizing photons and higher collisional excitation rates (e.g., \citealt{erb10,stark14}). However, at these low metallicities, the carbon abundance decreases as well, suppressing \ciii\ emission (e.g., \citealt{rigby15}).
Moreover, as also discussed in Sec.~\ref{sec:discussion-ne}, the strength of \ciii\ emission is not only affected by metallicity, but also by the ionization parameter, stellar age and gas density, with increasing equivalent widths predicted to reach values $\gtrsim 10-15 $~\AA\ at ages $<3$~Myr and log($U$)~$>-2$ \citep{jaskot16,nakajima18,ravindranath20}. 
This explains why J0934+5514 (i.e., I\,Zw\,18), characterized by the lowest metallicity among the CLASSY targets (12+log(O/H)~$\sim 6.98$), has EW(\ciii)~$\sim 1.10$~\AA\ despite a relative high ionization parameter (log($U$)~$\sim -2.20$), as already acknowledged by \citet{rigby15}.

Overall, the best-fit relation between log(EW(\ciii$\lambda\lambda$1907,9)) and the direct-method metallicity, valid at 12+log(O/H)~$> 7.5$, is shown in the right panel of Fig.~\ref{fig:temp-sn} by the red solid line and is represented by:
\begin{equation}\label{eq:ewciii-met}
\begin{aligned}
    12+\log(O/H) = (-0.51\pm0.13) & \times \log(EW(CIII]1907,9))\\
    & +(8.43\pm0.11)
    \end{aligned}
\end{equation}
with an intrinsic scatter of 0.18~dex. The dashed and dotted black lines indicate the 1$\sigma$ and $2.6\sigma$ scatter from the relation, respectively.

\begin{figure*}
\begin{center}
    \includegraphics[width=0.8\textwidth]{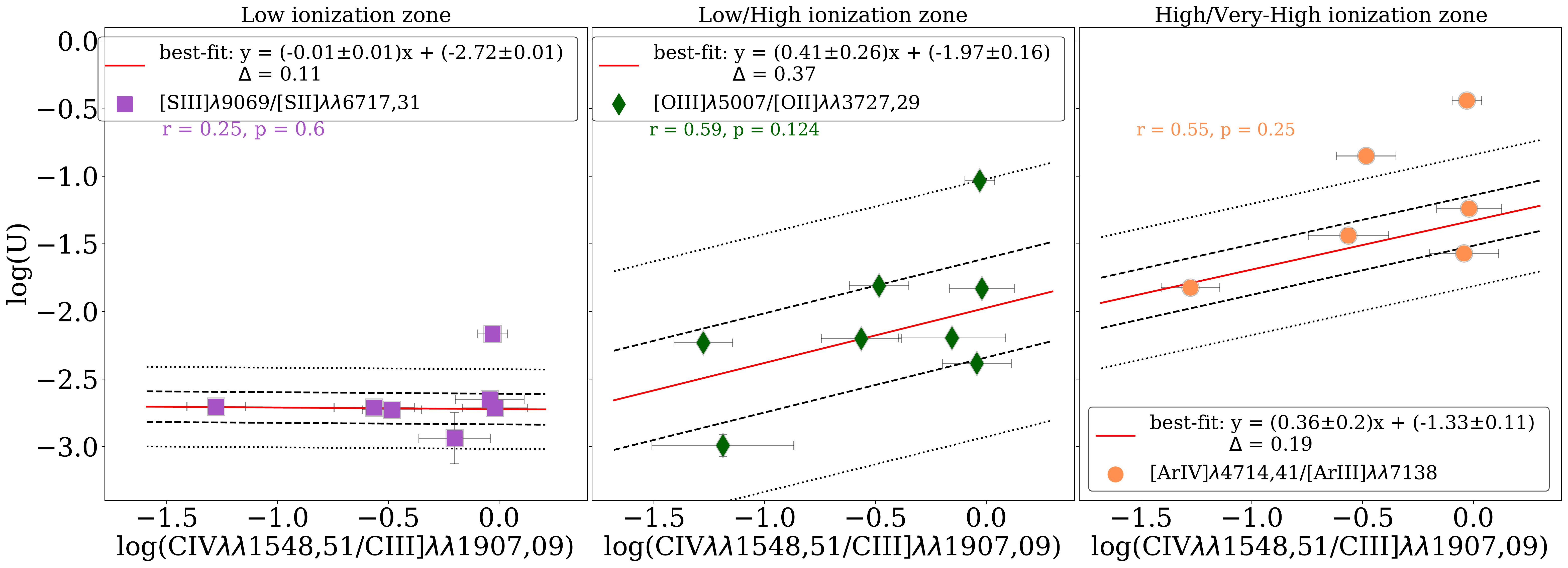}
    \includegraphics[width=0.8\textwidth]{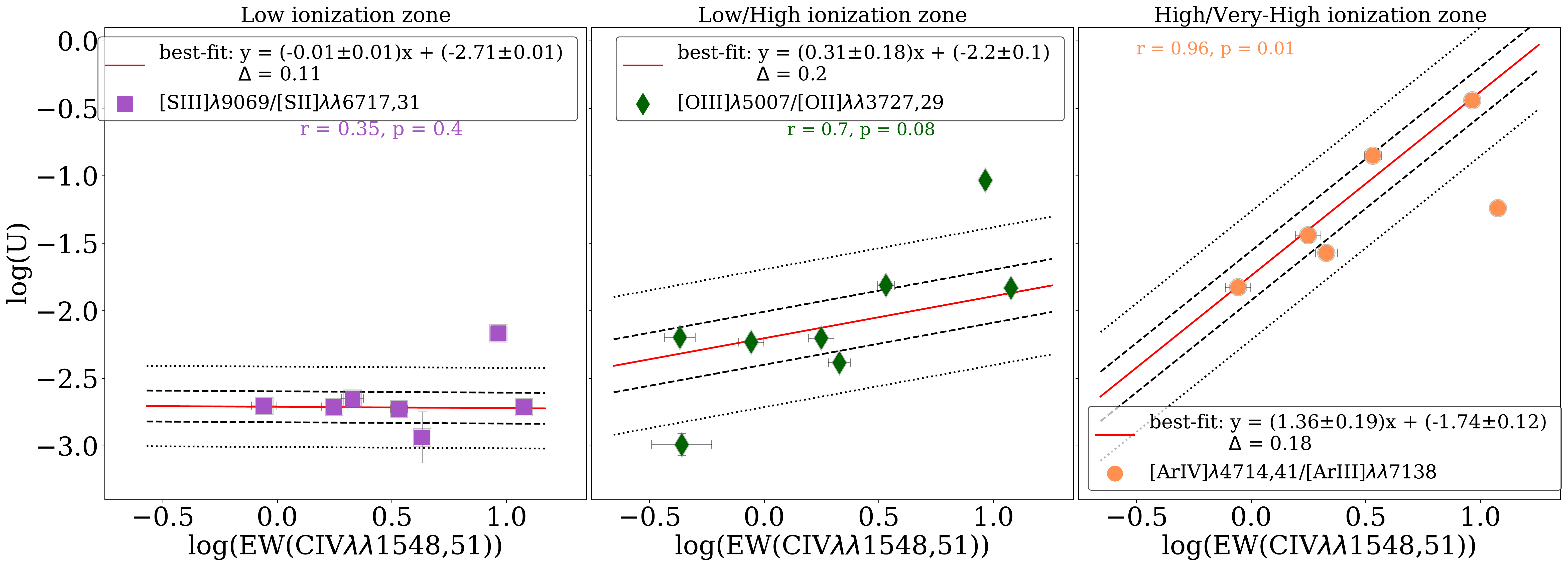}
    
    \includegraphics[width=0.8\textwidth]{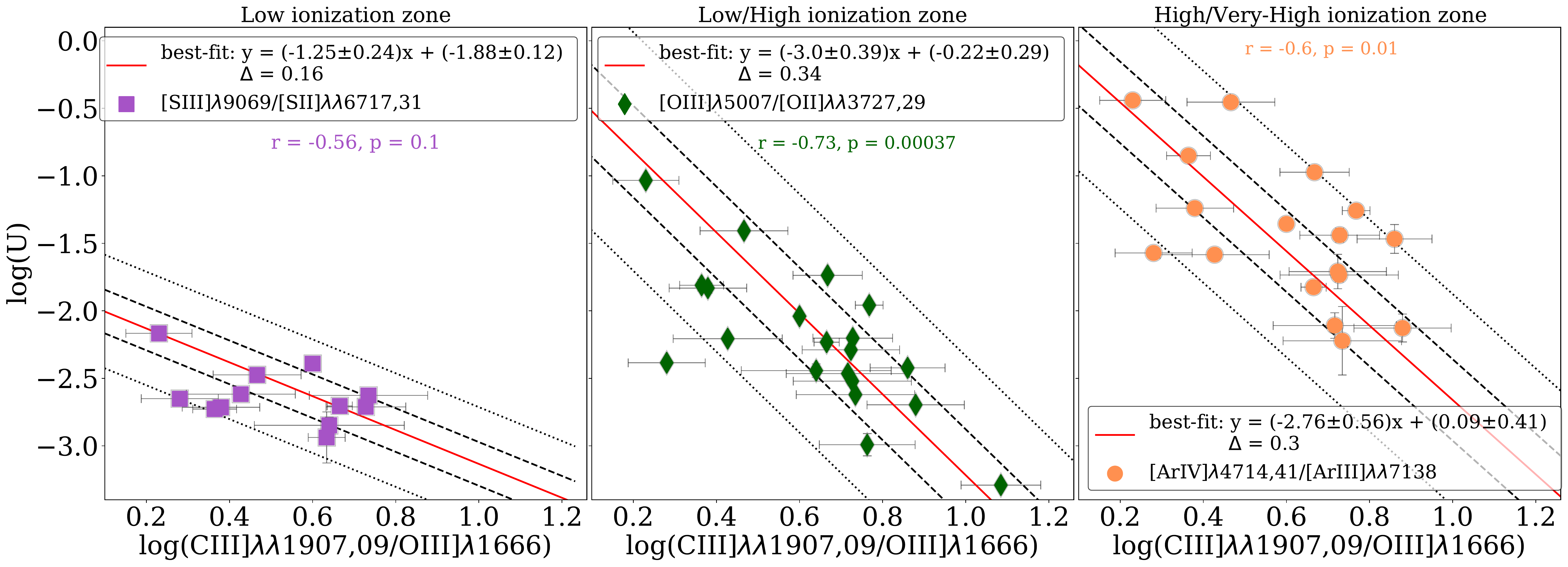}

    \includegraphics[width=0.8\textwidth]{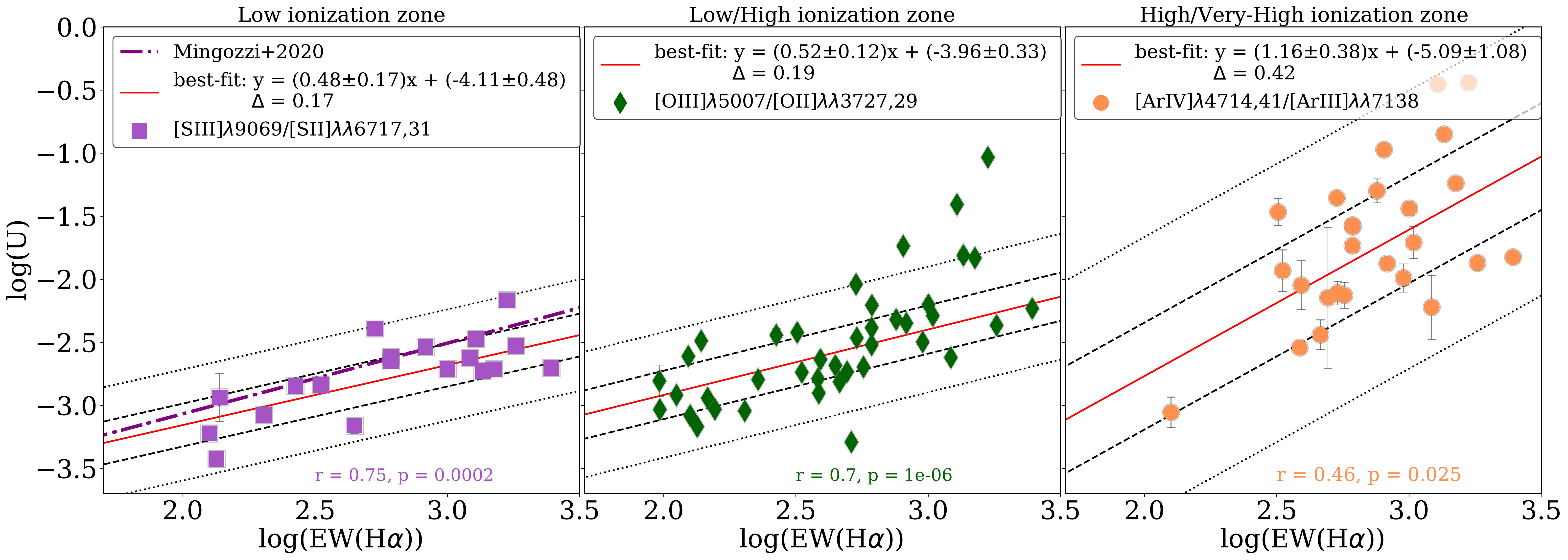}

\end{center}
\caption{log($U$) estimated for low- (purple squares; S3S2 line ratio), intermediate- (green diamonds; O3O2 line ratio) and high-ionization regions (orange circles; Ar4Ar3 line ratio) as a function of \civ~$\lambda\lambda$1548,51/\ciii~$\lambda\lambda$1907,09, log(EW(\civ~$\lambda\lambda$1548,51)), \ciii~$\lambda\lambda$1907,09/\oiiiuv~$\lambda$1666 and log(EW(H$\alpha$)). The uncertainties of the displayed quantities are shown in both axis. The red solid lines represent the best-fit that we calculated between the observables shown and the three log($U$) values, while the dashed and dotted lines show the $1\sigma$ and $2.6\sigma$ scatter around the relation. The best-fit equations as well as the intrinsic scatter are reported in the legend in each panel. The dash-dot purple line in the bottom figure, left panel, is the correlation found by \citet{mingozzi20} between log($U$) and log(EW(H$\alpha$)). This figure shows the most promising UV tracers of log($U$).}
\label{fig:ionu_diag2}
\end{figure*}
Returning to Fig.~\ref{fig:temp-sn-all}, log(\ciii/\oiiiuv), log(\civ/\ciii), log(EW(\oiiiuv)) and log(EW(\civ)) show the most promising relations with log($U$), with the highest Pearson factors for EW(\civ) ($r\sim0.78$).
Overall, many of the shown quantities in Fig.~\ref{fig:temp-sn-all} show a dependence on both 12+log(O/H) and log($U$), as typically happens for UV tracers (e.g., \citealt{maiolino19,ravindranath20}), with \ciii/\oiiiuv\ increasing with increasing 12+log(O/H) and decreasing log($U$), while EW(\oiiiuv) and EW(\ciii) show the opposite behavior.
We further investigate the most promising correlations with ionization parameter in Fig.~\ref{fig:ionu_diag2}, where we separate log($U$) estimates for the low- (purple squares; from S3S2), intermediate- (green diamonds; from O3O2) and high-ionization regions (orange circles; from Ar4Ar3) (as discussed in Sec.~\ref{sec:results_logu}).
The first three rows with three panels each show the relations of log(\civ/\ciii)\ (C4C3), log(EW(\civ)) and log(\ciii/\oiiiuv), with the S3S2, O3O2 and Ar4Ar3 log($U$), respectively. The solid red lines show the best-fit to each of the UV-based log($U$) measurements from CLASSY, while the dashed and dotted lines represent the $1\sigma$ and $2.6\sigma$ around the best-fit.
The equations for the best-fit correlations and their intrinsic scatter are reported in the legend of each panel, alongside the corresponding Pearson factors. In general, the UV emission line properties are more tightly correlated to O3O2 and Ar4Ar3 log($U$), and only slightly or mostly not related to S3S2 log($U$). This could be due to the fact that the three log($U$) diagnostics are tracing different ionizing regions (see Sec.~\ref{sec:methods-logu}), and the considered UV lines are coming from intermediate or high-ionization regions \citep{berg21a}.
Overall, the most promising UV-based diagnostics for log($U$) are the following. \\
Using log(\civ/\ciii):
\begin{equation}\label{eq:loguc3c4-1}
\begin{aligned}
    \log(U(O3O2)) & = (0.41\pm0.26) \times \log(C4C3)\\
    & -(1.97\pm0.16) \\
    \end{aligned}
\end{equation}
\begin{equation}\label{eq:loguc3c4-2}
\begin{aligned}
    \log(U(Ar4Ar3)) & = (0.36\pm0.2) \times \log(C4C3)\\
    & -(1.33\pm0.11) \\
    \end{aligned}
\end{equation}

Using log(EW(\civ)):

\begin{equation}\label{eq:loguciv-1}
\begin{aligned}
    \log(U(O3O2)) & = (0.31\pm0.18) \times log(EW(CIV1548,51))\\
    & -(2.2\pm0.1) \\
    \end{aligned}
\end{equation}
\begin{equation}\label{eq:uciv-2}
\begin{aligned}
    \log(U(Ar4Ar3)) & = (1.36\pm0.19) \times log(EW(CIV1548,51))\\
    & -(1.74\pm0.0.12) \\
    \end{aligned}
\end{equation}

Using log(\ciii/\oiiiuv):
\begin{equation}\label{eq:o3c3-1}
\begin{aligned}
    \log(U(S3S2)) & = (-1.25\pm0.24) \times log(CIII]1907,9/OIII]1666)\\
    & -(2.47\pm0.05) \\
    \end{aligned}
\end{equation}
\begin{equation}\label{eq:logo3c3-2}
\begin{aligned}
    \log(U(O3O2)) & = (-2.74\pm0.44) \times log(CIII]1907,9/OIII]1666)\\
    & -(0.22\pm0.29) \\
    \end{aligned}
\end{equation}
\begin{equation}\label{eq:o3c3-3}
\begin{aligned}
    \log(U(Ar4Ar3)) & = (-2.83\pm0.5) \times log(CIII]1907,9/OIII]1666)\\
    & +(0.09\pm0.41) \\
    \end{aligned}
\end{equation}
In practice, Eq.~\ref{eq:loguc3c4-1}-\ref{eq:o3c3-3} in combination with Eq.~\ref{eq:ciiioiii-met} and \ref{eq:ewciii-met} for 12+log(O/H) (see Fig.~\ref{fig:temp-sn}) could guide photoionization models to explore the relationship between log($U$) and 12+log(O/H), with the hope of breaking its degeneracy. 

\subsubsection{A Note regarding  \civ}\label{sec:discussion-civciii}
As a ratio of two UV emission lines from consecutive ionization states, C4C3, could trace the ionization structure of the nebula \citep{jaskot16}, similarly to S3S2, O3O2 and Ar4Ar3 in the optical. It is thought to be also a potential indicator of Lyman continuum (LyC) escape \citep{jaskot13,nakajima14,schaerer22}. 
Several authors have indeed proposed the use of C4C3 as a ionization parameter tracer (e.g., \citealt{perez-montero17,kewley19,schaerer22}). The first row of Fig.~\ref{fig:ionu_diag2} confirms a correlation between C4C3 and O3O2 and ArAr3 log($U$), with a large intrinsic scatter, reported in the legend. 
Also EW(\civ) correlates  well with log($U$) derived from both O3O2 and Ar4Ar3, as shown in the second row of Fig.~\ref{fig:ionu_diag2}. 

We stress that the fitting method that we are using performs a clipping of the points out from $2.6\sigma$ of the relation (black dotted lines), which means that the best-fit equations that we find are not able to take into account all the available galaxies. 
This limitation is probably related to the low statistics, given that it is rare to observe pure nebular \civ\ emission, while the remaining galaxies showing pure absorption or P-Cygni profiles. 
Indeed, nebular \civ\ emission is rarely observed in the literature, and comes uniquely from studies of systems with 12+log(O/H)$\lesssim8$, indicative of a rapid hardening of the ionizing spectrum at low metallities (see also \citealt{senchyna19a,berg21a,senchyna22a,schaerer22}).
This holds also for the galaxies of the CLASSY sample, where \civ\ nebular emission is coming only from objects with 12+log(O/H)$\lesssim8$ and generally high ionization parameter log($U$)~$\gtrsim -2.5$, as highlighted in Fig.~\ref{fig:ionu_diag2}. 

\citet{senchyna19a} suggested also a correlation between EW(\civ) and EW(\hb) (with galaxies characterized by \civ\ nebular emission having EW(\hb)~$\geq 100$~\AA). In turn, the equivalent width of the strongest hydrogen recombination lines measures the ratio of the young, ionising stars over the old, non-ionising population (e.g., \citealt{leitherer05}), and thus correlates with the sSFR and the degree to which the youngest stars dominate the optical \citep{kewley15,kaasinen18,mingozzi20}.
This partly explains the correlation between log(EW(\civ)) and log($U$).
Indeed, \citet{mingozzi20} found a tight correlation between log(EW(\ha)) and log($U$) (derived from S3S2), explained in terms of the good correlation between log($U$) and the age of the stellar population \citep{pellegrini20}, traced by EW(\ha) (a similar correlation holds for EW(\hb); e.g., \citealt{senchyna19b}). 
A comparable relation holds also for the CLASSY galaxies, as shown in the bottom panel of Fig.~\ref{fig:ionu_diag2}, that displays the optical log($U$) as a function of log(EW(H$\alpha$)).

It should be noted, however, that diagnostics using nebular \civ\ emission may be difficult to employ due to strong stellar \civ\ P-Cygni profiles, as highlighted also in \citet{jaskot16}.
As commented in Sec.~\ref{sec:data-analysis}, CLASSY data benefits from a combination of relatively high-S/N and high-spectral resolution which allowed us to separate the stellar and the nebular components of \civ\ emission via UV stellar continuum modeling, but this may not always be the case - especially for high-$z$ observations of faint galaxies. Another caveat to take into account for C4C3 is the different nature of \ciii\ and \civ\ emission, while both are collisional nebular lines, \civ\ is also a resonant line affected by radiation transfer effects. 
This means that \civ\ absorption along the line-of-sight, which depends on the gas kinematics and column density, could contribute in reducing the observed \civ\ emission (e.g., \citealt{steidel16}). 
We will investigate in detail any differences in the kinematics of these lines and the causes that contribute to their emission in \citetalias{mingozzi22} (in prep.), which is dedicated to the ionization mechanisms behind the UV emission lines presented here.

\section{Conclusions} \label{sec:conclusions}
In this work we investigated UV and optical diagnostics of ISM properties by exploiting the CLASSY survey presented by \citet{berg22,james22}, which represents the first high-quality, high-resolution and broad wavelength range ($\sim1200-2000$ \AA) FUV spectral database of 45 nearby ($0.002<z<0.182$) star-forming galaxies. 
Specifically we focused on the main UV emission lines visible in the COS spectra apart from \lya\ (i.e., \niv~$\lambda\lambda$1483,87, \civ~$\lambda\lambda$1548,51, \heii~$\lambda$1640, \oiiiuv$\lambda\lambda$1661,6, \siiii~$\lambda\lambda$1883,92, \ciii~$\lambda\lambda$1907,9), and combine them with emission lines comprised between \oii$\lambda\lambda$3727,9 and \siii$\lambda$9069 from optical spectroscopy of the same pointing. Our aim was to provide a \emph{ UV toolkit} of ISM diagnostics, i.e., a set of equations to diagnose $E(B-V)$, $n_e$, $T_e$, 12+log(O/H) and log($U$), that use only UV emission lines. By carefully assessing the stratified ionization structure of our targets, we accurately calculated the physical and chemical properties using diagnostic line ratios specific to each ionization zone (constituting a set of `direct' diagnostics), then subsequently comparing and calibrating each property with the well-known optical diagnostics to derive a series of `indirect' diagnostic UV-based equations.
In the following we summarize our main findings.

\begin{itemize}
\item  UV density diagnostics (\fciii~$\lambda$1907/\ciii~$\lambda$1909, \fsiiii~$\lambda$1883/\siiii~$\lambda$1892, \niv~$\lambda$1483/\niv$~\lambda$1487) give log($n_e$) $\sim 2$~dex higher than the optical counterparts \sii~$\lambda$6717/\sii~$\lambda$6731 and \oii~$\lambda$3729/$\lambda$3727 and $\sim 1$~dex than \cliii~$\lambda$5518/$\lambda$5538 and \ariv~$\lambda$4714/$\lambda$4741, as summarized in Fig.~\ref{fig:den_diag} and discussed in Sec.~\ref{sec:discussion-ne}. UV $n_e$ are consistent with $n_e$(\feiii~$\lambda$4701/$\lambda$4659), for which we can derive only upper limits.  
The UV-to-optical electron density offset enables us to derive a low-ionization zone $n_e$ (e.g., $n_e$(\sii)) for high-$z$ targets in case optical wavelengths are not available.


\item The hybrid UV/optical temperature diagnostic \oiiiuv~$\lambda$1666/$\lambda$5007 is not dominated by major systematic uncertainties, as shown in Fig.~\ref{fig:temp_diag}.  As such, we find this auroral line ratio to be as reliable a tracer of electron temperature as the most generally used optical ratio, \oiii~$\lambda$4363/$\lambda$5007. We also find that the \oiiiuv~$\lambda$1666 flux is on average $\sim$0.3~dex brighter than the \oiii~$\lambda$4363, and thus it could be more easily observed in high-$z$ galaxies than the weaker and less detectable \oiii~$\lambda$4363 emission.

\item On average, \oiiiuv~$\lambda$1666/$\lambda$5007 gives temperatures higher than \oiii~$\lambda$4363/$\lambda$5007 by $\sim$ 1000$~K$. Despite this offset, we discuss in Sec.~\ref{sec:discussion-Te} that $T_e$(\oiiiuv~$\lambda$1666/$\lambda$5007) allows us to estimate direct-method gas-phase metallicities with only a scatter of $\pm0.3$~dex with respect to the optically-derived 12+log(O/H), as shown in Fig.~\ref{fig:tempmet_diag}. 
We also investigated how the optical 12+log(O/H) compare with metallicities derived from UV-based He2-O3C3 and Si3-O3C3 metallicity calibrations by \citetalias{byler20} (see Fig.~\ref{fig:bylermet}). We find an offset correlating with the optical 12+log(O/H), and we provide expressions to re-calibrate the \citetalias{byler20} equations (i.e., Eq.~\ref{eq:metcorr1} and \ref{eq:metcorr2}).

\item We derive ionization parameters in the range $-3.5<\log(U) <0.$ using the optical diagnostics S3S2 (\siii~$\lambda\lambda$9069,9532/\sii~$\lambda\lambda$6717,31), O3O2 (\oiii~$\lambda$5007/\oii~$\lambda\lambda$3727,29) and Ar4Ar3 (\ariv~$\lambda\lambda$4714,71/\ariii~$\lambda$7138), as shown in Fig.~\ref{fig:ionu_diag}.
Specifically, we find that log($U$(S3S2))~$<$~log($U$(O3O2))~$<$~log($U$(Ar4Ar43)), in line with the presence of a clear ionization structure in the nebular environments of our targets, as described in Sec.~\ref{sec:methods} and in \citet{berg21a}.

\item In Fig.~\ref{fig:ebv-gas-stars}, we investigated how the gas attenuation, $E(B-V)$ (obtained with the Balmer decrement and the \citealt{cardelli89} attenuation law; see Sec.~\ref{sec:method-dust}), compares with the stellar attenuation, $E(B-V)_{UV}$ (obtained from the $\beta$ slope of stellar continuum fitting and the \citealt{reddy16} attenuation law; see Sec.~\ref{sec:data-analysis-stellar-cont-uv}). 
Our conclusion is that the gas and stellar attenuation factors are approximately the same at $ \log(sSFR) \gtrsim -8$~yr$^{-1}$, whereas at lower sSFRs the values tend towards the \citetalias{calzetti97b} relation, $E(B-V) \sim 2.27 \times E(B-V)_{UV}$. 
This relation subsequently allows us to use $E(B-V)_{UV}$ to derive $E(B-V)$ in galaxies if optical wavelengths are not accessible.

\item Possible UV diagnostics for 12+log(O/H) and log($U$) were explored in Sec.~\ref{sec:discussion-metlogu} and Fig.~\ref{fig:temp-sn-all}, where we also confirm that UV lines are usually sensitive to both parameters. We proposed \ciii~$\lambda\lambda$1907,9/\oiiiuv~$\lambda$1666 and EW(\ciii~$\lambda\lambda$1907,9) as 12+log(O/H) diagnostics (see Eq.~\ref{eq:ciiioiii-met} and \ref{eq:ewciii-met}, displayed in Fig.~\ref{fig:temp-sn}).
We also highlight that the presence of nebular \civ~$\lambda\lambda$1548,51 can be used as a metallicity indicator, since it is revealed uniquely in galaxies with 12+loG(O/H)~$\lesssim8$, in agreement with the few previous works that studied this rare spectral feature.
In Fig.~\ref{fig:ionu_diag2}, we include the most promising log($U$) diagnostics, providing expressions to derive the ionization parameter from \civ~$\lambda\lambda$1548,51/\ciii~$\lambda\lambda$1907,9, EW(\civ~$\lambda\lambda$1548,51) and \ciii~$\lambda\lambda$1907,9/\oiiiuv~$\lambda$1666 (see Eq.~\ref{eq:loguc3c4-1}-\ref{eq:o3c3-3}). Finally, in Sec.~\ref{sec:discussion-civciii}, we discuss about the limitations in using nebular \civ\ diagnostics.

\end{itemize}

In summary, we present an empirically calibrated \emph{ UV toolkit} of ISM diagnostics that will be fundamental in characterizing and interpreting the spectroscopic observations of high-$z$ systems in the upcoming JWST and ELT era, for which the optical wavelength range will not be available. 
Specifically, NIRISS and NIRSpec JWST instruments will provide spectra in the wavelength ranges of $0.8-2.2$ and $0.6-5.3$~$\mu$m, respectively, covering (in full or in part) the suite of optical emission lines (from \oii$\lambda\lambda$3727 to \ha) together with the \ciii$\lambda\lambda1907,9$ at $5<z<6$. At higher $z$, optical lines are progressively shifted out of the observed range, while at $z\geq10$ only UV lines from \lya\ to \ciii\ will be accessible. The low-resolution ($R\sim150$) spectroscopy of JWST/NIRISS is not enough to resolve the \ciii\ doublet, but is expected to detect the UV emission lines down to $\sim10^{-18}$~erg/s/cm$^{-2}$ \citep{treu22,jdox}. On the other hand, the higher spectral resolution of JWST/NIRSpec (up to $R\sim2700$) will provide a resolved \ciii\ doublet \citep{jdox}.
As demonstrated here, the CLASSY survey has provided the ideal ATLAS to create a powerful UV toolkit to explore these high-$z$ spectra in terms of specific star-formation rate, direct gas-phase metallicity, ionization level, reddening, and nebular density, since it represents a high-quality UV and optical database for local analogs to reionization-era systems.
To complete our toolkit, in the next paper on UV-based diagnostics, we will focus on the ionized gas kinematics, both of UV and optical emission, and on the diagnostics to inspect their main ionization mechanisms, to also understand the origin of the main UV emission lines taken into account.

\clearpage
We thank the anonymous referee for their comments and advice, which help improve the quality of the paper. MM, DAB and XX are grateful for the support for this program, HST-GO-15840, that was provided by 
NASA through a grant from the Space Telescope Science Institute, 
which is operated by the Associations of Universities for Research in Astronomy, 
Incorporated, under NASA contract NAS5-26555. 
Also, MM is grateful to Carlo Cannarozzo for inspiring conversations and advice.
BLJ, SH and NK are thankful for support from the European Space Agency (ESA).
AF acknowledges the support from grant PRIN MIUR2017-20173ML3WW\_001.
AW acknowledges the support of UNAM via grant agreement PAPIIT no. IN106922. RA acknowledges support from ANID Fondecyt Regular 1202007.

The CLASSY collaboration extends special gratitude to the Lorentz Center for useful discussions 
during the "Characterizing Galaxies with Spectroscopy with a view for JWST" 2017 workshop that led 
to the formation of the CLASSY collaboration and survey.
The CLASSY collaboration thanks the COS team for all their assistance and advice in the
reduction of the COS data.

Funding for SDSS-III has been provided by the Alfred P. Sloan Foundation, the Participating Institutions, the National Science Foundation, and the U.S. Department of Energy Office of Science. The SDSS-III web site is \href{http://www.sdss3.org/}{http://www.sdss3.org/}.
SDSS-III is managed by the Astrophysical Research Consortium for the Participating Institutions of the SDSS-III Collaboration including the University of Arizona, the Brazilian Participation Group, Brookhaven National Laboratory, Carnegie Mellon University, University of Florida, the French Participation Group, the German Participation Group, Harvard University, the Instituto de Astrofisica de Canarias, the Michigan State/Notre Dame/JINA Participation Group, Johns Hopkins University, Lawrence Berkeley National Laboratory, Max Planck Institute for Astrophysics, Max Planck Institute for Extraterrestrial Physics, New Mexico State University, New York University, Ohio State University, Pennsylvania State University, University of Portsmouth, Princeton University, the Spanish Participation Group, University of Tokyo, University of Utah, Vanderbilt University, University of Virginia, University of Washington, and Yale University.

This work also uses the services of the ESO Science Archive Facility,
observations collected at the European Southern Observatory under 
ESO programmes 096.B-0690, 0103.B-0531, 0103.D-0705, and 0104.D-0503, and observations obtained with the Large Binocular Telescope (LBT).
The LBT is an international collaboration among institutions in the
United States, Italy and Germany. LBT Corporation partners are: The
University of Arizona on behalf of the Arizona Board of Regents;
Istituto Nazionale di Astrofisica, Italy; LBT Beteiligungsgesellschaft,
Germany, representing the Max-Planck Society, The Leibniz Institute for
Astrophysics Potsdam, and Heidelberg University; The Ohio State
University, University of Notre Dame, University of
Minnesota, and University of Virginia.


This research has made use of the HSLA database, developed and maintained at STScI, Baltimore, USA.

\facilities{HST (COS), LBT (MODS), APO (SDSS), KECK (ESI), VLT (MUSE, VIMOS)}
\software{
astropy (The Astropy Collaboration 2013, 2018)
BEAGLE (Chevallard \& Charlot 2006), 
CalCOS (STScI),
dustmaps (Green 2018),
jupyter (Kluyver 2016),
LINMIX (Kelly 2007) 
MPFIT (Markwardt 2009),
MODS reduction Pipeline,
Photutils (Bradley 2021),
PYNEB (Luridiana 2012; 2015),
python,
pysynphot (STScI Development Team),
RASCAS (Michel-Dansac 2020),
SALT (Scarlata \& Panagia 2015), 
STARLIGHT (Fernandes 2005), 
TLAC (Gronke \& Dijkstra 2014),
XIDL}

\typeout{} 
\bibliography{CLASSY}

\clearpage


\newpage
\appendix

\section{UV-optical flux offset}\label{app:A}
Before comparing COS and optical data, it was necessary to check and correct optical spectra for potential UV-optical flux offsets. Below we explain in detail the procedure we carried out to check if there is alignment between the continuum flux of the two wavelength regimes.
Such an offset is expected since the UV and optical spectra have been obtained via different instruments and different apertures. 
Specifically, we have to take into account: (i) the different aperture size of COS (2.5"), SDSS (3"), MMT (10"$\times$1" slit) and LBT (2.5"$\times$1" slit), while the MUSE, VIMOS and KCWI data were extracted using an aperture of 2.5", identical to COS; (ii) a possible different orientation of the COS and SDSS apertures, and MMT and LBT slits, for extended sources and/or targets with multi-component light profiles throughout the aperture; (iii) COS vignetting of non-compact objects; (iv) flux calibration issues specific to the instrument.
Having said that, we stress that this is an issue \emph{only} for mixed UV-optical diagnostics, such as the \oiiiuv~$\lambda$1666/$\lambda$5007 line ratio, that can be used to measure the electron temperature of the gas (see Sec.~\ref{sec:methods-temp}, \ref{sec:results_temp} and \ref{sec:discussion-Te}).
Indeed, \citetalias{arellano-cordova22} analyzed the optical data of 12 CLASSY galaxies taking into account SDSS, VLT/MUSE, and, for a subsample, LBT/MODS and Keck/ESI longslit spectra, to assess the impact of using different aperture combinations on the determination of the ISM physical conditions (i.e., $n_e$, $T_e$, $E(B-V)$, log($U$) and 12+log(O/H)). Overall, \citetalias{arellano-cordova22} found that these measurements remained roughly constant with aperture size, indicating that the optical gas conditions are relatively uniform for this sample.

\begin{figure*}
\begin{center}
    \includegraphics[width=0.62\textwidth]{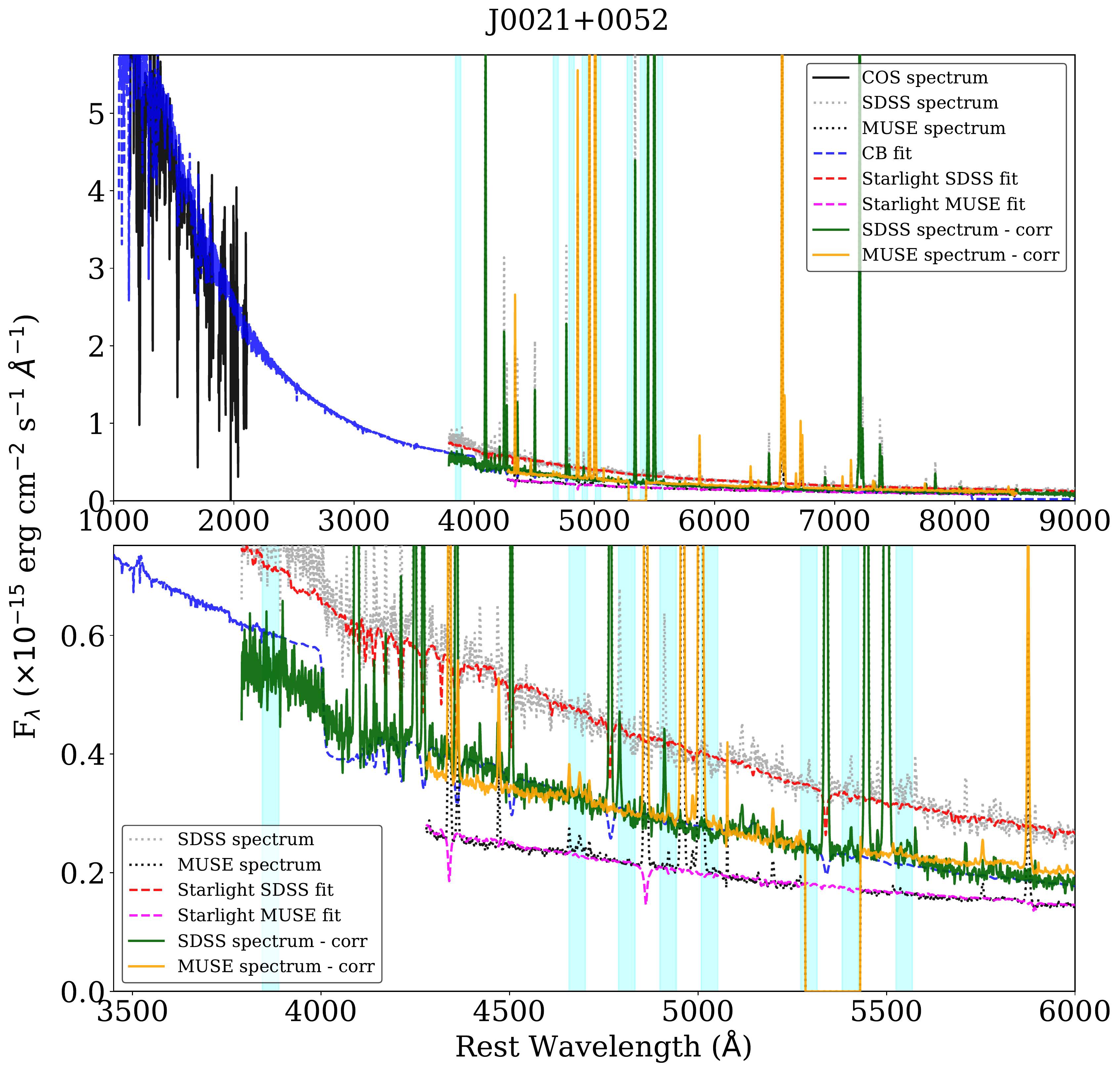}
    \includegraphics[width=0.62\textwidth]{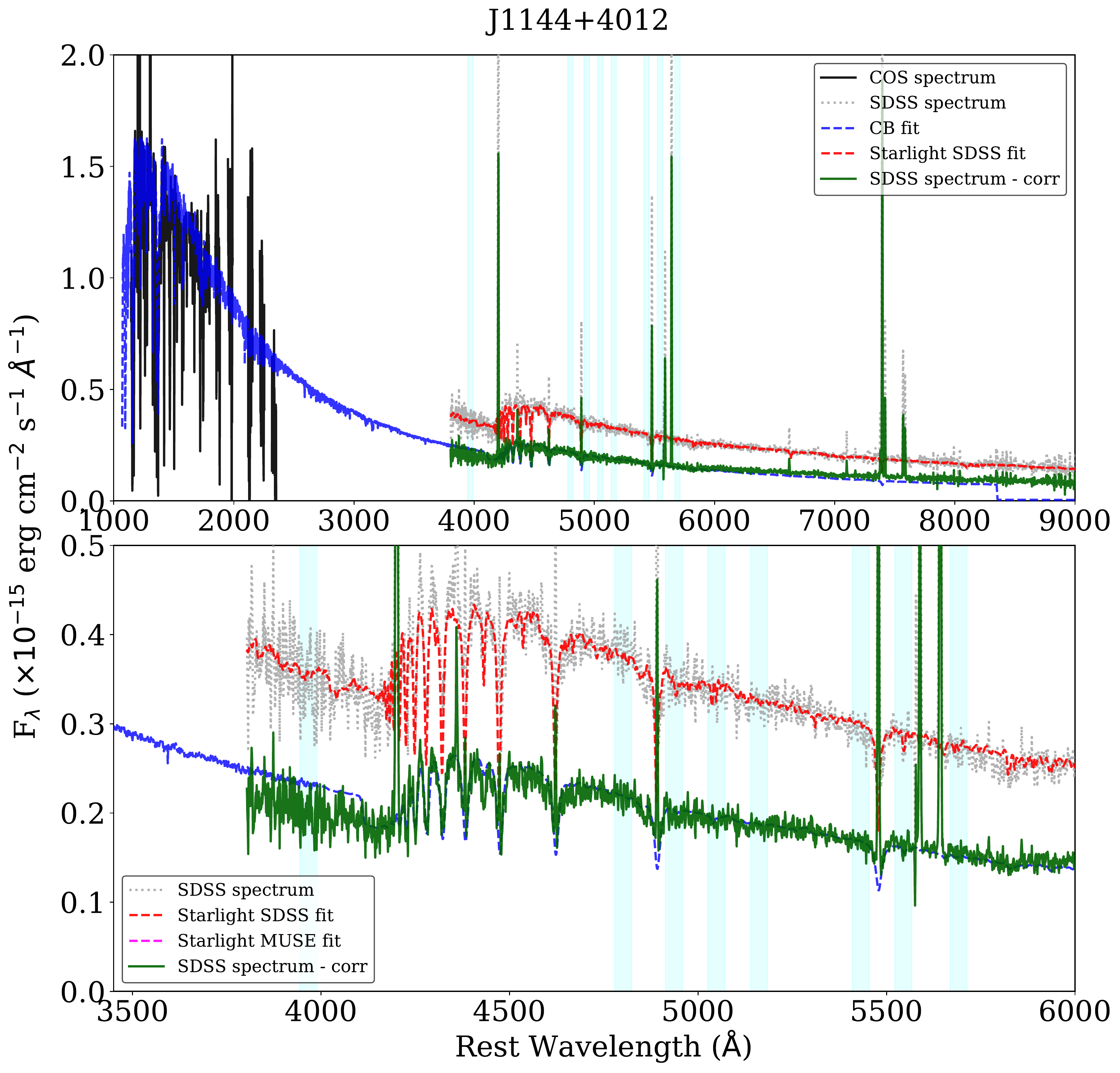}
\end{center}
\caption{Upper panels: COS, SDSS and MUSE spectra of the galaxy J0021+0052. The COS spectrum is shown with a black solid line, while the original SDSS and MUSE optical spectra with a dotted gray and black line, respectively. The UV-optical flux offset is estimated comparing the UV-extended stellar continuum fitting (dashed blue line) and optical stellar continuum fitting (dashed red line for SDSS and magenta for MUSE) in several featureless wavelength ranges of 40~\AA\  between 4000 and 5600 \AA, highlighted in cyan. The UV-optical flux offset corrected spectra are shown from the green and orange solid lines, for SDSS and MUSE, respectively. The bottom panel shows a zoom on the optical region of the spectra. Lower panels: same for the galaxy J1144+4012.}
\label{fig:optuv_offset}
\end{figure*}

To explore the UV-optical flux offset, we extended the UV stellar continuum best-fit performed in the range 1200-2000~\AA\ with the method from \citetalias{senchyna22} and C\&B models as described in Sec.~\ref{sec:data-analysis-stellar-cont-uv}, to optical wavelengths up to $\sim 9000$~\AA, taking into account a SMC attenuation law \citep{gordon03}. Here we stress that the choice of the attenuation law is not affecting the results in the optical wavelength range, since the different known attenuation laws (e.g., \citealt{cardelli89}, \citealt{calzetti00}, \citealt{gordon03}) are all superimposed at $\lambda \gtrsim 3800$~\AA\ (e.g., see Fig.~3 in \citealt{shivaei20}). This is not the case in the UV wavelength range, where we adopted the most suitable \citet{reddy16} attenuation law, as discussed in Sec.~\ref{sec:data-analysis-stellar-cont-uv} and Sec.~\ref{sec:method-dust}. 
The performed extrapolation of the UV stellar continuum best-fit towards the optical wavelengths allows us to perform a direct comparison with the optical spectrum fitted with \texttt{Starlight} as described in Sec.~\ref{sec:data-analysis-stellar-cont-opt}, by providing an overlapping wavelength window.
This is illustrated in Fig.~\ref{fig:optuv_offset}, which shows the observed COS (black solid line) and optical SDSS and MUSE (gray and black dotted line, respectively) spectra for the CLASSY galaxies J0021+0052 and J1144+4012.  Overlaid are the extended-UV stellar-continuum best-fit (dashed blue line) and the optical stellar-continuum best-fit (dashed red and magenta line for SDSS and MUSE spectra, respectively). 
The UV-optical flux offset is then estimated by dividing the extended UV stellar continuum fit by the optical stellar continuum best-fit in eight featureless wavelength windows (40~\AA\ each in width) between 3700 and 5600~\AA\ (highlighted in cyan), avoiding the region of the 4000~\AA\ break, and calculating their median value and using the standard deviation as corresponding error.
The final optical spectra, corrected for the UV-optical flux offset, are shown by the green and orange solid lines for SDSS and MUSE, respectively.

\begin{figure}
\begin{center}
    \includegraphics[width=0.47\textwidth]{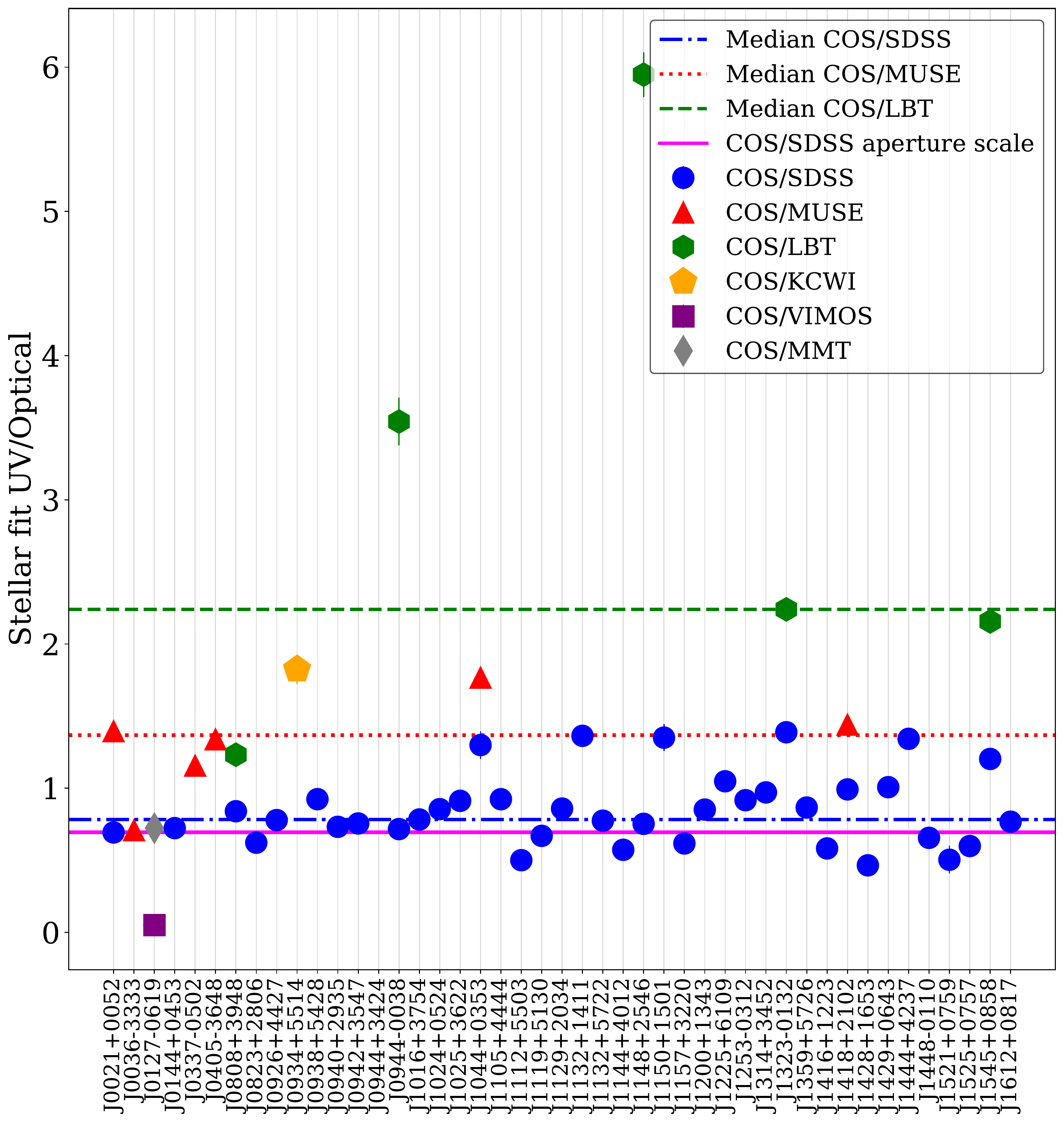}
    \includegraphics[width=0.5\textwidth]{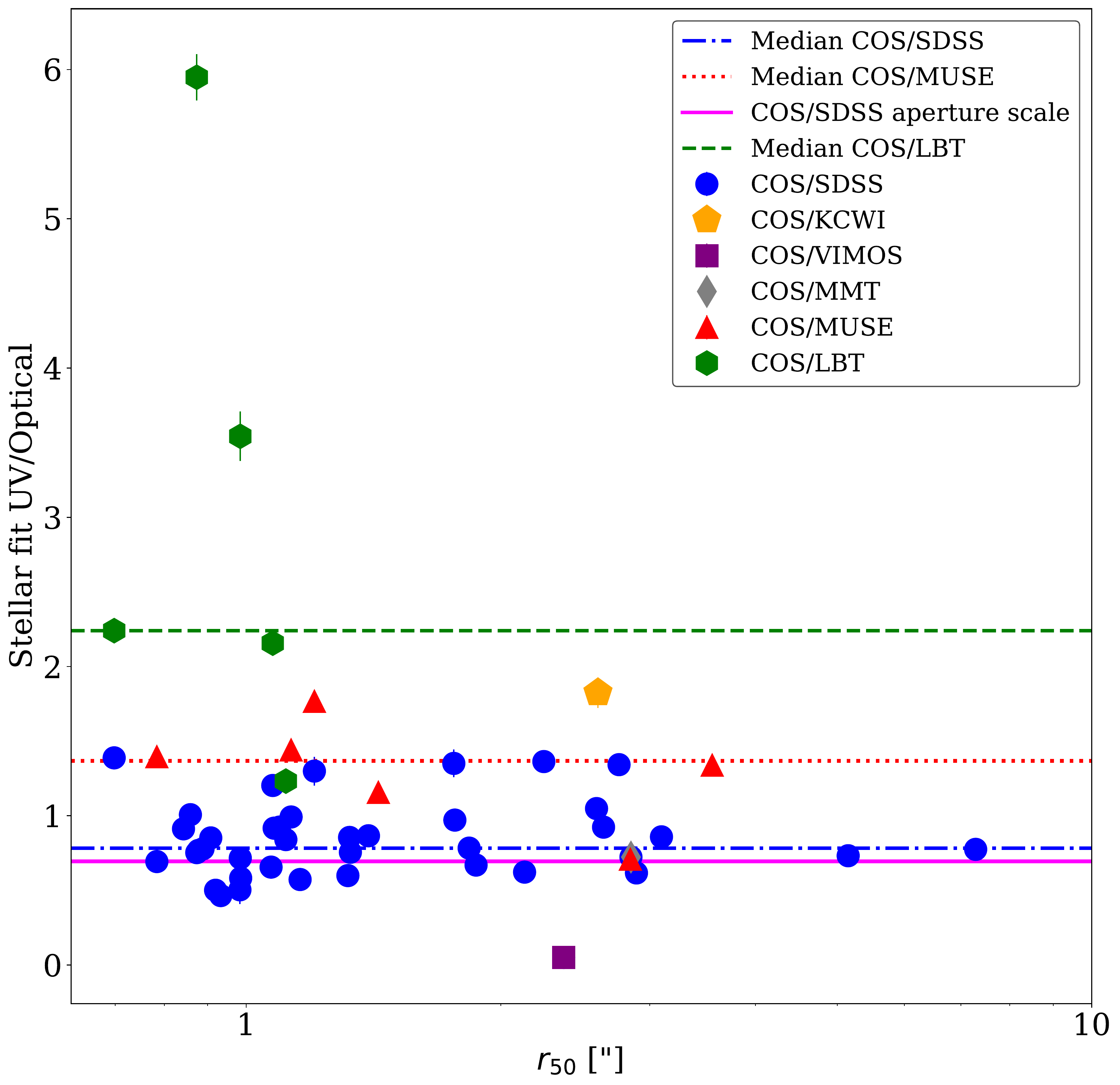}
\end{center}
\caption{Left panel: UV-optical flux offset for each CLASSY galaxy, obtained from the comparison of extended-UV stellar continuum fitting and optical stellar continuum fitting of SDSS (blue circles), MUSE (red triangles), KCWI (gold pentagon), VIMOS (purple square), LBT (green hexagons) and MMT (gray diamond) data. Right panel: UV-optical flux offset as a function of the average half light radius ($r_{50}$) taken from \citetalias{berg22} and calculated from PanSTARRS imaging.
The solid magenta horizontal lines are centred at the ratio of COS/SDSS aperture (i.e., (2.5"/3")$^2$), while the dash-dot blue, dotted red and dashed green horizontal lines show the median UV-optical flux offsets for the SDSS, MUSE and LBT spectra, respectively. The uncertainties are shown for all the displayed quantities.}
\label{fig:optuv_offset2}
\end{figure}
The left panel of Fig.~\ref{fig:optuv_offset2} shows the UV-optical flux offsets measured throughout the CLASSY sample, using different symbols to indicate the SDSS, MUSE, LBT, KCWI, VIMOS and MMT observations, as reported in the legend. 
An important aspect to explore is that there is no dependence of the UV-optical flux offset we measure with the galaxy extent, that for some targets can be larger than that covered by the COS and optical apertures that we are taking into account. 
In these cases, COS is not measuring the entire flux for these objects. For instance, in the hypothesis that the sources were filling uniformly the COS and SDSS apertures centered in the same position, then the optical fluxes should be scaled by the ratio of their apertures $(2.5"/3.0")^2$ (i.e., dashed magenta line). The dash-dot blue blue line in Fig.~\ref{fig:optuv_offset2} represents the median value of the SDSS UV-optical offset, taking into account the 24 galaxies with an average half light radius ($r_{50}$) larger than the optical extent, calculated by plotting the fraction of the optical flux from PanSTARRS imaging (taken from Table~6 in \citetalias{berg22} and reported in Tab.~\ref{tab:classyprop}). 
This value is close to the magenta line, but offsets obtained for the SDSS galaxies (blue dots) are spread in the range $\sim 0.5-1.3$, demonstrating that there are other factors at play. 
Furthermore, in the right panel of Fig.~\ref{fig:optuv_offset2} we show the UV-optical offset as a function of $r_{50}$, and reassuringly we notice that there is no correlation between the two (Pearson factor of $\sim 0.02$, with pvalue of $\sim0.90$).
The dotted red and dashed green horizontal lines in both the left and right panels of Fig.~\ref{fig:optuv_offset2} show the median UV-optical flux offsets for the MUSE and LBT spectra, respectively.
In particular, the COS and MUSE-KCWI-VIMOS data are aperture-matched. Indeed, these optical spectra were obtained by selecting the same aperture and position of COS (see Sec.~\ref{sec:sample}), and thus the difference in aperture size can be completely ruled out as the origin of the flux offset. 
Nevertheless, the flux offset calculated can be far different than 1, down to $\sim0.05$ for the VIMOS data and up to $\sim1.95$ for the KCWI, and with values closer to $\sim1$ for MUSE.
In these cases, apart from flux calibration issues, there are other effects within the COS aperture that may cause a flux offset here, such as vignetting (see \citetalias{james22} for details).

We stress that, as already reported in Sec.~\ref{sec:data-analysis-stellar-cont}, the C\&B models include the nebular continuum.
Specifically, the nebular continuum emits significantly in the UV blue-ward of the Balmer break, accounting for $\sim5-10$\% of the total flux in the range 900--1800~\AA\ and up to $\sim20$\% in the range 1800--4000~\AA\ \citep{byler17}, increasing the overall continuum level. Also, its impact increases if the gas is characterized by high ionization parameter and low metallicity, typical of CLASSY objects (see e.g., Fig.~12 of \citealt{byler17}).
Therefore, it is fundamental to take its contribution into account for our comparison. 
A caveat of our approach is that in the UV stellar-continuum best-fit, we fixed the ionization parameter to log($U$)~$=-2.5$, which sets the overall intensity of the nebular continuum spectrum, since recombination emission depends on the number of incident ionizing photons \citep{byler17}.
However, we tested to what extent the variation of the stellar population ionization parameter in the range [−3; −1] could influence the nebular continuum shape to estimate the UV-optical offset, finding minimal variations ($< 0.05$~dex compared to the case with log($U$)~$= - 2.5$) in the 4000--6000~\AA\ wavelength range considered for the comparison.
Overall, we conclude that the observed UV-optical flux offsets are primarily due to systematic effects of the instruments such as the COS aperture vignetting and/or uncertainties in the flux calibration. 

To summarize, the need to use hybrid line ratios such as \oiiiuv~$\lambda$1666/\oiii~$\lambda$4363 from UV and optical spectra, where each line is measured using different instruments, makes it essential to align the continuum fluxes using scaling factors. 
Deriving this scaling factor is not trivial, since it is not related to a single cause such as the aperture difference, but could instead be due to a combination of instrumental systematics beyond our control (e.g., flux calibration, vignetting). 
Therefore, we consider the method described above our best avenue in estimating this important flux offset. 
This offset is reported in Table~\ref{tab:lines-opt} for the first 5 galaxies of the CLASSY sample, as an example. The complete table covering all CLASSY galaxies is available on the \href{https://archive.stsci.edu/hlsp/classy}{CLASSY MAST Webpage.}


\section{UV Emission Lines Fits}\label{app:B}
\begin{figure}[b]
\begin{center}
    \includegraphics[width=1\textwidth]{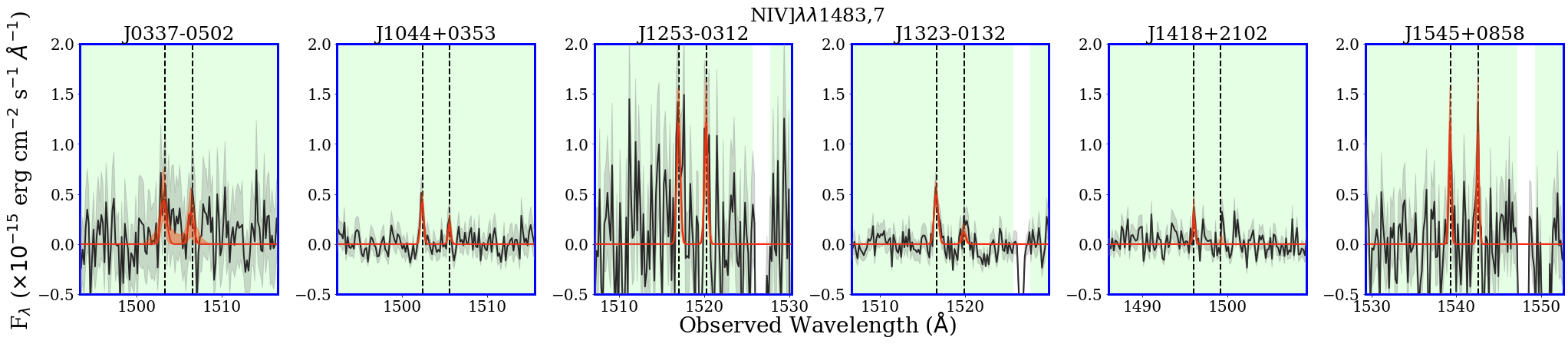}
\end{center}
\caption{Fit of the \niv~$\lambda\lambda$1483,87 emission line doublet, with one of the lines detected with a $S/N>3$ in 6 out of 44 galaxies. The observed flux and the best-fit model are shown in black and red, respectively, while their uncertainties are given by the gray and red shades. The black dashed vertical lines indicate the line positions, taking into account $z_{\rm lit.}$.} 
\label{fig:uvfitniv}
\end{figure}

\begin{figure}
\begin{center}
    \includegraphics[width=0.5\textwidth]{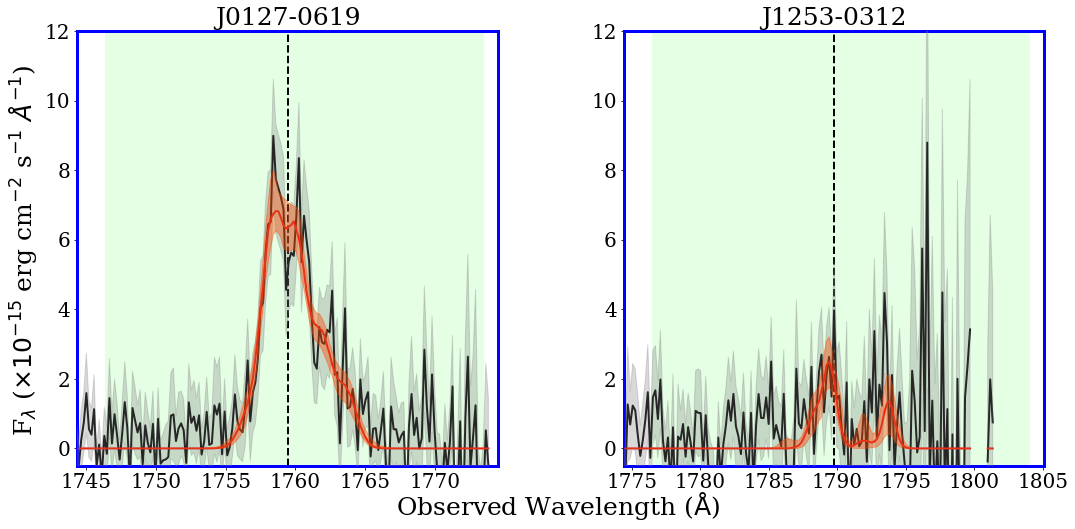}
\end{center}
\caption{Same as Fig.~\ref{fig:uvfitniv} for the \niii~$\lambda$1750 multiplet in the CLASSY survey detected with a $S/N>3$ in 2 out of 44 galaxies. }
\label{fig:uvniii}
\end{figure}

\begin{figure}
\begin{center}
    \includegraphics[width=1\textwidth]{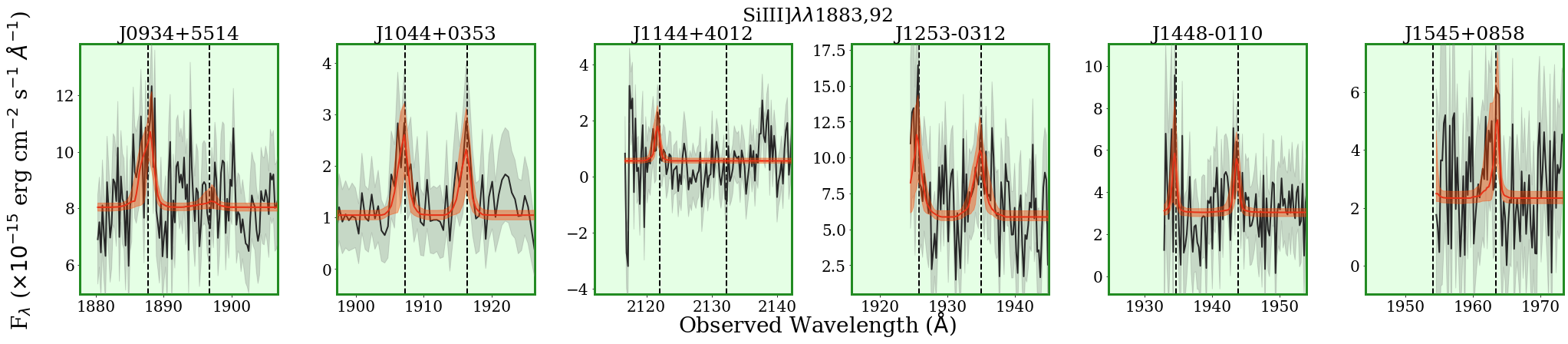}
\end{center}
\caption{Same as Fig.~\ref{fig:uvfitniv} for the \siiii~$\lambda\lambda$1883,92 doublet, with one of the lines detected with a $S/N>3$ in 6 out of 44 galaxies.}
\label{fig:uvfitsiiii}
\end{figure}

In Figures~\ref{fig:uvfitniv}--\ref{fig:uvfitsiiii} we show each of the UV emission line detections (i.e., corresponding to $S/N>3$) for \niv~$\lambda\lambda$1483,87, \heii~$\lambda$1640, \oiiiuv~$\lambda\lambda$1661,6, \niii~$\lambda$1750 and \siiii~$\lambda\lambda$1883,92. 
The analog figures for \civ~$\lambda\lambda$1548,51 and \ciii~$\lambda\lambda$1907,09 are shown in Sec.~\ref{sec:emission-lines-detection} (Fig.~\ref{fig:uvfitcarbon}). 
The observed flux and the best-fit model are shown in black and red, respectively, while their uncertainties are given by the gray and red shades. 
The dashed vertical lines indicate the position of the emission features according to the literature $z$, reported in Table~\ref{tab:classyprop}.
In each figure, the color of the frames indicates which dataset and binning are used for the fit: HR rebinned of 15 pixels spectra in blue, HR rebinned of 30 in cyan, MR rebinned of 6 in dark green, MR rebinned of 12 in light green. 
The dereddened fluxes of the fitted UV lines (expressed in 10$^{-15}$~erg/s/cm$^2$) and the equivalent widths (in \AA) shown in this paper are reported in Tab.~\ref{tab:lines-uv} for the first 5 galaxies of the CLASSY sample, as an example. The complete table covering all CLASSY galaxies is available on the \href{https://archive.stsci.edu/hlsp/classy}{CLASSY MAST Webpage}. The $E(B-V)$ taken into account (see Sec.~\ref{sec:method-dust} and Sec.~\ref{sec:results_ext}) is also reported. The fluxes for undetected lines are given as less than their 3$\sigma$ upper limits. 

\begin{figure}
\begin{center}
    \includegraphics[width=0.8\textwidth]{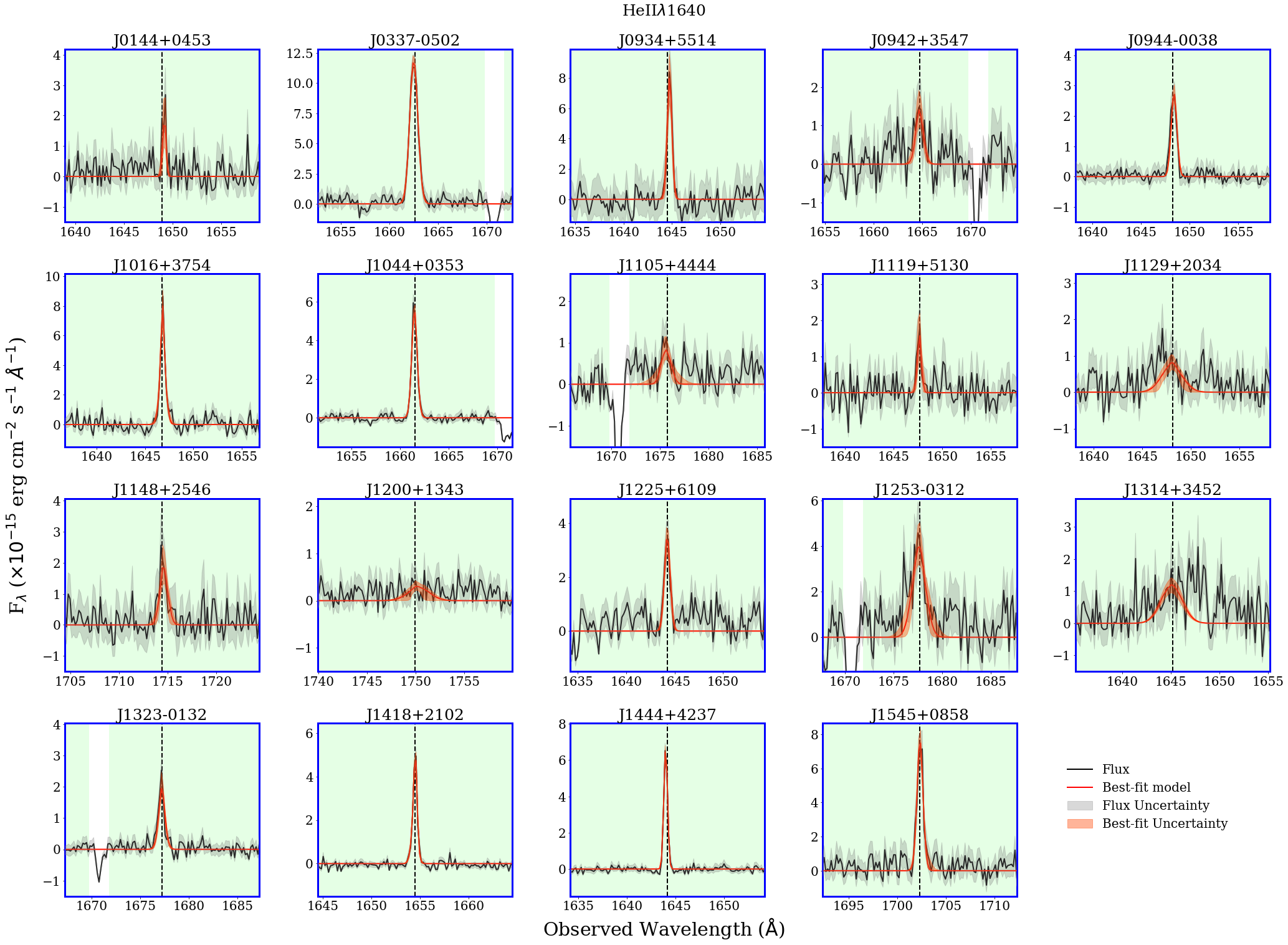}
\end{center}
\caption{Same as Fig.~\ref{fig:uvfitniv} for the \heii~$\lambda$1640 emission line in emission, detected with a $S/N>3$ in 19 out of 44 galaxies.}
\label{fig:uvfitheii}

\begin{center}
    \includegraphics[width=0.8\textwidth]{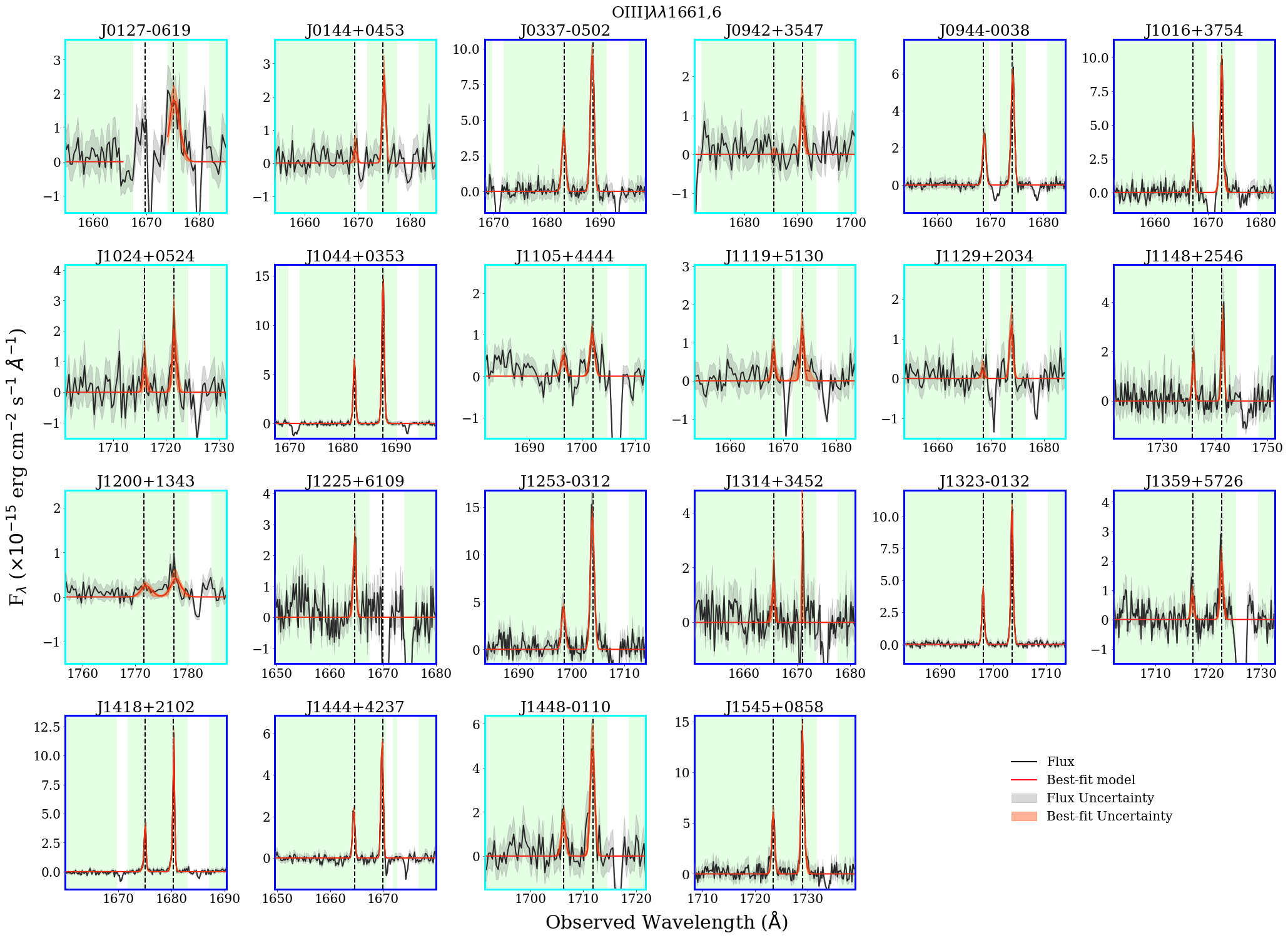}
\end{center}
\caption{Same as Fig.~\ref{fig:uvfitniv} for \oiii~$\lambda\lambda$1661,6 emission line in the CLASSY survey detected with a $S/N>3$ in 22 out of 44 galaxies. }
\label{fig:uvfitoiii}
\end{figure}

\newpage

\begin{deluxetable*}{lccccccc}
\setlength{\tabcolsep}{4pt}
    \tablewidth{0pt}
    \tablecaption{FUV Emission line fluxes from HST/COS for the first 5 galaxies of the CLASSY survey. The complete table covering all CLASSY galaxies is available on the \href{https://archive.stsci.edu/hlsp/classy}{CLASSY MAST Webpage}.}
\tablehead{
Ion                     & J0021+0052 & J0036-3333 & J0127-0619 & J0144+0453 & J0337-0502 }
\startdata
\ion{Si}{2} \W1265.00   & $5.84\pm2.07$ & $144.04\pm18.59$ & \nodata & $0.98\pm0.28$ & \nodata \\ 
\ion{Si}{4} \W1393.76   & \nodata & \nodata & \nodata & \nodata & \nodata \\ 
\ion{O}{4} \W1401.16    & \nodata & \nodata & \nodata & \nodata & $< 1.31$ \\ 
\ion{Si}{4} \W1402.77   & \nodata & \nodata & \nodata & \nodata & $1.23\pm0.39$ \\ 
\ion{O}{4} \W1404.81    & \nodata & \nodata & \nodata & $< 0.42$ & $2.18\pm0.53$ \\ 
\ion{S}{4} \W1406.02    & \nodata & \nodata & \nodata & \nodata & \nodata \\ 
\ion{O}{4} \W1407.38    & \nodata & $< 23.98$ & \nodata & \nodata & $0.98\pm0.32$ \\ 
\ion{O}{4} \W1410.00    & \nodata & $33.76\pm12.23$ & \nodata & \nodata & \nodata \\ 
\ion{S}{4} \W1416.89    & \nodata & \nodata & \nodata & \nodata & \nodata \\ 
\ion{S}{4} \W1423.85    & \nodata & \nodata & \nodata & \nodata & $< 0.60$ \\ 
\ion{N}{4}{]} \W1483.33    & \nodata & \nodata & \nodata & \nodata & $0.89\pm0.34$ \\ 
\ion{N}{4}{]} \W1486.50    & \nodata & \nodata & \nodata & \nodata & \nodata \\ 
\ion{Si}{2}* \W1533.43  & $3.53\pm1.22$ & $21.59\pm6.90$ & $33.66\pm9.75$ & $< 0.80$ & \nodata \\ 
\ion{C}{4} \W1548.19    & \nodata & \nodata & \nodata & $< 0.45$ & $15.99\pm2.20$ \\ 
\ion{C}{4} \W1550.77    & \nodata & \nodata & \nodata & $< 0.51$ & $6.86\pm1.26$ \\ 
\ion{He}{2} \W1640.42   & \nodata & \nodata & $< 51.99$ & $1.27\pm0.37$ & $18.30\pm3.17$ \\ 
\ion{O}{3}{]} \W1660.81    & \nodata & \nodata & \nodata & $< 1.03$ & $5.53\pm1.67$ \\ 
\ion{O}{3}{]} \W1666.15    & \nodata & $< 24.27$ & $161.28\pm35.14$ & $2.56\pm0.61$ & $13.21\pm2.70$ \\ 
{[}\ion{Si}{3}{]} \W1883.00   & \nodata & \nodata & \nodata & \nodata & \nodata \\ 
\ion{Si}{3}{]} \W1892.03   & $< 4.24$ & \nodata & $< 182.77$ & \nodata & \nodata \\ 
\ion{C}{3}{]} \W1906.68    & \nodata & \nodata & $1185.31\pm146.28$ & $8.61\pm2.66$ & $14.73\pm3.40$ \\ 
{[}\ion{C}{3}{]} \W1908.72    & $< 4.04$ & $< 105.54$ & $773.70\pm68.09$ & $6.22\pm1.69$ & $10.46\pm1.55$ \\ 
\hline 
EW(\ion{C}{4})          & \nodata & \nodata & \nodata & \nodata & $2.13\pm0.24$ \\ 
EW(\ion{He}{2})         & \nodata & \nodata & \nodata & $0.63\pm0.18$ & $1.97\pm0.34$ \\ 
EW(\ion{O}{3}{]}\W1666) & \nodata & \nodata & $1.23\pm0.27$ & $1.33\pm0.31$ & $1.49\pm0.31$ \\ 
EW(\ion{C}{3})          & \nodata & \nodata & $20.58\pm1.68$ & $12.44\pm2.61$ & $4.02\pm0.60$ \\
\hline 
\enddata
\tablecomments{The fluxes are expressed in 10$^{-15}$~erg/s/cm$^2$ and are corrected for the dust attenuation using the \citet{reddy16} law. Fluxes for undetected lines are given as less than their 3$\sigma$ upper limits. The EWs are in \AA. }
\label{tab:lines-uv}
\end{deluxetable*}

\section{Optical Emission Lines Fits}\label{app:C}
In Table~\ref{tab:lines-opt} we report the fluxes of the fitted optical lines (expressed in 10$^{-15}$~erg/s/cm$^2$) and the \ha\ equivalent widths (in \AA) for the first 5 galaxies of the CLASSY sample. The complete table covering all CLASSY galaxies is available on the \href{https://archive.stsci.edu/hlsp/classy}{CLASSY MAST Webpage}. The fluxes for undetected lines are given as less than their 3$\sigma$ upper limits. We also report the Telescope and instrument taken into account for each galaxy, and the UV-optical flux offsets calculated and used to correct the optical line fluxes (see Sec.~\ref{app:A}). The $E(B-V)$ taken into account (see Sec.~\ref{sec:method-dust} and Sec.~\ref{sec:results_ext}) is reported in Table~\ref{tab:lines-uv}.
\begin{deluxetable*}{lccccccc}
\setlength{\tabcolsep}{4pt}
    \tablewidth{0pt}
    \tablecaption{Optical emission line fluxes for the first 5 galaxies of the CLASSY survey. The complete table covering all CLASSY galaxies is available on the \href{https://archive.stsci.edu/hlsp/classy}{CLASSY MAST Webpage}.}
\tablehead{
Ion                     & J0021+0052 & J0036-3333 & J0127-0619 & J0144+0453 & J0337-0502 }
\startdata
{[}\ion{O}{2}{]} \W3727.092   & $14.55\pm0.45$ & \nodata & \nodata & \nodata & \nodata \\ 
{[}\ion{O}{2}{]} \W3729.875   & $16.39\pm0.05$ & \nodata & \nodata & \nodata & \nodata \\ 
{[}\ion{S}{2}{]} \W4069.749    & $0.52\pm0.05$ & \nodata & \nodata & \nodata & \nodata \\ 
H$\delta$ \W4102.892   & $5.20\pm0.16$ & \nodata & \nodata & \nodata & \nodata \\ 
H$\gamma$ \W4341.691    & $9.64\pm0.19$ & \nodata & $155.11\pm2.04$ & \nodata & \nodata \\ 
{[}\ion{O}{3}{]} \W4364.436    & $0.50\pm0.10$ & \nodata & \nodata & \nodata & \nodata \\ 
{[}\ion{Fe}{3}{]} \W4659.35    & $0.32\pm0.04$ & $0.61\pm0.00$ & $9.78\pm0.60$ & \nodata & $0.10\pm0.00$ \\ 
\ion{He}{2} \W4687.015    & $0.34\pm0.04$ & $0.25\pm0.01$ & $4.94\pm0.54$ & $< 0.40$ & $1.42\pm0.00$ \\ 
{[}\ion{Fe}{3}{]} \W4702.85    & $< 0.12$ & $0.12\pm0.00$ & $4.80\pm0.48$ & \nodata & $0.03\pm0.00$ \\ 
{[}\ion{Ar}{4}{]} \W4712.69    & \nodata & $0.02\pm0.01$ & \nodata & \nodata & $0.41\pm0.00$ \\ 
{[}\ion{Ar}{4}{]} \W4741.49    & \nodata & \nodata & \nodata & \nodata & $0.34\pm0.00$ \\ 
H$\beta$ \W4862.691        & $22.71\pm0.30$ & $30.04\pm0.01$ & $423.38\pm3.33$ & $< 5.22$ & $38.64\pm0.02$ \\ 
{[}\ion{O}{3}{]} \W5008.240  & $110.54\pm1.00$ & $94.62\pm0.02$ & $1553.65\pm4.44$ & $26.42\pm5.62$ & $133.51\pm0.04$ \\ 
{[}\ion{Cl}{3}{]} \W5519.25    & \nodata & $0.12\pm0.00$ & $1.86\pm0.28$ & \nodata & $0.03\pm0.00$ \\ 
{[}\ion{Cl}{3}{]} \W5539.43    & $< 0.09$ & $0.15\pm0.00$ & $< 1.39$ & \nodata & $0.01\pm0.00$ \\ 
{[}\ion{N}{2}{]} \W5756.240    & \nodata & $0.69\pm0.00$ & \nodata & \nodata & \nodata \\ 
{[}\ion{O}{1}{]} \W6302.046    & $1.13\pm0.06$ & $2.70\pm0.00$ & $18.77\pm0.67$ & \nodata & $0.28\pm0.00$ \\ 
{[}\ion{S}{2}{]} \W6313.8   & $0.29\pm0.05$ & $0.37\pm0.00$ & $9.66\pm1.13$ & \nodata & $0.27\pm0.00$ \\ 
H$\alpha$ \W6564.632        & $71.82\pm0.75$ & $125.23\pm0.03$ & $1934.23\pm19.79$ & $50.76\pm2.07$ & $89.08\pm0.03$ \\ 
{[}\ion{N}{2}{]} \W6585.271    & $3.76\pm0.11$ & $21.81\pm0.01$ & $125.97\pm3.15$ & \nodata & $0.44\pm0.00$ \\ 
{[}\ion{S}{2}{]} \W6718.294    & $4.13\pm0.10$ & $13.85\pm0.01$ & $108.59\pm0.77$ & $2.24\pm0.15$ & $0.79\pm0.00$ \\ 
{[}\ion{S}{2}{]} \W6732.674   & $2.98\pm0.09$ & $9.05\pm0.01$ & $99.22\pm0.71$ & $1.50\pm0.32$ & $0.64\pm0.00$ \\ 
{[}\ion{O}{2}{]} \W7322.01   & $0.58\pm0.06$ & $1.18\pm0.00$ & $24.29\pm1.51$ & \nodata & $0.22\pm0.00$ \\ 
{[}\ion{O}{2}{]} \W7332.0    & $0.45\pm0.07$ & $0.90\pm0.00$ & $20.00\pm1.48$ & \nodata & $0.17\pm0.00$ \\ 
{[}\ion{Ar}{3}{]} \W7137.8    & $1.96\pm0.07$ & $3.77\pm0.00$ & $61.35\pm1.29$ & \nodata & $0.81\pm0.00$ \\ 
{[}\ion{S}{3}{]} \W9071.1   & \nodata & $9.64\pm0.01$ & \nodata & \nodata & $1.28\pm0.00$ \\ 
\hline
EW(H$\alpha$) & $385.94\pm4.06$ & $125.95\pm0.03$ & $512.35\pm5.24$ & $0.00\pm0.00$ & $611.40\pm0.22$ \\ 
Instrument/Telescope & APO/SDSS & VLT/MUSE & VLT/VIMOS & MMT & VLT/MUSE \\ 
UV-Opt offset & $0.69\pm0.05$ & $0.71\pm0.01$ & $0.05\pm0.01$ & $0.73\pm0.02$ & $1.16\pm0.01$ \\ 
\enddata
\tablecomments{The fluxes are expressed in 10$^{-15}$~erg/s/cm$^2$ corrected for dust attenuation using the \citet{cardelli89} law, but not multiplied yet by the UV-optical offset, which is also reported. Fluxes for undetected lines are given as less than their 3$\sigma$ upper limits. The EW(\ha) is expressed in \AA. The last line of each block indicates from which instrument and telescope the data are taken.}
\label{tab:lines-opt}
\end{deluxetable*}

\section{Interstellar medium properties}\label{app:D}
In Table~\ref{tab:ism-prop-calc} we report the densities, temperatures, ionization parameters of the low, high and intermediate ionization zones, as well as the $E(B-V)$ used to correct the optical and UV emission lines, and optical and UV metallicities for the first 5 galaxies of the CLASSY sample. 
\begin{deluxetable*}{lccccccc}
\setlength{\tabcolsep}{4pt}
    \tablewidth{0pt}
    \tablecaption{ISM properties for the first 5 CLASSY galaxies. The complete table covering all CLASSY galaxies is available on the \href{https://archive.stsci.edu/hlsp/classy}{CLASSY MAST Webpage}.}
\tablehead{
                    & J0021+0052 & J0036-3333 & J0127-0619 & J0144+0453 & J0337-0502 }
\startdata
\hline
Low Ionization Zone & & & & &  \\
$n_e$([\ion{S}{2}],[\ion{N}{2}])   & $ 21\pm  2$ & \nodata & $488\pm  1$ & \nodata & $155\pm  1$ \\ 
$n_e$([\ion{O}{2}],[\ion{N}{2}])   & \nodata & \nodata & \nodata & \nodata & \nodata \\ 
$n_e$([\ion{S}{2}],[\ion{S}{2}])   & \nodata & \nodata & $394\pm  1$ & \nodata & $155\pm  1$ \\ 
$n_e$([\ion{O}{2}],[\ion{S}{2}])   & \nodata & \nodata & \nodata & \nodata & \nodata \\ 
$n_e$([\ion{S}{2}],[\ion{O}{2}])   & $ 39\pm  2$ & \nodata & $394\pm  1$ & \nodata & $155\pm  1$ \\ 
$n_e$([\ion{O}{2}],[\ion{O}{2}])   & \nodata & \nodata & \nodata & \nodata & \nodata \\ 
$T_e$([\ion{S}{2}],[\ion{N}{2}])   & $23763\pm140$ & \nodata & \nodata & \nodata & \nodata \\ 
$T_e$([\ion{O}{2}],[\ion{N}{2}])   & \nodata & \nodata & \nodata & \nodata & \nodata \\ 
$T_e$([\ion{S}{2}],[\ion{S}{2}])   & \nodata & \nodata & $15854\pm149$ & \nodata & \nodata \\ 
$T_e$([\ion{O}{2}],[\ion{S}{2}])   & \nodata & \nodata & \nodata & \nodata & \nodata \\ 
$T_e$([\ion{S}{2}],[\ion{O}{2}])   & $12123\pm 28$ & \nodata & $15854\pm149$ & \nodata & \nodata \\ 
$T_e$([\ion{O}{2}],[\ion{O}{2}])   & \nodata & \nodata & \nodata & \nodata & \nodata \\ 
$T_{e, low-G92}$   & $11610\pm11610$ & \nodata & $15854\pm15854$ & \nodata & \nodata \\ 
log$(U)$(S3S2)   & \nodata & $-3.22\pm0.00$ & \nodata & \nodata & $-2.65\pm0.00$ \\ 
\hline
Intermediate Ionization Zone & & & & &  \\
$n_e$([\ion{Cl}{3}],[\ion{S}{3}])   & \nodata & $4119\pm 12$ & $272\pm2019$ & \nodata & \nodata \\ 
$n_e$([\ion{Fe}{3}],[\ion{S}{3}])   & $31819\pm461332$ & \nodata & \nodata & \nodata & $2440\pm507$ \\ 
$n_e$([\ion{Si}{3}],[\ion{S}{3}])   & \nodata & \nodata & \nodata & \nodata & \nodata \\ 
$n_e$([\ion{C}{3}],[\ion{S}{3}])   & \nodata & \nodata & \nodata & $3308\pm4532$ & $4258\pm1067$ \\ 
$T_e$([\ion{Cl}{3}],[\ion{S}{3}])   & \nodata & $11961\pm  1$ & $15854\pm149$ & \nodata & \nodata \\ 
$T_e$([\ion{Fe}{3}],[\ion{S}{3}])   & $11610\pm142$ & \nodata & \nodata & \nodata & $21466\pm  8$ \\ 
$T_e$([\ion{Si}{3}],[\ion{S}{3}])   & \nodata & \nodata & \nodata & \nodata & \nodata \\ 
$T_e$([\ion{C}{3}],[\ion{S}{3}])   & \nodata & \nodata & $15854\pm149$ & \nodata & $21850\pm  9$ \\ 
$T_{e,int-G92}$   & $11610\pm142$ & \nodata & $15854\pm149$ & \nodata & \nodata \\ 
log$(U)$(O3O2)   & $-2.90\pm0.01$ & $-3.08\pm0.00$ & $-3.29\pm0.02$ & $-2.99\pm0.08$ & $-2.38\pm0.00$ \\ 
\hline
High Ionization Zone & & & & &  \\
$n_e$([\ion{S}{2}],[\ion{O}{3}])   & $ 39\pm  1$ & \nodata & $415\pm  1$ & \nodata & $155\pm  1$ \\ 
$n_e$([\ion{S}{2}],[\ion{O}{3}]$_{UV}$)   & \nodata & \nodata & $645\pm  2$ & \nodata & $180\pm  1$ \\ 
$n_e$([\ion{N}{4}],[\ion{O}{3}])   & \nodata & \nodata & \nodata & \nodata & \nodata \\ 
$n_e$([\ion{N}{4}],[\ion{O}{3}]$_{UV}$)   & \nodata & \nodata & \nodata & \nodata & \nodata \\ 
$n_e$([\ion{Ar}{4}],[\ion{O}{3}])   & \nodata & \nodata & \nodata & \nodata & $1107\pm  5$ \\ 
$n_e$([\ion{Ar}{4}],[\ion{O}{3}]$_{UV}$)   & \nodata & \nodata & \nodata & \nodata & $1003\pm  7$ \\ 
$T_e$([\ion{S}{2}],[\ion{O}{3}])   & $12300\pm 12$ & \nodata & \nodata & \nodata & \nodata \\ 
$T_e$([\ion{S}{2}],[\ion{O}{3}]$_{UV}$)   & \nodata & \nodata & $55513\pm849$ & $23172\pm148$ & $19330\pm 63$ \\
$T_e$([\ion{N}{4}],[\ion{O}{3}])   & \nodata & \nodata & \nodata & \nodata & \nodata \\ 
$T_e$([\ion{N}{4}],[\ion{O}{3}]$_{UV}$)   & \nodata & \nodata & \nodata & \nodata & $19330\pm 62$ \\ 
$T_e$([\ion{Ar}{4}],[\ion{O}{3}])   & \nodata & \nodata & \nodata & \nodata & \nodata \\ 
$T_e$([\ion{Ar}{4}],[\ion{O}{3}]$_{UV}$)   & \nodata & \nodata & \nodata & \nodata & $19323\pm 63$ \\ 
log$(U)$(Ar4Ar3)   & $-2.54\pm0.03$ & $-3.06\pm0.12$ & \nodata & \nodata & $-1.57\pm0.01$ \\ 
\hline 
E(B-V)(100 cm$^{-3}$; 10$^{4}$ K)   & $0.12\pm0.02$ & $0.38\pm0.00$ & $0.47\pm0.01$ & $0.04\pm0.03$ & $0.05\pm0.01$ \\ 
E(B-V)   & $0.14\pm0.01$ & $0.38\pm0.00$ & $0.50\pm0.01$ & $0.04\pm0.01$ & $0.05\pm0.00$ \\ 
\hline 
(12+log(O/H))([\ion{S}{2}],[\ion{O}{3}]$_{UV}$)   & \nodata & \nodata & \nodata & \nodata & $7.37\pm0.01$ \\ 
(12+log(O/H))([\ion{Ar}{4}],[\ion{O}{3}])  & \nodata & \nodata & \nodata & \nodata & \nodata \\ 
(12+log(O/H))([\ion{Ar}{4}],[\ion{O}{3}]$_{UV}$)  & \nodata & \nodata & \nodata & \nodata & $7.37\pm0.01$ \\ 
(12+log(O/H))([\ion{N}{4}],[\ion{O}{3}])  & \nodata & \nodata & \nodata & \nodata & \nodata \\ 
(12+log(O/H))([\ion{N}{4}],[\ion{O}{3}]$_{UV}$)  & \nodata & \nodata & \nodata & \nodata & \nodata \\ 
(12+log(O/H))$_{He2}$  & \nodata & $ 8.47$ & $ 8.53$ & $ 8.05$ & $ 7.24$ \\ 
(12+log(O/H))$_{Si3}$  & \nodata & \nodata & \nodata & \nodata & \nodata \\ 
\hline
\enddata
\tablecomments{$n_e$, $T_e$ and log($U$) derived for the different ionization zones, where the first and second species listed in parenthesis indicate the diagnostics used to calculate $n_e$ and $T_e$, respectively (see Sec.~\ref{sec:methods} for details). $T_{e,low-G92}$ and $T_{e_,int-G92}$ are the low and intermediate ionization temperatures derived with the \citet{garnett92} relations, as explained in Sec.~\ref{sec:methods-temp}.
$E(B-V)$ derived assuming $n_e$=100~cm$^{-3}$ and $T_e = 10^4$~K, $E(B-V)$ used to correct the optical and UV emission lines, and the different estimates of 12+log(O/H).}
\label{tab:ism-prop-calc}
\end{deluxetable*}

\end{document}